\documentclass[10pt]{report}
\usepackage[french,english]{babel}
\usepackage{amsmath}
\usepackage{amsfonts}
\usepackage{amssymb}
\usepackage{graphicx}                              
\usepackage{times}
\usepackage{version}
\usepackage{framed,color}
\begin{document}
%
%

November 15th 2017 \hfill 

\vskip 5cm
{\baselineskip 10pt
\begin{center}
{\bf UNLOCKING THE STANDARD MODEL;}
\end{center}
\begin{center}
\bf {$\boldsymbol{N =1\ \text{AND}\ 2}$ GENERATIONS OF QUARKS :
                        SPECTRUM,  MIXING AND SYMMETRIES}
\end{center}
}
\baselineskip 16pt
\arraycolsep 3pt  
%
\vskip .2cm
\centerline{
B.~Machet
     \footnote{Sorbonne Universit\'es, UPMC Univ Paris 06, UMR 7589,
LPTHE, F-75005, Paris, France}
     \footnote{CNRS, UMR 7589, LPTHE, F-75005, Paris, France.}
     \footnote{Postal address:
LPTHE tour 13-14, 4\raise 3pt \hbox{\tiny \`eme} \'etage,
          UPMC Univ Paris 06, BP 126, 4 place Jussieu,
          F-75252 Paris Cedex 05 (France)}
    \footnote{machet@lpthe.jussieu.fr}
     }
\vskip .5cm

{\bf Abstract:} The Glashow-Salam-Weinberg model for N=2 generations is
extended to 8 composite Higgs multiplets by using a one-to-one
correspondence between its complex Higgs
doublet and very specific quadruplets of bilinear quark operators.
This is the minimal number  required to
suitably account, simultaneously,
for the pseudoscalar mesons that can be built  with 4 quarks
and for the masses of the $W$ gauge bosons.
They are used as  input, together with
elementary low energy considerations, from which
all other parameters, masses and couplings can be calculated. We focus in
this work on the spectrum of the 8 Higgs bosons, on the mixing angles, and
on the set of ``horizontal'' and ``vertical'' entangled symmetries that,
within the chiral $U(4)_L \times U(4)_R$ group,
 strongly frame this extension of the Standard Model.
In particular, the $u-c$ ($\theta_u$) and $d-s$ ($\theta_d$) mixing angles
satisfy the robust relation
$\tan(\theta_d+\theta_u)\tan(\theta_d-\theta_u)
= \Big(\frac{1}{m_{K^\pm}^2}-\frac{1}{m_{D^\pm}^2}\Big) \big/
\Big(\frac{1}{m_{\pi^\pm}^2}-\frac{1}{m_{D_s^\pm}^2}\Big)$.
Light scalars (below $90 MeV$) arise and the mass of (at least)
one of the Higgs bosons
grows like that of the heaviest $\bar q\gamma_5 q$ bound state.
$\theta_u$ cannot be safely tuned to zero and several
parameters have no reliable expansion in terms of ``small'' parameters like
$m_\pi$ or the mixing angles.
This study does not call for extra species of fermions.
The effective couplings of scalars, which depend on the non-trivial
normalization of their kinetic terms, can be extremely weak. For the sake
of (relative) brevity, their rich content of non-standard physics
(including astrophysics), the inclusion of the 3rd generation and the taming
of quantum corrections are left for a subsequent work.

\smallskip


\newpage


\tableofcontents
\listoffigures

%
%
%


\chapter{Overview}
\label{chapt:overview}

\section{Introduction} \label{section:intro}

The Higgs boson of the Glashow-Salam-Weinberg (GSW) \cite{GSW} model
may be a fundamental scalar and the only one of this sort. 
There however exist in nature scalar mesons which are most probably
quark-antiquark  composites. It is therefore natural to wonder whether
the Higgs boson could  be such a  particle, that is,
just one member of the family of scalar mesons.  

Previous tentatives led to the introduction of super-heavy quarks
(techniquarks) \cite{Susskind}. This was  thought to be the only solution to the
mismatch between the electroweak scale (the mass of the $W$ or the vacuum
expectation value of the Higgs boson) and the chiral scales $m_\pi, f_\pi$,
that unavoidably led, otherwise, to $m_W \simeq m_\pi$.

We shall follow here an orthogonal way and only interpret this mismatch as
the need for (at least) two different scales, and thus for (at
least) 2 Higgs bosons with vacuum expectation values (VEV's)
 of the order of $m_W$ and $m_\pi$.
Instead of introducing extra fermions, we therefore prefer to introduce
extra scalars, and we do it in a natural way.

In this work, the underlying definition of naturality is the simplest possible: all known particles,
presently mesons and quarks 
\footnote{though quarks are not particles.}
 should be described in agreement with observations,
including ``the'' Higgs boson (presumably the
$125\,GeV$ state discovered at the LHC \cite{LHCHiggs}). All parameters (VEV's, couplings)
should be calculable in terms of ``physical'' quantities (masses of
pseudoscalar mesons, $W$ mass, quark masses). Their number 
we shall reduce from the start as much as possible by simple and sensible
physical arguments,
like the absence of coupling between scalar and pseudoscalar mesons, that of
flavor changing neutral currents (FCNC), the need for 3 true Goldstone
bosons to provide 3 longitudinal degrees of freedom for the $W$'s, the
role of Yukawa couplings to provide soft masses to the other
``Goldstones'' of the broken chiral symmetry {\em etc}.
In this perspective, a model built to extend or complete the Standard Model should:
 first, not get in contradiction with present observations;
secondly,  be able to predict the properties and couplings
 of all states which have not been observed yet.

Along this path,
one is unavoidably led to introduce several Higgs multiplets. The $2N$
quarks of $N$ generations are the building blocks of $(2N)^2$ pseudoscalar
mesons and of $(2N)^2$ scalar mesons. The total of $8N^2$ such composite
states should fit into $2N^2$ quadruplets (or complex doublets). In
particular, for 1 generation, 2 Higgs multiplets are expected. They
involve 4 pseudoscalar mesons (2 neutral and 2 charged),
 2 Higgs bosons (neutral scalars) and 2 charged scalars.\newline
This would only be phraseology without the one-to-one correspondence that
we demonstrate, concerning the transformations by the weak group $SU(2)_L$,
between the complex Higgs doublet of the GSW model and two sets of $N^2$
quadruplets of bilinear quark operators. The first set is made of quadruplets of
the type $(\bar q_i q_j, \overrightarrow{\bar q_i\gamma_5 q_j})$ that is, one
scalar and 3 pseudoscalars, and the second set of quadruplets of the type
$(\bar q_i\gamma_5 q_j, \overrightarrow{\bar q_i q_j})$.
All quadruplets transform alike by $SU(2)_L$ and each one includes both
parities.  The two sets are parity-transformed of each other.

Each quadruplet has to be normalized. The normalization factors must in
particular make the transition between bilinear quark operators of
dimension $[mass]^3$ and bosonic fields of dimension $[mass]$. Since each
quadruplet includes one scalar $\bar q q$ operator with $<\bar q q>=\mu^3
\not=0$, its natural normalization is realized through the factor
\begin{equation}
\frac {v}{\sqrt{2}\mu^3}.
\label{eq:norm1}
\end{equation}
 In this way, the corresponding Higgs boson 
$\frac {v}{\sqrt{2}\mu^3}\bar q q$ gets a ``bosonic'' VEV
$\frac{v}{\sqrt{2}}$ reminiscent of the GSW model.
One normalizes all 4 elements of the same quadruplet by the same factor.
Thus,
to each quadruplet will be accordingly associated one ``$v$'' and one
``$\mu^3$'', which makes, for 2 generations,
 a total of 8 ``bosonic'' VEV's and 8 ``fermionic'' VEV's.  We shall
suppose in the following
 that $<\bar q_i \gamma_5 q_j>=0$ and that $<\bar q_i q_j> =
<\bar q_j q_i>$, though both statements can only be approximations in a
theory that violates parity and also, eventually, $C$ and $CP$.

The dual nature of the components of the Higgs quadruplets will
be extensively used. The simple example below shows the principle of the method.
Let a given quadruplet $X$ (see (\ref{eq:Xq2gen})) include the charged pseudoscalar
 bilinear quark operator (after the normalization explained above has been
implemented) 
\begin{equation}
X^+= \frac{v_X}{\sqrt{2}\mu^3}\frac{1}{\sqrt{2}}(2\bar u \gamma_5 d),
\label{eq:example1}
\end{equation}
 and, at the same time, a
``Higgs boson'' (scalar with non-vanishing VEV) $X^0=
\frac{v_X}{\sqrt{2}\mu^3}\frac{\bar u u+\bar d
d}{\sqrt{2}}$ which has a VEV $\frac{<\bar u u+\bar d d>}{\sqrt{2}}=\mu^3$.
After the mixing of $d$ and $s$ quarks has been accounted for,
$X^+$  can be expressed in terms of quark mass states, which yields
$X^+= \frac{v_X}{\sqrt{2}\mu^3}\sqrt{2}
(\cos\theta_c \bar u_m\gamma_5 d_m + \sin\theta_c \bar
u_m\gamma_5 s_m)$. Now, PCAC \cite{Dashen} \cite{Lee} for the $\pi^\pm$ mesons yields
$i(m_u+m_d)\bar u_m\gamma_5 d_m = \sqrt{2} f_\pi m_{\pi^+}^2 \pi^+$, in
which $\pi^+$  is the charged pion (mesonic) interpolating
 field with dimension $[mass]$.
This makes that, at least at low energy, one can also write
$X^+ = \frac{v_X}{\sqrt{2}\mu^3}\sqrt{2}\left(
\cos\theta_c \frac{-i\sqrt{2} f_\pi m_{\pi^+}^2}{m_u+m_d}\pi^+ +\ldots
\right)$, which therefore now appears as a bosonic field like the
components of the Higgs doublets of the GSW model. Eventually, one
can also use the Gell-Mann-Oakes-Renner (GMOR) relation \cite{GMOR}
\cite{dAFFR}
$(m_u+m_d)<\bar u_m u_m+\bar d_m d_m> = 2f_\pi^2 m_{\pi^+}^2$
to relate $\mu^3$ to pionic parameters, which leads finally to
\begin{equation}
[X^+] = -i\frac{v_X}{f_\pi}(\cos\theta_c \pi^+ +\ldots).
\label{eq:example2}
\end{equation}
In (\ref{eq:example2}) we used the notation $[X^+]$ for $X^+$ when it is
expressed in terms of a bosonic field. This notation we shall use
throughout the paper: for any Higgs multiplet, $\Delta_i$ stands for its
expression in terms of bilinear quark operators, and $[\Delta_i]$ stands
for its ``bosonic'' form.

The calculations have been performed for N=1 and N=2 generations of quarks.
It turns out that no simple argument or general principle could have
 anticipated the results, though they can be understood in simple terms {\it a
posteriori}.
It is also evident that a suitable solution cannot exist
with a number of Higgs multiplets smaller that the one that we have
introduced because, in particular, it could not fit observed pseudoscalar
mesons. In this respect, the extension that we propose for the GSW model is
minimal. 

Two among its main features  are the following:\newline
* it is very fine-tuned;\newline
* the $d-s$ and $u-c$ mixing angles $\theta_d$ and $\theta_u$
 are independent parameters and, though $\theta_u \approx \sqrt{m_u/m_c}
\ll1$, it cannot be turned safely to $0$.

\section{Main results for N=1 generation} \label{section:sum1gen}

Only  2 quarks ($u,d$) are present. They build up 4 pseudoscalar mesons
(the 3 pions and the  $\eta$ flavor singlet) and 4 scalars. The latter include 2
neutral states which are 2 Higgs bosons, and 2 charged scalars.
These 8 states fit into 2 quadruplets.

The $W$ and $\pi$ masses are inputs.  The 3 longitudinal $W_\parallel$'s
are built in this case from the 2 charged scalars and from the neutral
pseudoscalar singlet; the three of them  accordingly disappear from the physical
spectrum.

The 2 Higgs bosons have respective masses $\sqrt{2} m_\pi\approx 197\,MeV$
 and
$m_\pi\frac{f_\pi}{\sqrt{2}m_W/g}\approx 69\,KeV$.
They have the same ratio as the corresponding  VEV's, respectively
$\frac{f_\pi}{\sqrt{2}}$ and $\frac{\sqrt{2}m_W}{g}$,  exhibiting a
large hierarchy $\approx 2858$.

PCAC provides the usual correspondence between pseudoscalar
bilinear and pions and their leptonic decays through $W$'s
are suitably described.

Very light scalars start to spring out.
For one generation there is only one such particle.

While $<\bar u u+\bar d d>$ is determined by the GMOR relation, one gets
\footnote{We use the same notations as in the bulk of the paper.}
$\hat r_X\equiv\frac{<\bar u u-\bar d d>}{<\bar u u+\bar d d>}= -\frac12
\frac{m_u+m_d}{m_u-m_d}$, which already points out at a negative $d$ quark
mass parameter $m_d=-|m_d|$ as will be confirmed for 2 generations.
Then, $\hat r_X=\frac12 \frac{|m_d|-m_u}{|m_d|+m_u}\approx \frac16$.

The price to pay for this drastic truncation  of the physical world
 is threefold:\newline
* a very large hierarchy between the 2 bosonic VEV's;\newline
* a  very small mass $\sqrt{2}m_\pi$ for the heaviest Higgs boson  which
 cannot compete with the expected $125\,GeV$ \cite{LHCHiggs}
 and could naively look
 like the revival of the  mismatch between $m_W$ and $m_\pi$ that led to
technicolor models;\newline
* the disappearance of the singlet $\eta$ pseudoscalar meson in favor of the
 neutral longitudinal $W^3_\parallel$.

These issues  get on their way to a solution when one
increases by 1 the number of generations. In particular,
the mass of the ``quasi-standard'' Higgs boson
becomes comparable to that of the heaviest
pseudoscalar meson, $D_s$ instead of $\pi$.

It is therefore reasonable to think that extending this approach to
 N=3 generations can bring a suitable agreement with the observed world.
Unfortunately, the number of equations  and constraints to fulfill
increases so dramatically that a convenient method to solve them could not
 be found, yet.

\section{Main results for N=2 generations} \label{section:sum2gen}

4 quarks ($u,c,d,s$) are now involved, which build up 32 $\bar q_i q_j$ and
$\bar q_i \gamma_5 q_j$ composite states. These fit into $2N^2=8$ quadruplets.
There are therefore in particular 8 Higgs bosons.

There are 2 mixing angles: $\theta_u$ describes the mixing between $u$ and
$c$ flavor eigenstates while $\theta_d$ concerns $d$ and $s$.
Flavor and gauge symmetries are tightly entangled in this extension and the
freedom to tune $\theta_u$ to $0$ by a flavor rotation no longer exists.
As a consequence, the Cabibbo
angle $\theta_c \equiv \theta_d-\theta_u$ cannot describe alone correctly 
the physics under concern.
These features are exhibited by  studying successively the case when one
approximates $\theta_u$ to $0$ and the one when both $\theta_d\not=0$ and
$\theta_u \not=0$.

\subsection{The case $\boldsymbol{\theta_d\not=0, \theta_u=0}$}

The $2N^2=8$ Higgs bosons  split into 1 triplet, 2 doublets and 1 singlet.
Inside each of these, they  are close to degeneracy.
 3 have masses $\approx \sqrt{2}m_{D_s}$, more precisely
$2.79\,GeV, 2.796\,GeV, 2.80\,GeV$,
2 have intermediate masses $1.23\,GeV, 1.26\,GeV$,
1 has a very small mass $19\,MeV$ and the last two only get massive by
quantum corrections.

The hierarchies between VEV's stay below
$151$ (instead of $2858$ for 1 generation).

The situation has substantially improved  with respect to 1 generation; indeed,
the masses of
the quasi-standard Higgs boson(s) suitably increase and depart from
$m_\pi$ and the hierarchies between VEV's decrease  towards more reasonable values.

$\theta_d$, that one identifies with the Cabibbo angle is expressed
by the 2 formul{\ae}
\begin{equation}
\begin{split}
\tan^2\theta_d
&=\displaystyle\frac{1/m_{K^+}^2 -1/m_{D^+}^2}{1/m_{\pi^+}^2
-1/m_{D_s}^2}
\approx \frac{m_{\pi^+}^2}{m_{K^+}^2}\Big(
1-\frac{m_{K^+}^2}{m_{D^+}^2}+\frac{m_{\pi^+}^2}{m_{D_s}^2}
\Big)+{\cal O}\left(\Big(\frac{m_\pi^2}{m_{K,D,D_s}^2}\Big)^2\right),\cr
& \cr
\tan^2\theta_d &=\frac{|m_d|+m_u}{m_s-m_u} \approx \frac{|m_d|}{m_s},
\quad m_d=-|m_d|<0.
\end{split}
\label{eq:cabb1}
\end{equation}
The first equation in (\ref{eq:cabb1}), yields
\begin{equation}
\theta_d \approx .26685,
\label{eq:tetadnum0}
\end{equation}
$15\,\%$ off the experimental value of the Cabibbo angle
\begin{equation}
\theta_c^{exp} \approx .22759 \simeq \sqrt{\frac{|m_d|}{m_s}}. 
\label{eq:numtetac}
\end{equation}

A negative sign for $m_d$ is needed, like for 1 generation.
 Since $|m_d|>m_u$, it yields,  by the GMOR relation, a negative
sign for $<\bar u u+\bar d d>=2f_\pi^2 m_\pi^2/(m_u+m_d)$.

One however still faces problematic issues :

$\bullet$\ the nice description of $\pi^+$ leptonic decays that we had
found for 1 generation  gets totally spoiled;  the situation
could only be improved if $b_\Omega= (v_\Omega/\hat v_H)^2$ was very small
\footnote{$v_\Omega$ and $\hat v_H$ are the vacuum expectation values of
the neutral scalars of the quadruplets $\Omega$ and $\hat H$ respectively,
which are defined in section \ref{section:quadruplets}; see also subsection
\ref{subsec:notations}};

$\bullet$\ taking the masses of the charged pseudoscalar mesons as inputs,
the mass of the neutral $K$ mesons is off by  $140\,MeV$, unless
one goes to a  value larger than $1$ for $\hat b_X \equiv
(\hat v_X/\hat v_H)^2$,  in conflict with most needed orthogonality relations 
\footnote{$\hat v_X$ is the vacuum expectation value of
the neutral scalar of the quadruplet $\hat X$,
which is defined in section \ref{section:quadruplets}; see also subsection
\ref{subsec:notations}};

$\bullet$\ defining the interpolating field of the $\eta$ meson as 
proportional to $\bar u\gamma_5 u + \bar d\gamma_5 d$, the latter
cannot be set orthogonal to $K^0+\bar K^0$ (actually, we do not get too
worried by this problem because of the  mixing between
neutral pseudoscalars);

$\bullet$\ 2  ratios of bosonic VEV's,  $b_\Omega$ and
$\hat b_\Omega = (\hat v_\Omega/\hat v_H)^2$
\footnote{$\hat v_\Omega$ is the vacuum expectation value of
the neutral scalar of the quadruplet $\hat \Omega$,
which is defined in section \ref{section:quadruplets}; see also subsection
\ref{subsec:notations}}
 come out too large
to match intuitive arguments concerning (non-diagonal) quark condensates;

$\bullet$\ the last problem concerns mixing, and  proves later to be
correlated with the previous one.
On one side eqs.~(\ref{eq:cabb1}) give fairly good estimates of the mixing
angle; the result is independent of the so-called $b$ parameters (ratios of bosonic
VEV's) and looks robust.
On another side, Yukawa couplings provide diagonal and non-diagonal
mass terms for the $d$ and $s$ quarks:  with intuitive notations
\begin{equation}
\tan 2\theta_c = -\frac{2\mu_{ds}}{\mu_d-\mu_s},
\label{eq:cabb2}
\end{equation}
 in which $\mu_{ds}, \mu_d, \mu_s$
depend on the $b$ parameters through the normalizing
coefficients (\ref{eq:norm1}) of the 8 Higgs quadruplets.
The paradox is that, at the values of the $b$
parameters which fit all other data, in particular pseudoscalar meson
masses, $\mu_{ds}$ comes very close to a pole, like if the ``fermionic mixing
angle''  was close to maximal ($\pi/4$).\newline
So, either the quark mixing  exhibits a dual nature (maximal mixing or
close to maximal  only concerns leptons at present), or one must find a way
out of this paradox. It would be feasible at very small values of the
parameters $b_\Omega$ and $\hat b_\Omega$, which seems
excluded at $\theta_u=0$.

\subsection{The case $\boldsymbol{\theta_d\not=0,\theta_u\not =0}$}

The first equation in (\ref{eq:cabb1}) is only the approximation at
$\theta_u=0$ of the exact formula
\begin{equation}
\boxed{
\tan(\theta_d+\theta_u)\tan(\theta_d-\theta_u)
=\displaystyle\frac
{\displaystyle\frac{1}{m_{K^\pm}^2}-\displaystyle\frac{1}{m_{D^\pm}^2}}
{\displaystyle\frac{1}{m_{\pi^\pm}^2}-\displaystyle\frac{1}{m_{D_s^\pm}^2}},
}
\label{eq:cab0}
\end{equation}
which shows that $\theta_d$ and $\theta_u$ cannot be dealt with
independently. Using the experimental value of
Cabibbo angle (\ref{eq:numtetac}), (\ref{eq:cab0})
yields
\begin{equation}
\boxed{
\theta_u \approx .04225,
\quad \theta_d \approx .2698.}
\label{eq:tetaudnum}
\end{equation}
The values that we find for the mixing angles
 correspond to a good approximation to
\footnote{We take $m_u=2.5\,MeV, |m_d|= 5\,MeV, m_s=100\,MeV,
m_c=1.2755\,GeV$.}
$\theta_d\; (\approx \theta_d-\theta_u) \sim \sqrt{|m_d|/m_s}\approx .2236$
 and $\theta_u \sim \sqrt{m_u/m_c} \approx .044$
\footnote{Among  first attempts to calculate the Cabibbo
angle are the ones by Oakes \cite{Oakes} and by Weinberg \cite{Weinberg}.
Since then, it has been  a most sought for goal
of calculating the mixing angles from basic principle (see for example
\cite{Fritzsch} and \cite{WilczekZee} 
in which specific hypotheses are made concerning the symmetries involved
and/or the mass
matrices, and the estimate for $\theta_d-\theta_u$ in \cite{MachetPetcov}
based on the sole existence of mass hierarchies among
quarks).\label{foot:oldmix}}.

Despite its very small value,  switching on $\theta_u$ has very important
consequences, for example on the values of the $b$ parameters,
and brings a very good agreement between the model and
the basis of meson physics :

$\bullet$\ leptonic decays of $\pi^+$ and $K^+$ are  well described;

$\bullet$\ the parameters $b_\Omega$ and $\hat b_\Omega$ become very small
 which opens the way to a matching between bosonic and fermionic  mixing;

$\bullet$\ the masses of the neutral pion and kaon are now well accounted for,
and the $D^0$ is only off by $20\,MeV$.

The spectrum of the Higgs bosons is  changed to
$m_{\hat H^3} \approx 3.24\,GeV, m_{H^0} \approx 1.65\,GeV,
m_{X^0} \approx 3.24\,GeV, m_{\Omega^0}\approx 86\,MeV$.

$\Xi^0$ and $\hat \Xi^3$ are still expected to be very light, and so does
$\hat \Omega^3$ because $\hat b_\Omega$  is expected to be of the same
order of magnitude as $b_\Omega$.
$\hat b_X ={\cal O}(1)$ is preferred, though it is difficult to give yet a
precise value. The positive improvement is that it does not need any longer
to be larger than $1$.

All parameters are very fine tuned. The importance of the small $\theta_u$
is just one among the symptoms of this; one
often deals with rapidly varying functions which furthermore have poles,
 parameters that have no trustable expansions at the chiral limit,
``unlucky'' coincidences {\em etc}.
It is probably the price to pay for naturalness as we define it: it is indeed very
unlikely that some  general principle or god-given symmetry
miraculously tunes the  values of physical observables
up to many digits after the decimal dot. Nature is obviously fine tuned and a
model that pretends to describe it accurately
 has many chances to be fine-tuned, too.

\section{Principle of the method}\label{section:principle}

One works at two levels,  bosonic and fermionic.

$\bullet$\ Bosonic considerations rely on few statements.

$\star$\  The mass of the $W$ gauge bosons, which, in this framework, comes
 from the VEV's of several Higgs bosons, is known.

$\star$\  The masses of all charged pseudoscalar mesons is also known with high
 precision. One should be more careful about some neutral pseudoscalars
that can mix and the definition of which in terms of quark bilinears can be
unclear.

$\star$\  The effective Higgs potential to be minimized is built from the genuine
scalar potential, suitably chosen, to which is added the bosonised form of the
Yukawa couplings. Its minima are constrained to occur at the set
of bosonic VEV's $v/\sqrt{2}$'s.

$\star$\  The VEV's are supposed to be real and, therefore, there squares to be
  positive.

$\star$\ Among the components of the Higgs 8 quadruplets:\newline
\quad - there must exist 3 true Goldstone bosons related to the breaking of the local
 $SU(2)_L$;\newline
\quad - all other scalar and pseudoscalar fields that do not have non-vanishing
 VEV's are pseudo-Goldstone bosons that get ``soft'' masses via the Yukawa
couplings at the same time as quarks get massive. This restricts and
simplifies the scalar potential.

$\star$\ The $mass^2$ of the known pseudoscalar mesons will  be
calculated as the ratios of the corresponding quadratic terms in the
bosonised Yukawa Lagrangian and in the kinetic terms. They depend on the
VEV's,  on the mixing angle(s), and of course on the set of Yukawa couplings.
Their number  is reduced by a suitable and motivated choice for the
Yukawa potential.

$\star$\ Additional relations among Yukawa couplings arise from various sets of
constraints:\newline
\quad -  no transition should occur between scalar and pseudoscalar
 states;\newline
\quad -  likewise,  no transition should occur between charged
 pseudoscalar mesons;\newline
\quad -  similar orthogonality relations are explored among neutral pseudoscalars
 and, for 2 generations, most of them (but not all of them) can be satisfied.

$\bullet$\ Fermionic considerations use the genuine (not bosonised)
form of the Yukawa
Lagrangian, which  provides mass terms for the 4 quarks, both diagonal and
non-diagonal. We mainly use them at $\theta_u=0$.

$\star$\  A first set of constraints comes when turning to $0$ the mixing
between the $u$ and $c$ quarks; then the mixing angle $\theta_d$
 becomes the Cabibbo angle $\theta_c$;\newline
$\star$\  A second set of constraints comes from requirements of reality for the
quarks masses;\newline
$\star$\  A third set of constraints comes when studying the $s$ quark at 
the chiral limit $m_u, m_d\to 0$.

All these constraints are checked by evaluating the masses of pseudoscalar
mesons, the leptonic decays of charged pseudoscalars \ldots and used to
predict unknown quantities in particular the masses of the scalar mesons
(Higgs bosons).

\section{Contents}\label{section:contents}

$\bullet$\ {\bf Chapter \ref{chapt:gener}} is dedicated to general
considerations.

*\ Section \ref{section:bijection} establishes, in the general case of $N$
generations, the basic one-to-one relation between
the complex Higgs doublet of the Glashow-Salam-Weinberg model and $2N^2$ very
specific quadruplets of bilinear $\bar q_i q_j$ and $\bar q_i\gamma_5 q_j$
quark operators including either 1 scalar and 3 pseudoscalars, or 1
pseudoscalar and 3 scalars.
To this purpose, the group $SU(2)_L$ of weak interactions is trivially
 embedded  into the chiral group $U(2N)_L \times U(2N)_R$.

The normalization of the quadruplets
is then explained, which introduces $2N^2$ ``bosonic''
VEV's of the form $v/\sqrt{2}$ and $2N^2$ ``fermionic'' VEV's which are
$<\bar q q>$ condensates.

The connection is made between parity and the 2 generators  ${\mathbb I}_L$
or ${\mathbb I}_R$.

*\ Section \ref{section:genyuk} presents general
considerations concerning the Yukawa couplings ${\cal L}_{Yuk}$.
Arguments will be given concerning how and why they can be simplified.
They are chosen as the most straightforward generalization
 to $N$ generations of the most
general Yukawa couplings for 1 generation, in which 2 quarks are coupled
to 2 Higgs doublets.
Yukawa couplings are  no longer passive in determining the VEV's of the Higgs bosons.
This leads to introducing their bosonised form. Subtracting it from
the scalar potential yields an effective potential that can be used to find
the (bosonic) VEV's of the Higgs bosons. A first set of constraints arises from
 the condition that no transitions should occur between
scalars and pseudoscalar mesons.

*\ Section \ref{section:genpot} presents and motivates
 our simple choice for the Higgs potential. The minimization of
the corresponding effective potential (see above)  leads to another set of
relations between its parameters, Yukawa couplings and bosonic VEV's.
Goldstones and pseudo-Goldstones are investigated, in relation with the
concerned broken symmetries. The (soft) masses of the pseudo-Goldstones
can be calculated from the bosonised form of the Yukawa Lagrangian.

*\ Section \ref{section:higgsmass} gives general formul{\ae} for the masses of the
Higgs bosons.

$\bullet$\ {\bf Chapter \ref{chapt:1gen}} deals with the simplest case
of 1 generation.
 Using as input the masses of the $W$'s, pions, $u$ and $d$
quarks, bosonic and fermionic equations are solved which yield the spectrum of the 2
 Higgs bosons, the values of the 4 VEV's and all couplings.
The leptonic decays of $\pi^+$ are shown to be in agreement with the usual PCAC
estimate.

$\bullet$\ {\bf Chapter \ref{chapt:2gengen}} gives general results in the case
of 2 generations.

*\ Section \ref{section:quadruplets} displays the 8 Higgs
quadruplets.
The choice of the quadruplet that contains
the 3 Goldstones of the spontaneously broken $SU(2)_L$ is motivated.
Notations that will be used throughout the paper are given.

*\ Sections \ref{section:gaugemass} and \ref{section:yukscal}
 give generalities concerning
the kinetic terms,  Yukawa couplings and the Higgs potential.
The mass of the $W$'s  is expressed in terms of the bosonic VEV's.

*\ Section \ref{section:quadsym} identifies the group of transformations that
moves inside the space of quadruplets. Its generators commute with the ones
of the gauge group.

*\ Section \ref{section:charged} is devoted to charged pseudoscalar mesons. Their
masses and orthogonality relations are explicitly written.

*\ Section \ref{section:mix} shows how the formula
(\ref{eq:cab0}) relating $\theta_d$, $\theta_u$ and the
masses of charged pseudoscalar mesons is obtained very simply.
It does not depend on low energy
theorems like GMOR (which are  badly verified for heavy mesons) and
only relies on the statement that $\bar
u_m\gamma_5 d_m, \bar u_m\gamma_5 s_m, \bar c_m\gamma_5 d_m, \bar
c_m\gamma_5 s_m$ are proportional to the interpolating fields of,
respectively $\pi^+, K^+, D^+, D_s^+$. There is no need to know the
proportionality constants, which makes this result specially robust.

*\ Section \ref{section:neutralgen} gives the general formul{\ae}
 for the masses and orthogonality conditions for 
$\pi^0, \eta, K^0$ and $ D^0$.

$\bullet$\ {\bf Chapter \ref{chapt:2gentetaunul}} deals with the
approximation $\theta_u=0$, keeping of course $\theta_d=\theta_c \not=0$.

*\ In section \ref{section:cab}, the value of the Cabibbo angle $\theta_c$
 is extracted from the general formula (\ref{eq:cab0}).
Its value falls within $15\%$ of the experimental $\theta_c^{exp}$.

*\ In section \ref{section:4bs}, we write a basic set of equations
that will determine the $b$ ratios of bosonic VEV's, and
we show how, from the sole spectrum of charged pseudoscalar mesons,
one already gets a lower bound for the mass of the ``quasi-standard'' Higgs
boson $m_{\hat H^3} \geq \sqrt{2}m_{D_s}$.

*\ Section \ref{section:neutralpseudos} is dedicated to
neutral pseudoscalar mesons and to the constraints given by their masses
and orthogonality. We find a tension concerning $\hat b_X=(<\hat
X^3>/<X^0>)^2$, because its value obtained from the mesonic mass spectrum
is slightly larger than 1 is in contradiction with  orthogonality relations
of $K^0$ to $\bar K^0$, and of $\pi^0$ to $K^0+\bar K^0$.

*\ Section \ref{section:chargedscal} studies  charged scalar
mesons. Their orthogonality relations cannot be satisfied unless they
align with flavor eigenstates. This is not surprising since the two of them
which coincide with the charged Goldstone bosons of the broken $SU(2)_L$
gauge symmetry are by construction flavor eigenstates.

*\ Section \ref{section:sumbos} summarizes all the bosonic constraints.

*\ In section \ref{section:neutmass}, we study the masses of $\pi^0, K^0$
and $D^0$. In particular,  $\hat b_X >1$ is needed to correctly
account  for the mass of $K^0$; the latter is otherwise off by $140\,MeV$.

The next 3 sections deal with fermionic constraints.

*\ Section \ref{section:genferm} lists  the equations coming from
the definition of quark masses in terms of Yukawa couplings and VEV's.
Additional constraints are given by  using the freedom (as we already did
for bosons) to turn $\theta_u$ to $0$.

*\ Section \ref{section:reality} displays the constraints coming from
the reality of the quark masses. Among the outcomes are: - the knowledge of 
$|\hat r_H|=\left\vert\frac{<\bar c c-\bar s s>}{<\bar u u+ \bar d
d>}\right\vert$; - the expression of
$\tan^2\theta_c$ in terms of quark masses as given by the second line
 of (\ref{eq:cabb1}), which requires in particular $m_d<0$.

*\ In section \ref{section:fermix} we calculate  the
``fermionic mixing angle'' from Yukawa couplings and show that it tends to
be maximal, in contrast with the small value of the Cabibbo angle.

*\ Section \ref{section:chilim} studies $m_s$ at the chiral
limit $m_u, m_d \to 0$. This determines in particular the sign of $\hat
r_H$ and the value of the mass of the ``quasi-standard'' Higgs bosons $\hat
H^3$.

*\ Section \ref{section:solution} summarizes the solution of all equations.
It is shown how $\Xi^0$ and $\hat\Xi^3$, which are classically massless,
are expected to get soft masses from quantum corrections. Hierarchies
between VEV's are shown to be much smaller than for 1 generation.

*\ Section \ref{section:piKlept1} is dedicated to leptonic decays of
$\pi^+$ and $K^+$. We show that they cannot be suitably described for
$\theta_d\not=0$ and $\theta_u=0$. The situation is therefore, at the
moment, worse than for 1 generation.

*\ Section \ref{section:endtetaunul} is a brief summary of the case
$\theta_u=0$, mainly pointing at the problems that arise.

$\bullet$\ {\bf Chapter \ref{chapt:2gentetau}} concerns the general case
$\theta_u\not=0$ and $\theta_d \not=0$.

*\ In section \ref{section:valtetau}, we use the experimental value of the
Cabibbo angle to calculate $\theta_u$, which is very close to
$\sqrt{m_u/m_c}$.

*\ In section \ref{section:newpiKlept} we re-analyze the leptonic decays of
the charged pions and kaons. Unlike at $\theta_u=0$, a nice
agreement can be obtained. Then, the parameters of the model are
updated, a large set of them  being very sensitive to $\theta_u$. In particular,
$b_\Omega$, and presumably $\hat b_\Omega$ too, are now very small.

*\ In section \ref{section:newneutral} we re-investigate  the masses of
$\pi^0, K^0$ and $D^0$. We find that, even for $\hat b_X\leq 1$, they can
now be quite well accounted for. The  $D^0$ is the worst, but its mass
is only off by $20\,MeV$. One however needs a fairly large value of the $<\bar c
c>$ condensate, which coincides with what we already suspected at
$\theta_u=0$ namely that, in one way or another, ``some'' heavy quark should
have a large condensate.

*\ In section \ref{section:higgsspectrum}, we update the Higgs spectrum.
The masses of the 3 heaviest Higgs bosons $\hat H^3, X^0, \hat X^3$
 has increased to $2.9-3.2\,GeV$,
$H^0$ has risen to an intermediate mass of $1.65\,GeV$ and the 4 others
are light (they should not exceed $90\,MeV$).

*\ Section \ref{section:conclus2}  concludes  the case $\theta_d\not=0,
\theta_u \not=0$. The tensions that occurred at $\theta_u=0$ have been
mostly removed or on their way to be (like the paradox of the fermionic ``maximal
mixing''). Including the 3rd generation
of quarks is of course highly wished for, but goes technically largely beyond the
limits of this work.
We emphasize the impressive ability of this multi-Higgs model to
account for the physics of both the broken weak symmetry and that of
mesons. We also largely comment on the very fined-tuned character of all
the physical outputs. Their sensitivity to the small $\theta_u$ is just one
example among a list of parameters which have no trustable expansions at
the chiral limit, or at the limit of small mixing angles.

$\bullet$\ {\bf Chapter \ref{chapt:conclusion}} is  a general conclusion,
mostly focused on symmetries

*\ In section \ref{section:symmetries}
we study how the chiral/gauge group $SU(2)_L \times SU(2)_R$
acts inside each quadruplet, which shows that the third generator $T^3$ of
the custodial $SU(2)$ is identical to the electric charge
\footnote{This had already been noticed in \cite{Machet0}.}. 

We then study which generators of the diagonal $U(4)$ annihilate the Higgs
states, which provides the properties of invariance of the vacuum.

Next, we show that this extension of the GSW model can be a right-handed
 gauge $SU(2)_R$ theory as well as a left-handed $SU(2)_L$ one, and that it
is in principle ready to be a left-right gauge symmetry. This requires 6
true Goldstone bosons, which cannot be achieved with only 2 generations
because some states should be absent from the physical spectrum which have
in reality been observed.

Then, we make some more remarks concerning parity and its breaking.

We study the ``generation'' chiral group of transformations and show how
the 8 Higgs quadruplets fall into 4 doublets of $SU(2)^g_R$ or $SU(2)^g_L$, and
1 triplet + 1 singlet of its diagonal $SU(2)^g$ subgroup.

*\ In section \ref{section:scalpattern},
 we explain why the spectrum of the 8 Higgs bosons, 1 triplet, 2
doublets and 1 singlet can be considered to fall into representations
of this generation $SU(2)^g$ subgroup orthogonal to the custodial
$SU(2)$.

*\ In section \ref{section:conclusion} we make miscellaneous remarks
and  give prospects for forthcoming works.
We emphasize the very important role of the normalization of  bosonic
asymptotic states to determine their  couplings  to quarks; we
outline in particular why present bounds on the masses of
 light scalars have to be revised. We give a list of topics to be
investigated and their spreading to other domains of physics.

$\bullet$\ in {\bf appendix} \ref{section:flavmass}, we collate the expressions of 
bilinear flavor quark operators in terms of their mass counterparts and of
the mixing angles $\theta_u$ and $\theta_d$. These formul{\ae} are used
throughout the paper.

\bigskip

\underline{Note} : to help the reader, formul{\ae} which are valid for the
whole paper, including definitions, and final results, have been boxed.


\chapter{General results} \label{chapt:gener}

Embedding the gauge group $SU(2)_L$ into the chiral group $U(2N)_L \times
U(2N)_R$, we start by establishing a one-to-one correspondence
between the Higgs doublet of the GSW model and quadruplets of bilinear
 quark operators. We then proceed to constructing our multi-Higgs 
 extension  of the standard model. We introduce  Yukawa couplings, then
the genuine and effective scalar potentials. The latter plays an important
role because
the Yukawa couplings are no longer passive in defining the vacuum of the
theory. Last we give general formul{\ae} for the masses of the Higgs
bosons.

\section{A one-to-one correspondence}
\label{section:bijection}

\subsection{The Higgs doublet of the GSW model}
\label{subsec:GSWhiggs}

We give below the laws of transformations of the components of the
Higgs doublet of the GSW model, and, by a very simple change of variables,
put them in a form that  matches the ones of specific 
 bilinear fermion operators that we shall introduce later.

The generators of the group $SU(2)_L$ are  the three hermitian
$2\times 2$ matrices
\begin{equation}
\vec T_L = \frac{\vec \tau}{2},
\end{equation}
where the $\tau$'s are the Pauli matrices
\begin{equation}
\tau^1=\left(\begin{array}{rr}0 & 1\cr 1 & 0\end{array}\right),\quad
\tau^2=\left(\begin{array}{rr}0 & -i\cr i & 0\end{array}\right),\quad
\tau^3=\left(\begin{array}{rr}1 & 0\cr 0 & -1\end{array}\right).
\label{eq:Pauli}
\end{equation}
The Higgs doublet $H$ is generally written
\begin{equation}
H=\frac{1}{\sqrt{2}}\left(\begin{array}{c} \chi^1+i\chi^2 \cr
\chi^0 - \chi^3\end{array}\right), \quad \chi^3=ik^3,
\end{equation}
in which $\chi^{0,1,2}$ and $k^3$ are considered to be real. The vacuum
expectation value of $H$ arises from $<\chi^0>=v$ such that
$<H>= \left(\begin{array}{c} 0\cr v/\sqrt{2}\end{array}\right)$.
$H$ lies in the fundamental representation of $SU(2)_L$ such that 
generators $\vec T_L$ act according to
\begin{equation}
T^i_L.H= T^i_L H.
\label{eq:action}
\end{equation}
The transformed $T^i_L.\chi^\alpha, i=1,2,3,\alpha=0,1,2,3$ of
the components $\chi^\alpha$ are naturally defined by
\begin{equation}
T^i_L.H = \frac{1}{\sqrt{2}}\left(\begin{array}{c} 
T^i_L.\chi^1 + i T^i_L.\chi^2\cr
T^i_L.\chi^0 - T^i_L.\chi^3\end{array}\right),
\label{eq:transcomp1}
\end{equation}
such that the law of transformation (\ref{eq:action}) is equivalent to
\medskip

\vbox{
\begin{equation}
\begin{array}{lll}
 T^1_L\,.\,\chi^0=+\frac{i}{2}\,\chi^2,&
T^2_L\,.\,\chi^0=+\frac{i}{2}\,\chi^1,&
T^3_L\,.\,\chi^0=+\frac12\, \chi^3,\cr
 T^1_L\,.\,\chi^1=-\frac{1}{2}\,\chi^3,&
T^2_L\,.\,\chi^1=-\frac{i}{2}\,\chi^0,&
T^3_L\,.\,\chi^1=+\frac{i}{2}\, h^2,\cr
 T^1_L\,.\,\chi^2=-\frac{i}{2}\,\chi^0,&
T^2_L\,.\,\chi^2=+\frac{1}{2}\,\chi^3,&
T^3_L\,.\,\chi^2=-\frac{i}{2}\, \chi^1,\cr
 T^1_L\,.\,\chi^3=-\frac{1}{2}\,\chi^1,&
T^2_L\,.\,\chi^3=+\frac{1}{2}\,\chi^2,&
T^3_L\,.\,\chi^3=+\frac{1}{2}\, \chi^0.
\end{array}
\label{eq:rule}
\end{equation}
}

It takes the form desired for later considerations
\begin{equation}
\begin{array}{rcl}
T^i_L\,.\,h^j&=&-\frac{1}{2}\left(i\,\epsilon_{ijk}h^k +
\delta_{ij}\,h^0\right),\cr
T^i_L\,.\,h^0 &=& -\frac{1}{2}\, h^i,
\end{array}
\label{eq:ruleL0}
\end{equation}
when one makes the substitutions
\begin{equation}
\chi^0 = -h^3,\quad \chi^1= h^1, \quad \chi^2= -h^2,\quad
\chi^3= h^0 \quad(\Leftrightarrow k^3=-i h^0).
\label{eq:subst}
\end{equation}
$H$ then rewrites
\begin{equation}
H=\frac{1}{\sqrt{2}}\left(\begin{array}{c} h^1-ih^2\cr
-(h^0+h^3)\end{array}\right),
\label{eq:Hbis}
\end{equation}
and $<\chi^0>=v$ is thus tantamount to $<h^3>=-v$.

Later, we shall often, instead of complex Higgs doublets, consider 
indifferently quadruplets, for example, in this case
\begin{equation}
H= \frac{1}{\sqrt{2}}(h^0, h^3, h^+, h^-),\quad h^\pm=h^1\pm i\,h^2,
\end{equation}
keeping in mind that, to any such quadruplet is associated a
complex doublet in the fundamental representation of $SU(2)_L$ given
by (\ref{eq:Hbis}).

\subsection{Embedding the gauge group into the chiral group}
\label{subsec:embed}

For $N$ generations of quarks, that is, $2N$ quarks, let us embed $SU(2)_L$
into the chiral group $U(2N)_L \times U(2N)_R$ by representing its three
generators as the following $2N\times 2N$ matrices
\begin{equation}
T^3 = \frac 12 \left(\begin{array}{ccc} {\mathbb I} & \vline & \\
\hline
                      & \vline & -{\mathbb I}\end{array}\right),
\quad
T^+ = T^1+iT^2= \left(\begin{array}{ccc}  & \vline &{\mathbb I} \\
\hline
                      & \vline & \end{array}\right),
\quad
T^- = T^1-iT^2= \left(\begin{array}{ccc}  & \vline & \\
\hline
                      {\mathbb I}& \vline & \end{array} \right),
\label{eq:Ts}
\end{equation}
in which $\mathbb I$ is the $N\times N$ identity matrix.
They act trivially on $2N$ vectors of flavor quark eigenstates
$\psi=(u,c,t,\ldots, d,s,b,\ldots)^t$
\footnote{The superscript $(\ )^t$ means
``transpose''.\label{footnote:transpose}}.

\subsection{Quadruplets of bilinear quark operators}
\label{subsec:isomorphism}

$\mathbb M$ being any $2N\times 2N$ matrix, we now consider bilinear quark
operators of the form $\bar\psi{\mathbb M}\psi$ and $\bar\psi \gamma_5
{\mathbb M}\psi$. ${\cal U}_R$ and ${\cal U}_L$ being transformations of
$SU(2)_R$ and $SU(2)_L$ respectively, these bilinears transform by the chiral group
according to
\begin{equation}
\begin{split}
({\cal U}_L \times {\cal U}_R)\,.\,\bar\psi\frac{1+\gamma_5}{2}{\mathbb
M}\psi
&= \bar \psi\; {\cal U}_L^{-1}\,{\mathbb M\;{\cal
U}_R\;\frac{1+\gamma_5}{2}}\psi,\cr
& \cr
({\cal U}_L \times {\cal U}_R)\,.\,\bar\psi\frac{1-\gamma_5}{2}{\mathbb
M}\psi
&= \bar \psi\; {\cal U}_R^{-1}\,{\mathbb M\;{\cal
U}_L\;\frac{1-\gamma_5}{2}}\psi.
\end{split}
\label{eq:group}
\end{equation}
Writing ${\cal U}_R$ and ${\cal U}_L$  as
\begin{equation}
{\cal U}_{L,R} = e^{-i \alpha_i T^i_{L,R}},\quad i=1,2,3,
\end{equation}
eq.~(\ref{eq:group}) entails
\begin{equation}
\begin{split}
{ T}^j_L\,.\,\bar\psi{\mathbb M} \psi &=
-\frac12\,\left(\bar\psi\,[{ T}^j,{\mathbb M}]\,\psi
                 +\bar\psi\, \{{ T}^j,{\mathbb
M}\}\,\gamma_5\psi\right),\cr
& \cr
{ T}^j_L\,.\,\bar\psi{\mathbb M}\gamma_5 \psi &=
-\frac12\,\left(\bar\psi\,[{ T}^j,{\mathbb
M}]\,\gamma_5\psi
                 +\bar\psi\, \{{ T}^j,{\mathbb M}\}\,\psi\right),\cr
& \cr
{ T}^j_R\,.\,\bar\psi{\mathbb M} \psi &=
-\frac12\,\left(\bar\psi\,[{ T}^j,{\mathbb M}]\,\psi
                 -\bar\psi\, \{{ T}^j,{\mathbb
M}\}\,\gamma_5\psi\right),\cr
& \cr
{ T}^j_R\,.\,\bar\psi{\mathbb M}\gamma_5 \psi &=
-\frac12\,\left(\bar\psi\,[{ T}^j,{\mathbb
M}]\,\gamma_5\psi
                 -\bar\psi\, \{{ T}^j,{\mathbb M}\}\,\psi\right),
\end{split}
\label{eq:trans2}
\end{equation}
in which $[\ ,\ ]$ and $\{\ ,\ \}$ stand respectively for the commutator
and anticommutator.

Let us now define the specific $2N\times 2N$ matrices
\begin{equation}
{\mathbb M}^0= \left(
\begin{array}{ccc} M & \vline & 0 \cr
\hline
0 & \vline & M \end{array}\right),
{\mathbb M}^3 = \left(
\begin{array}{ccc} M & \vline & 0 \cr
\hline
0 & \vline & -M \end{array}\right),
 {\mathbb M}^+ =2 \left(
\begin{array}{ccc} 0 & \vline & M \cr
\hline
0 & \vline & 0 \end{array}\right),
{\mathbb M}^-= 2\left(
\begin{array}{ccc} 0 & \vline & 0 \cr
\hline
M & \vline & 0 \end{array}\right),
\label{eq:generquad}
\end{equation}
in which ${\mathbb M}^\pm= {\mathbb M}^1 \pm i{\mathbb M}^2$, and $M$ is a
real $N\times N$ matrix.
Let us call  $(a^0, a^3, a^+, a^-)$ the generic components
of the two sets of $N^2$ quadruplets
\begin{equation}
\bar\psi \Big({\mathbb M}^0, \gamma^5{\mathbb M}^3, \gamma^5{\mathbb
M}^+, \gamma^5{\mathbb M}^-\Big)\psi,
\label{eq:SP}
\end{equation}
of the type $({\mathfrak s}^0, \vec {\mathfrak p})$,
made with one scalar and three pseudoscalars,
and
\begin{equation}
\bar\psi \Big(\gamma^5{\mathbb M}^0, {\mathbb M}^3, {\mathbb M}^+,
{\mathbb M}^-\Big)\psi,
\label{eq:PS}
\end{equation}
of the type $({\mathfrak p}^0, \vec {\mathfrak s})$,
made with one pseudoscalar and three scalars. By (\ref{eq:trans2}),
the $a^i$'s transform by $SU(2)_L$ and $SU(2)_R$ according to
\begin{equation}
\boxed{
\begin{array}{rcl}
T^i_L\,.\,a^j&=&-\frac{1}{2}\left(i\,\epsilon_{ijk}a^k +
\delta_{ij}\,a^0\right)\cr
T^i_L\,.\,a^0 &=& -\frac{1}{2}\, a^i
\end{array}
}
\label{eq:ruleL}
\end{equation}
or, equivalently, since it is often convenient to manipulate states with
given electric charge,
\begin{equation}
\begin{split}
&T^3_L.a^0 =-\frac12 a^3,\quad T^3_L.a^3 =-\frac12 a^0,\quad
T^3_L.a^+=-\frac12 a^+,\quad
T^3_L.a^-=+\frac12 a^-,\cr
&T^+_L.a^0 =-\frac12 a^+,\quad T^+_L.a^3 =+\frac12 a^+,\quad
T^+_L.a^+=0,\quad
T^+_L.a^-=- a^0-a^3,\cr
&T^-_L.a^0 =-\frac12 a^-,\quad T^-_L.a^3 =-\frac12 a^-,\quad
T^-_L.a^+=-a^0+a^3,\quad
T^-_L.a^-=0,
\end{split}
\end{equation}
and
\begin{equation}
\boxed{
\begin{array}{rcl}
T^i_R\,.\,a^j&=&-\frac{1}{2}\left(i\,\epsilon_{ijk}a^k -
\delta_{ij}\,a^0\right)\cr
T^i_R\,.\,a^0 &=& +\frac{1}{2}\, a^i
\end{array}
}
\label{eq:ruleR}
\end{equation}
or, equivalently
\begin{equation}
\begin{split}
&T^3_R.a^0 =+\frac12 a^3,\quad T^3_R.a^3 =+\frac12 a^0,\quad
T^3_R.a^+=-\frac12 a^+,\quad
T^3_R.a^-=+\frac12 a^-,\cr
&T^+_R.a^0 =+\frac12 a^+,\quad T^+_R.a^3 =+\frac12 a^+,\quad
T^+_R.a^+=0,\quad T^+_R.a^-=+ a^0-a^3,\cr
&T^-_R.a^0 =+\frac12 a^-,\quad T^-_R.a^3 =-\frac12 a^-,\quad
T^-_R.a^+=+a^0+a^3,\quad T^-_R.a^-=0.
\end{split}
\end{equation}

The laws of transformations (\ref{eq:ruleL0}) and (\ref{eq:ruleL}) being
identical, we have therefore found $2N^2$ quadruplets isomorphic, for their
law of transformation by $SU(2)_L$, to the complex Higgs doublet of the
GSW model. We shall deal later with their normalizations.

The two sets respectively of the $({\mathfrak s}^0,\vec {\mathfrak p})$ type
and of the $({\mathfrak p}^0,\vec {\mathfrak s})_)$
 type are, up to their normalizing factors,
 parity transformed of each other. As can be easily checked from
eq.~(\ref{eq:trans2}) the operators that switch
parity of bilinear quark operators
${\mathfrak s}=\bar q_i q_j$ and ${\mathfrak p}= \bar q_i\gamma_5 q_j$
 are the generators ${\mathbb I}_L$ and ${\mathbb I}_R$ of the
transformations $U(1)_L$ and $U(1)_R$ such that, for each matrix $M$, the
corresponding pair of quadruplets $({\mathfrak s}^0, \vec {\mathfrak p})$
and $({\mathfrak p}^0, \vec {\mathfrak s})$
transform by
\begin{equation}
\boxed{
{\mathbb I}_L.({\mathfrak s}^0,\vec {\mathfrak p})=-({\mathfrak p}^0,
 \vec {\mathfrak s}),\quad
{\mathbb I}_L.({\mathfrak p}^0,\vec {\mathfrak s})=-({\mathfrak s}^0,\vec
{\mathfrak p}),\quad
{\mathbb I}_R.({\mathfrak s}^0,\vec {\mathfrak p})=+({\mathfrak p}^0, \vec
{\mathfrak s}),\quad
 {\mathbb I}_R.({\mathfrak p}^0,\vec
{\mathfrak s})=+({\mathfrak s}^0,\vec {\mathfrak p})}
\label{eq:par1gen}
\end{equation}

In the following, we shall note generically $\Delta_i$ the quadruplets of
the type $({\mathfrak s}^0, \vec {\mathfrak p})$, and $\hat\Delta_i$ the
ones of the type $({\mathfrak p}^0,
\vec {\mathfrak s})$. For 2 generations (see section \ref{section:quadruplets}),
 the index $i$ can take 4 values and spans the
set $\{X,H,\Omega,\Xi\}$. All $\Delta_i, \hat\Delta_i$ are expressed in terms
of bilinear quark operators.

\section{Generalities on Yukawa couplings}\label{section:genyuk}

Yukawa couplings are, in the GSW model, $SU(2)_L$ invariant couplings
between the quarks and the Higgs doublet $H$, tailored to give masses
to the quarks by  spontaneous breaking of the gauge symmetry. In there, 
$H$ is supposed to be unique. By coupling to left-handed flavor
doublets and to right-handed flavor singlets it gives masses to the quarks
with charge $-1/3$ and the same procedure operated with $H$ replaced by
$i\tau^2 H^\ast$ gives masses to the quarks with charge $+2/3$.

We  saw in subsection \ref{subsec:isomorphism}
 that, as soon as one considers the possibility that the
Higgs fields are bilinear quark-antiquark operators,
 $2N^2$ complex Higgs doublets become available. There is no reason to
discard any of them such that any extension of the Standard Model with
composite Higgses should include $2N^2$ complex Higgs doublets. 
This is what we shall do here.

The Yukawa Lagrangian that we consider for $N$ generations is
\begin{equation}
\boxed{
{\cal L}_{Yuk}= \sum_{i=1}^{N^2}
-\frac12 \delta_i \big(\Delta_i^\dagger [\Delta_i]+ h.c.\big)
-\frac12\delta_{i\hat\imath }\big(\Delta_i^\dagger [\hat\Delta_i]+h.c.\big)
 -\frac12\kappa_{\hat\imath i}\big(\hat\Delta_i^\dagger[\Delta_i]+h.c.\big)
-\frac12\big(\hat\delta_i\hat\Delta_i^\dagger[\hat\Delta_i]+ h.c.\big)
}
\label{eq:Lyuk0}
\end{equation}
where the notations are as follows. As we already mentioned, the $2N^2$
quadruplets that we consider are split into $N^2$ pairs $(\Delta_i,
\hat\Delta_i)$ in which $\Delta_i$ and $\hat\Delta_i$ are parity
transformed of each other. The index $i$ in (\ref{eq:Lyuk0})
spans this set of pairs (in the case of 1 generation (see
(\ref{eq:genyuk2}) in section \ref{section:yuk1gen}), $i=1$ and one
deals with only one pair of parity transformed Higgs quadruplets; for 2
generations $i$ goes from $1$ to $4$ since one has 4 pairs of parity
transformed quadruplets (see section \ref{section:quadruplets})).
 Moreover, according to the notation that we introduced in
(\ref{eq:example2}) each of them can be 
expressed either in terms of bilinear quark operators, in which case we
write it $\Delta_i$ or $\hat\Delta_i$, or in terms of bosonic fields
(Higgs bosons,  mesons \ldots), in which case we write it $[\Delta_i]$
or $[\hat\Delta_i]$.

With respect to the most general Yukawa Lagrangian,
the choice (\ref{eq:Lyuk0})  drastically reduces the number of Yukawa couplings
down to $4N^2$. It has the following properties:\newline
* it is the simplest and most
 straightforward generalization of the case of 1 generation.
 For 1 generation, as we shall see in chapter \ref{chapt:1gen},
 (\ref{eq:Lyuk0}) for $N=1$ describes the most general $SU(2)_L$ invariant
 couplings (\ref{eq:genyuk1})
 between the 2 quarks $u,d$ and the two parity transformed
 Higgs quadruplets $X$ and $\hat X$ (eqs.(\ref{eq:Xq1gen}) and
(\ref{eq:hXq1gen});\newline
* it is diagonal in the index $i$, which means that it is a sum over pairs
$(\Delta_i, \hat\Delta_i)$ of parity transformed doublets.
It accordingly discards all crossed couplings $\Delta_i^\dagger[\Delta_j],
\Delta_i^\dagger[\hat\Delta_j], \hat\Delta_i^\dagger[\hat\Delta_j],
\hat\Delta_i^\dagger [\hat\Delta_j]$ with $i\not = j$, which 
 forbids  tree level FCNC's.

Since it involves couplings between quadruplets of opposite parity, it a
priori allows classical transitions between scalars and pseudoscalars.
Requesting that they vanish will provide constraints on the couplings.

By embedding the $SU(2)_L$ group into the chiral group, we have in
particular established a connection between the spontaneous breaking of
these
two symmetries, which, due to the composite nature of the Higgs bosons,
become ``dynamical'' because they are triggered by quark condensation.
Now, the Yukawa couplings involve couplings of the Higgs bosons to quark pairs
for example $-\delta_X \left(\frac{v_X}{\sqrt{2}\mu_X^3}\frac{\bar u u+\bar d
d}{\sqrt{2}}\right) X^0$,
where $<X^0>=\frac{v_X}{\sqrt{2}}, \mu_X^3=
\frac{<\bar u u+\bar d d>}{\sqrt{2}}$, such that the quark condensate $<\bar
u u+\bar d d>\not=0$ gives birth to a term that is linear in the Higgs boson
$X^0$. The first derivative of such a Yukawa coupling with respect to $X^0$
yields, thus, a non-vanishing constant, like does the first derivative of
the term in the scalar potential that is quadratic in $X^0$ when one sets
 $X^0$ to its bosonic VEV (see subsection \ref{subsec:Vgen}). So, unlike in the
genuine GSW model, Yukawa couplings are no longer passive in the
construction of the vacuum, and shift its minimum with respect to what would
be obtained from the sole consideration of the scalar potential. This is
why, in relation to the twofold nature of the quadruplets, it is convenient to
also consider, for further minimization of an ``effective potential'' built
from the genuine one and Yukawa couplings,  a bosonised form of
 ${\cal L}_{Yuk}$ that writes 
\begin{equation}
{\cal L}_{Yuk}^{bos}= \sum_{i=1}^{N^2}
-\delta_i [\Delta_i]^\dagger [\Delta_i]
-\frac12(\delta_{i\hat\imath}+\kappa_{\hat\imath i})
\Big([\Delta_i]^\dagger [\hat\Delta_i]
+[\hat\Delta_i]^\dagger[\Delta_i]\Big)
- \hat\delta_i [\hat\Delta_i]^\dagger[\hat\Delta_i].
\label{eq:Lyukbos1}
\end{equation}

In the present approach, Yukawa couplings have three roles:\newline
* they give masses to fermions;\newline
* they give ``soft''  masses to scalar and pseudoscalar mesons; 
exceptions are the three Goldstones of the broken $SU(2)_L$ that
 should not become massive, which provides additional constraints;\newline
* they modify the scalar potential and shift accordingly
the position of its minima.

The bosonised form (\ref{eq:Lyukbos1}) of the Yukawa couplings
 can be further simplified by the requirement that no transition occurs
between scalars and pseudoscalars.
$[\Delta_i]^\dagger[\hat\Delta_i] = [\Delta^0][ \hat\Delta^0]
-[\vec\Delta].[\vec{\hat\Delta}] + i([\Delta^1][ \hat\Delta^2] -[ \Delta^2]
[\hat\Delta^1])+ [\Delta^0][ \hat\Delta^3] -[ \Delta^3][\hat\Delta^0]$
and $[\hat\Delta_i]^\dagger[\Delta_i] = -[\Delta^0][ \hat\Delta^0]
+[\vec\Delta].[\vec{\hat\Delta}] + i([\Delta^1][ \hat\Delta^2] -[ \Delta^2]
[\hat\Delta^1])+ [\Delta^0][ \hat\Delta^3] -[ \Delta^3][\hat\Delta^0]$
such that the middle terms in (\ref{eq:Lyukbos1}) write
$-(\delta_{i\hat\imath }+\kappa_{\hat{\imath }i})
\Big( i([\Delta^1][ \hat\Delta^2] -[ \Delta^2]
[\hat\Delta^1])+[ \Delta^0][ \hat\Delta^3] -[ \Delta^3][\hat\Delta^0]\Big)$.
$[\Delta^{1,2,3}],[ \hat\Delta^0]$ are pseudoscalars while
$[\hat\Delta^{1,2,3}],[ \Delta^0]$
 are scalars such that the first two terms are charged
 scalar-pseudoscalar crossed terms. We {\it a priori} declare them unwanted at
the classical level, which requires
\begin{equation}
\boxed{
\delta_{i\hat\imath }+\kappa_{\hat{\imath }i}=0} \qquad
\Rightarrow\qquad
\boxed{{\cal L}_{Yuk}^{bos}= \sum_{i=1}^{N^2}
-\delta_i [\Delta_i]^\dagger [\Delta_i]
- \hat\delta_i [\hat\Delta_i]^\dagger[\hat\Delta_i]}
\label{eq:Lyukbos2}
\end{equation}

\section{Generalities on the Higgs potential}\label{section:genpot}

\subsection{The  scalar potential $\boldsymbol
V$}\label{subsec:Vgen}

There, too, we shall use the simplest possible extension of the scalar
potential of the GSW model to the case of $2N^2$ Higgs multiplets
\begin{equation}
\boxed{
V=\sum_{i=1}^{N^2} -\frac{m_H^2}{2}\Big([\Delta_i]^\dagger [\Delta_i] +
[\hat\Delta_i]^\dagger [\hat\Delta_i]\Big)
+\frac{\lambda_H}{4}\Big(([\Delta_i]^\dagger [\Delta_i])^2 +
([\hat\Delta_i]^\dagger [\hat\Delta_i])^2\Big)
}
\label{eq:Vgen}
\end{equation}
(\ref{eq:Vgen}) is written as a sum over the $N^2$ pairs $(\Delta_i, \hat\Delta_i)$
of Higgs multiplets.  It only depends on 2 parameters $m_H^2$ and $\lambda_H$.

In there, all Higgs multiplets are written in their bosonic form 
$[\Delta]=([\Delta]^0, [\Delta]^3, [\Delta]^+, [\Delta]^-)$, this is
why we have used the notation with square brackets $[\ ]$. 

(\ref{eq:Vgen}) is not only invariant by $SU(2)_L$, but,
since all Higgs quadruplets are also complex doublets of $SU(2)_R$ (see
(\ref{eq:ruleR})),  by $SU(2)_R$, too.  It is thus invariant by
the chiral $SU(2)_L \times SU(2)_R$.
Would the normalization factors be the same for all multiplets, $V$ would be
invariant by the larger chiral group $U(2N)_L \times U(2N)_R$. This is why
we choose from the beginning to only make $V$ depend on 2 coupling
constants $\lambda_H$ and $m_H^2$, such that the underlying
$U(2N)_R\times U(2N)_L$
symmetry  gets spontaneously broken down to $SU(2)_L \times
SU(2)_R$ by the non-equality of the bosonic VEV's which break the gauge
symmetry, $v_i\not=v_j\not=\hat
v_k\not= \hat v_l, i,j,k,l \in [1,N^2]$ and, likewise, by the non-equality
of the fermionic VEV's ($<\bar q q>$) $\mu_i^3 \not=
\mu_j^3 \not=\hat\mu_k^3 \not=\hat\mu_l^3$ (the non-vanishing of the
fermionic VEV's  is usually attributed
``chiral symmetry breaking''). Since the gauge group has been embedded in
the chiral group, the situation is of course very intricate.

Additional remarks concerning the Higgs potential can be found in
section \ref{section:scalpot1gen} for 1 generation of quarks.

\subsection{The effective scalar potential
$\boldsymbol{V_{eff}}$}\label{subsec:Veff}

$V_{eff}$ we define as the difference between the
 genuine potential (\ref{eq:Vgen}) and the bosonised Yukawa Lagrangian
(\ref{eq:Lyukbos2})
\begin{equation}
\boxed{
V_{eff} = V - {\cal L}_{Yuk}^{bos}
}
\label{eq:Veffdev}
\end{equation}
Like $V$, it is invariant by the chiral group $SU(2)_L \times SU(2)_R$.

Each quadruplet $\Delta_i$ and $\hat\Delta_i$ includes a priori one Higgs
boson, that is, one scalar with a non-vanishing VEV: $\Delta_i^0 \in
\Delta_i$ and $\hat\Delta_i^3 \in \hat\Delta_i$.
It is usual matter, like in the GSW model, to find that the minimum of the
{\em genuine} potential $V$ given in (\ref{eq:Vgen})
occurs at $<\Delta_i^0>^2= \displaystyle\frac{m_H^2}{\lambda_H}
=<\hat\Delta_i^3>^2$ for all quadruplets. 
Since we  write the bosonic VEV's $<\Delta_i^0>=\frac{v_i}{\sqrt{2}}$
and $<\hat\Delta_i^3>=\frac{\hat
v_i}{\sqrt{2}}$, it is then natural to define
\begin{equation}
\boxed{
v_0^2 = \frac{2m_H^2}{\lambda_H}
}
\label{eq:v0def}
\end{equation}
For later use we shall also define the parameter $\delta$ such that
\begin{equation}
\boxed{
\lambda_H =\frac{4\delta}{v_0^2} \Rightarrow m_H^2=2\delta
}
\label{eq:deltadef}
\end{equation}

Because, as we saw in section \ref{section:genyuk},
Yukawa couplings are non longer passive in the
definition of the (spontaneously broken) vacuum of the theory, the latter
 is now defined by the minimum of $V_{eff}$ with respect to the $2N^2$ Higgs bosons.
It is obtained from the one from $V$ with the simple shift
$\frac{m_H^2}{2} \to \frac{m_H^2}{2} -\delta_i$ for a Higgs boson
$\Delta_i^0$ and $\frac{m_H^2}{2} \to \frac{m_H^2}{2} -\hat\delta_i$ for a
Higgs boson $\Delta_i^3$. The equations tuning to zero the corresponding
first derivatives are thus  of the type
\begin{equation}
m_H^2 - 2\delta_i = \lambda_H <\Delta_i^0>^2,\quad
m_H^2 - 2\hat\delta_i= \lambda_H <\hat\Delta_i^3>^2,
\end{equation}
that is, using (\ref{eq:deltadef}) and (\ref{eq:v0def})
\begin{equation}
\boxed{
\delta_i = \delta (1-b_i), \quad \hat\delta_i =\delta(1-\hat b_i)
}
\label{eq:delrel1}
\end{equation}
in which we have defined
\begin{equation}
\boxed{
b_i =\frac{v_i^2}{v_0^2}, \quad \hat b_i = \frac{\hat v_i^2}{v_0^2}
}
\label{eq:bdef}
\end{equation}

\subsection{Goldstones and pseudo-Goldstones}\label{subsec:golds}

In the absence of  Yukawa couplings, the spontaneous breaking of
$SU(2)_L$ would produce 3 Goldstone bosons inside each Higgs quadruplet.

Physically, only three of them can become the longitudinal
$W_\parallel$'s, and the rest is therefore doomed to get
``soft'' masses from the Yukawa
couplings, becoming what is commonly called ``pseudo-Goldstone bosons''.
So, the spectrum of the theory is expected to be composed, after symmetry
breaking, of $2N^2$ Higgs bosons and $6N^2-3$ pseudo-Goldstone bosons.

The $mass^2$ of the pseudo-Goldstones must be calculated from the
second derivative of $V-{\cal L}_{Yuk}$. The result, as we show below on
the same simple example of the $X^\pm$ bosons as we used in section
\ref{section:genyuk}, turns out to be the same as
if it were calculated from the sole term $\delta_X X^\dagger X$ of $(-1)
\times$ the bosonised Yukawa Lagrangian (\ref{eq:Lyukbos2}).

$-{\cal L}_{Yuk} + V$ involve the couplings
\begin{equation}
\begin{split}
& -\frac12\delta_X \left(X^+ \left(\frac{v_X}{\sqrt{2}\mu_X^3}\frac{1}{\sqrt{2}}2\bar
d\gamma_5 u \right)+ X^-\left(
\frac{v_X}{\sqrt{2}\mu_X^3}\frac{1}{\sqrt{2}}2\bar u\gamma_5 d
\right)+\ldots\right)\cr
& -\frac{m_H^2}{2}(-X^+ X^- + \ldots)
 +\frac{\lambda_H}{4}
\Big(-X^+X^--(X^3)^2 +(X^0)^2 \Big)^2,
\end{split}
\end{equation}
in which we have only written the terms which are relevant for the first and
second derivatives when $<X^0>=\frac{v_X}{\sqrt{2}}$. This yields, using
(\ref{eq:deltadef}), (\ref{eq:delrel1}) and (\ref{eq:bdef})
\begin{equation}
\frac{\partial^2 (-{\cal L}_{Yuk}+V)}{\partial X^+ \partial X^-}
\stackrel{<X^0>=v_X/\sqrt{2}}{=}\frac{m_H^2}{2}
-\frac{\lambda_H}{4}v_X^2 = \delta(1-b_X)=\delta_X,
\end{equation}
which is the result announced above.

This could be naively interpreted by  saying that the $SU(2)_L$
Goldstone bosons that occur in $V$ after symmetry breaking
get, afterwards, their soft masses from the bosonised Yukawa couplings.
This maybe  right in practice but conceptually erroneous and misleading.
To see this it is enough to minimize the effective potential
$V_{eff}=V-{\cal L}_{Yuk}^{bos}$: at the appropriate minimum, all $6N^2$
Goldstones are true massless Goldstones.

The consequence is that, if we call $\hat H$ the quadruplet that
contains the 3 Goldstones doomed to become the 3 longitudinal gauge bosons,
its Yukawa couplings should satisfy
\begin{equation}
\boxed{
\hat\delta_H=0
}
\label{eq:truegol}
\end{equation}
which adds to the $N^2$ constraints in the l.h.s. of (\ref{eq:Lyukbos2}).
Because of (\ref{eq:truegol}), the effective potential $V_{eff}$
 (\ref{eq:Lyukbos2})
 for the Higgs multiplet $\hat H$ is identical to the genuine one $V$.
 It has accordingly its minimum at
\begin{equation}
\boxed{
\hat v_H = v_0}
\label{eq:vhv0}
\end{equation}
where $v_0$ has been defined in (\ref{eq:v0def}).

\section{The masses of the Higgs bosons}\label{section:higgsmass}

Let us call generically $Higgs$ a Higgs boson, which has a  VEV
$\frac{v_{Higgs}}{\sqrt{2}}$. The corresponding quantum Higgs field $h$ is
defined by the shift $Higgs = \frac{v_{Higgs}}{\sqrt{2}} + h$ and 
its squared mass $m_h^2$ is given by
\begin{equation}
m_h^2 =\frac12 \frac{\partial^2 V_{eff}}{(\partial
\Delta_0)^2}\Big|_{<\Delta_0>=\frac{v_{Higgs}}{\sqrt{2}}}.
\end{equation}
One gets accordingly for the ``quasi-standard'' Higgs boson $\hat H^3$
\begin{equation}
\boxed{
m^2_{\hat H^3} = 2\delta = m_H^2
\label{eq:mhh3}
}
\end{equation}
and for the others
\begin{equation}
\boxed{
m^2_{Higgs} = 2 \delta\, b_{Higgs}=  m^2_{\hat H^3}\,b_{Higgs}
}
\label{eq:mhiggs}
\end{equation}
where, according to (\ref{eq:bdef}), we have defined
$b_{Higgs}=(v_{Higgs}/\hat v_H)^2$ and we have used (\ref{eq:vhv0}).

This shows in particular that the set of parameters $(\delta, b's)$ is
 specially relevant for the spectrum of the theory.


\chapter{The case $\boldsymbol{N=1}$ generation}
\label{chapt:1gen}

This toy model is already a 2-Higgs doublet model. We show how all
parameters can be determined from pion physics, from the $W$ mass and
from fermionic constraints, and  how the leptonic decays of
charged pions are suitably described. The ``quasi-standard'' Higgs boson 
has mass $\sqrt{2}m_\pi$, and the second Higgs boson is very light
($69\,KeV$).

\section{The 2 Higgs quadruplets}\label{section:2Hquad}

This case only involves 2 parity-transformed  Higgs multiplets
\footnote{Among all the 2-Higgs doublets models that have been
investigated \cite{2HDM},  none involve parity-transformed
multiplets. This is probably due to the fact that the laws of
transformations (\ref{eq:rule}) and (\ref{eq:ruleL}) were never written
before.}, 4 VEV's, 3 pions, 2 Higgs bosons and 3 Goldstone bosons. The number of
independent Yukawa couplings reduces to 2 and the equations are easy to
solve, also simplified by the absence of mixing.

The 2 Higgs multiplets are the following
\begin{equation}
\begin{split}
 X &=\frac{v_X}{\sqrt{2}\,\mu_X^3}\frac{1}{\sqrt{2}}\ \bar\psi\left(
\left(\begin{array}{rr}
1 &  \cr
 & 1  
 \end{array}\right),
\gamma^5\left(\begin{array}{rr}
1 &  \cr
 &  -1 
\end{array}\right),
2\gamma^5\left(\begin{array}{rr}
  & 1   \cr
0 &  \end{array}\right),
2\gamma^5\left(\begin{array}{rr}
 & 0 \cr
1 & 
 \end{array}\right)
\right)\psi\cr
&= (X^0,X^3,X^+,X^-),
 \quad with\quad \mu_X^3 = \frac{<\bar u u+\bar d d>}{\sqrt{2}},
\end{split}
\label{eq:Xq1gen}
\end{equation}
and
\begin{equation}
\begin{split}
 \hat X &=\frac{\hat v_X}{\sqrt{2}\,\hat \mu_X^3}\frac{1}{\sqrt{2}}\ \bar\psi\left(
\gamma_5\left(\begin{array}{rr}
1 &  \cr
 & 1  
 \end{array}\right),
\left(\begin{array}{rr}
1 &  \cr
 &  -1 
\end{array}\right),
2\left(\begin{array}{rr}
  & 1   \cr
0 &  \end{array}\right),
2\left(\begin{array}{rr}
 & 0 \cr
1 & 
 \end{array}\right)
\right)\psi\cr
&= (\hat X^0,\hat X^3,\hat X^+,\hat X^-),
 \quad with\quad \hat\mu_X^3 = \frac{<\bar u u-\bar d d>}{\sqrt{2}},
\end{split}
\label{eq:hXq1gen}
\end{equation}
in which $\psi=\left(\begin{array}{c} u\cr d \end{array}\right)$.
The 2 Higgs bosons are $X^0$ and $\hat X^3$.
Since $X$ contains a triplet of pseudoscalars which we wish to identify
with the three pions, $\hat X$ is the quadruplet doomed to incorporate the
3 Goldstone bosons of the broken $SU(2)_L$ that become the 3 longitudinal
gauge bosons. This is why we shall call $\hat X^3$ the ``quasi-standard''
Higgs boson, and we shall keep this terminology in the rest of the paper.
 We see that, while the two charged Goldstones are the two
scalars $[\hat X]^+, [\hat X]^-$, the neutral one is the pseudoscalar
$[\hat X]^0$. Parity is thus, there, obviously broken.

\section{The Yukawa couplings}\label{section:yuk1gen}

We start writing them in the ``standard'' form, in which, as usual,
a given complex Higgs doublet $H$ must be associated to its complex
conjugate $i\tau^2 H^\ast$ to provide masses for both $d$-type  and
$u$-type quarks. This means that we use $[X], [\hat X]$ together with
$i\tau^2 [X]^\ast, i\tau^2 [\hat X]^\ast$ and write the Yukawa couplings as
\begin{equation}
\begin{split}
{\cal L}_{Yuk}= &+\rho_d
\left(\begin{array}{cc}\overline{u_L}\
\overline{d_L}\end{array}\right) [X]
\, d_R
- \rho_u
\left(\begin{array}{cc}\overline{u_L}\
\overline{d_L}\end{array}\right)
(i\tau^2 [X]^\ast)
\, u_R\cr
&+\lambda_d
\left(\begin{array}{cc}\overline{u_L}\
\overline{d_L}\end{array}\right) [\hat X]
\, d_R
+\lambda_u \left(\begin{array}{cc}\overline{u_L}\
\overline{d_L}\end{array}\right)
(i\tau^2 [\hat X]^\ast)
\, u_R\cr
&\hskip 3cm+   h.c.,
\end{split}
\label{eq:genyuk1}
\end{equation}
in which the Higgs multiplets $[X]$ and $[\hat X]$ are written as $SU(2)_L$
doublets, for example $[X]= \left(\begin{array}{c} [X^1]-i[X^2]\cr
-[X^0]-[X^3]\end{array}\right)$ and, therefore, $i\tau^2 [X]^*= \left(\begin{array}{c}
[X^0]-[X^3]\cr -[X^1]-i[X^2]\end{array}\right)$.

The expression
(\ref{eq:genyuk1}) represents the most general Yukawa couplings between 2
quarks and 2 Higgs doublets.  After suitably grouping terms, it gives, explicitly
\footnote{As can be easily verified
\begin{equation}
\begin{split}
\frac12 (X^\dagger [X] + h.c.)&=
X^0[X^0]-X^3[X^3]-\frac12\big(X^+[X^-]+X^-[X^+]\big),\cr
\frac12 \big(X^\dagger [\hat X]+h.c.\big)&=X^0[\hat X^3]-X^3[\hat
X^0]+\frac12 \big(X^-[\hat X^+]-X^+[\hat X^-] \big),\cr
\frac12 \big(\hat X^\dagger [X]+h.c.\big)&=-\hat X^0[X^3]+\hat X^3[ X^0]
-\frac12 \big(\hat X^-[X^+]-\hat X^+[X^-] \big),\cr
\frac12 (\hat X^\dagger [\hat X] + h.c.)&=
-\hat X^0[\hat X^0]+\hat X^3[\hat X^3]+\frac12\big(\hat X^+[\hat X^-]+\hat
X^-[\hat X^+]\big),
\end{split}
\end{equation}
in which the scalar or pseudoscalar nature of the different fields has been
taken into account for hermitian conjugation and all VEV's have been
supposed to be real. },
\begin{equation}
{\cal L}_{Yuk}= -\frac12\delta_X (X^\dagger [X]+ h.c.)
 -\frac12 \delta_{X\hat
X}\big(X^\dagger[\hat X] + h.c.\big)
-\frac12 \kappa_{\hat X X}\big(\hat X^\dagger [X]+h.c.\big)
-\frac12\hat\delta_X (\hat X^\dagger[\hat X] + h.c.).
\label{eq:genyuk2}
\end{equation}
In (\ref{eq:genyuk1}) and (\ref{eq:genyuk2}) the signs have been set such
that
for positive $<{\mathfrak s}^0>$ and $<{\mathfrak s}^3>$,
the fermion masses are positive for positive
$\rho_{u,d}$ and $\lambda_{u,d}$ (given that a fermion mass term is of the
form $-m\bar\psi \psi$) and,
in (\ref{eq:genyuk2}), we have introduced the  parameters with dimension
$[mass]^2$
\begin{eqnarray}
\delta_X &=& \displaystyle\frac{\rho_u +
\rho_d}{\sqrt{2}}\,\frac{\mu_X^3}{v_X/\sqrt{2}}  ,\cr
&&\cr
\kappa_{\hat X X} &=& \displaystyle\frac{\rho_u -
\rho_d}{\sqrt{2}}\,\frac{\hat\mu_X^3}{\hat v_X/\sqrt{2}}  ,\cr
&&\cr
\delta_{X\hat X} &=&
\displaystyle\frac{\lambda_u+\lambda_d}{\sqrt{2}}\,\frac{\mu_X^3}{v_X/\sqrt{2}}  ,\cr
&&\cr
\hat\delta_X &=&
\displaystyle\frac{\lambda_u-\lambda_d}{\sqrt{2}}\,\frac{\hat\mu_X^3}{\hat
v_X/\sqrt{2}}.
\label{eq:params1gen}
\end{eqnarray}
It is then trivial matter, by replacing, in (\ref{eq:genyuk2}), the bilinear
quark operators by the corresponding bosonic fields, to get the bosonised
form of the Yukawa couplings
\begin{equation}
{\cal L}_{Yuk}^{bos} = -\delta_X [X]^\dagger [X]
-\frac12\underbrace{ (\delta_{X\hat X}+\kappa_{\hat X X})}_{0}
\big([X]^\dagger[\hat X] + [\hat X]^\dagger [X]\big)
-\underbrace{\hat\delta_X}_{0} [\hat X]^\dagger[\hat X],
\end{equation}
in which the second and third term vanish from the general constraints
(\ref{eq:Lyukbos2}) (\ref{eq:truegol}) that we have
established $\delta_{X\hat X}+\kappa_{\hat X X}=0$, $\hat\delta_X=0$
\footnote{In this case, $[\hat X]$ is the quadruplet that contains the
three $SU(2)_L$ Goldstones such that (\ref{eq:truegol}) writes
$\hat\delta_X=0$.\label{foot:hdelx}}.
This 2-Higgs doublet model has thus only 2 Yukawa couplings $\delta_X$ and
$\delta_{X\hat X}$.

As a side remark, notice that a term
proportional to $[X]^\dagger [\hat X] +[\hat X]^\dagger [X]$, which can be
potentially present in ${\cal L}_{Yuk}^{bos}$, includes terms $[X^0][\hat X^3]-[\hat X^0][X^3]$, the last
being crossed couplings between the 2 pseudoscalars $[\bar u\gamma_5 u+\bar
d\gamma_5 d] \propto \eta$ and $[\bar u\gamma_5 u-\bar d\gamma_5 d]\propto
\pi^0$. Since $\hat\delta_X=0$ for the preservation of the 3 $SU(2)_L$
Goldstones, this binary system would have a matrix for its squared masses
proportional to $\left(\begin{array}{cc} \delta_X & \frac{\delta_{X\hat
X}+\kappa_{\hat X X}}{2} \cr  
\frac{\delta_{X\hat X}+\kappa_{\hat X X}}{2} & 0\end{array}\right)$ and
would exhibit a tachyonic state without the condition $\delta_{X\hat X} +
\kappa_{\hat X X}=0$.

\section{Normalizing the pions}\label{section:norm1gen}

As was already written in (\ref{eq:example1}) and (\ref{eq:example2})
 and since there is no mixing, PCAC
\footnote{We use PCAC with a factor $\sqrt{2}$ in the r.h.s. of
(\ref{eq:PCACm}), such that the corresponding value of $f_\pi$ is $f_\pi
\approx 93\,MeV$.}
\begin{equation}
\partial_\mu(\bar u\gamma_5\gamma^\mu d)=i(m_u+m_d)\bar u\gamma_5 d =
\sqrt{2}f_\pi m_\pi^2 \pi^+
\label{eq:PCAC}
\end{equation}
and the GMOR relation
\begin{equation}
(m_u+m_d)<\bar u u+\bar d d> = 2f_\pi^2 m_\pi^2
\label{eq:GMOR}
\end{equation}
entail that
\begin{equation}
[X^\pm] = \mp i \frac{v_X}{f_\pi} \pi^\pm.
\end{equation}

The kinetic terms are written in the standard way
\begin{equation}
{\cal L}_{kin} = (D_\mu[X])^\dagger D^\mu[X] + (D_\mu[\hat X])^\dagger
 D^\mu [\hat X],
\label{eq:kin1gen}
\end{equation}
where $D_\mu$ is the covariant derivative with respect to the $SU(2)_L$
group.
Since $[X]^\dagger [X] \ni -[X^+][X^-]$ the charged pions will be
normalized in a standard way if
\begin{equation}
v_X =f_\pi,
\label{eq:vx1gen}
\end{equation}
and, according to subsection \ref{subsec:golds},
 the masses of these pseudo-Goldstones will correspond to the pion
mass if
\begin{equation}
\delta_X = m_\pi^2.
\label{eq:deltax1gen}
\end{equation}

\section{The mass of the gauge bosons}\label{section:mw1gen}

The $W$ gauge bosons get their masses from the VEV's of the 2 Higgs bosons
$[\hat X^3]$ and $[X^0]$;
from the kinetic terms (\ref{eq:kin1gen}) one gets
\begin{equation}
m_W^2 = \frac{g^2}{2}\big(<[X^0]>^2 + <[\hat X^3]>^2\big)=g^2\frac{v_X^2 +\hat
v_X^2}{4},
\label{eq:mwigen}
\end{equation}
in which $g$ is the $SU(2)_L$ coupling constant (in (\ref{eq:gdef}) below
$G_F$ is the Fermi constant)
\begin{equation}
\frac{g^2}{8m_W^2} = \frac{G_F}{\sqrt{2}} 
\Rightarrow g\approx .61.
\label{eq:gdef}
\end{equation}
Because $v_X$ given by (\ref{eq:vx1gen}) is $\ll m_W$,
\begin{equation}
\hat v_X \approx \frac{2m_W}{g}.
\label{eq:hvx1gen0}
\end{equation}
The hierarchy of the two bosonic VEV's for 1 generation is thus very large
\begin{equation}
\frac{<[\hat X^3]>}{<[X^0]>}= \frac{\hat v_X/\sqrt{2}}{v_X/\sqrt{2}}
 \approx \frac{2m_W/g}{f_\pi} \approx 2858.
\label{eq:hier1gen}
\end{equation}

\section{The scalar potential and the masses of the Higgs bosons}
\label{section:scalpot1gen}

$\bullet$\ The genuine potential $V$ we take as
\begin{equation}
V([X],[\hat X]) = -\frac{m_H^2}{2}\big(
[X]^\dagger [X] + [\hat X]^\dagger [\hat X]\big)
+\frac{\lambda_H}{4}\Big(
([X]^\dagger[X])^2 + ([\hat X]^\dagger[\hat X])^2
\Big).
\end{equation}
With respect to the most general potential for two Higgs doublets
the following terms
have accordingly been discarded:\newline
*\  $(m^2 X^\dagger \hat X +h.c)$, with $m \in {\mathbb C}$
would mediate in particular transitions between scalars and pseudoscalars
that should not occur classically;\newline
*\  $\lambda_4(X^\dagger X)(X^\dagger \hat X) + h.c.$,
$\lambda_5(\hat X^\dagger \hat X)(X^\dagger \hat X) + h.c.$
with $\lambda_4, \lambda_5 \in {\mathbb C}$ would also mediate
unwanted classical transitions between scalars and pseudoscalars;\newline
*\  $\lambda_3(X^\dagger \hat X)^2 + h.c.$ with $\lambda_3 \in
{\mathbb C}$
would in particular  contribute to the mass of the neutral pion and not
to that of the charged pions. Such a classical $\pi^+-\pi^0$ mass
difference which is not electromagnetic nor due to $m_u \not= m_d$ is
unwelcome;\newline
*\  $\lambda_1(X^\dagger X)(\hat X^\dagger \hat X)$,
$\lambda_2(X^\dagger \hat X)(\hat X^\dagger X)$,
with $\lambda_1, \lambda_2 \in {\mathbb R}$ would also spoil the
Goldstone nature of the pions and $\eta$, the first because of terms
proportional to $<[\hat X^3]>^2 \vec\pi^2$ and $<[X^0]>^2
\eta^2$, the second because of terms proportional to $<[X^0]>^2
\eta^2, <[\hat X^3]>^2 {\pi^0}^2$ and $<[X^0]><[\hat X^3]> \pi^0 \eta$.

$\bullet$\ Owing to (\ref{eq:Veffdev}), (\ref{eq:Lyukbos2}) and (\ref{eq:truegol})
 the effective potential for the Higgs bosons is
\begin{equation}
\begin{split}
V_{eff}([X],[\hat X]) &= V([X],[\hat X]) -{\cal L}_{Yuk}^{bos}([X],[\hat
X])\cr
&= -\frac{m_H^2}{2}\big(
[X]^\dagger [X] + [\hat X]^\dagger [\hat X]\big)
+\frac{\lambda_H}{4}\Big(
([X]^\dagger[X])^2 + ([\hat X]^\dagger[\hat X])^2
\Big)
+\delta_X [X]^\dagger [X].
\end{split}
\end{equation}

$\bullet$\ It minimization with respect to $[\hat X^3]$ yields as expected
\begin{equation}
<[\hat X^3]>^2 = \frac{v_0^2}{2} \Leftrightarrow \hat v_X = v_0,
\label{eq:hvx1gen}
\end{equation}
which entails in particular
\begin{equation}
b_X \stackrel{(\ref{eq:bdef})}{\equiv} \left(\frac{v_X}{v_0}\right)^2
 = \left(\frac{v_X}{\hat v_X}\right)^2=\left(\frac{f_\pi}{2m_W/g}\right)^2 \approx
\frac{1}{2858^2}.
\label{eq:bx1gen}
\end{equation}
Then, by (\ref{eq:delrel1})
\begin{equation}
\delta =\frac{\delta_X}{1-b_X} \approx \delta_X = m_\pi^2.
\label{eq:delta1gen}
\end{equation}

The second derivative of $V_{eff}$  yields the mass squared of the
corresponding
Higgs boson $\hat x^3$ defined by $[\hat X^3] = \frac{\hat v_X}{\sqrt{2}}+\hat
x^3$
\begin{equation}
 m_{\hat x^3}^2 = 2\delta\approx 2\delta_X =2m_\pi^2 \approx (194\,MeV)^2.
\label{eq:mhx31gen}
\end{equation}

$\bullet$\ The minimization of $V_{eff}$
 with respect to $[X^0]$ yields at expected
\begin{equation}
<[X^0]>^2 = b_X \frac{v_0^2}{2},
\end{equation}
and the mass of the second Higgs boson $x^0$ defined by
$[X^0]=\frac{v_X}{\sqrt{2}} + x^0$ is
\begin{equation}
m_{x^0}^2 = b_X m_{\hat x^3}^2 \approx (68\,KeV)^2.
\label{eq:mx01gen}
\end{equation}
As a side product of (\ref{eq:hvx1gen}) one gets from (\ref{eq:deltadef})
\begin{equation}
\lambda_H =\frac{4\delta}{\hat v_X^2} \approx g^2 \frac{m_\pi^2}{m_W^2}
\ll1.
\label{eq:lambda1gen}
\end{equation}

Several remarks are in order.\newline
* the ratio of the masses of the 2 Higgs bosons is the same as the  ratio
of their VEV's $\frac{m_{\hat x^3}}{m_{x^0}} = \frac{\hat
v_X/\sqrt{2}}{v_X/\sqrt{2}}=\frac{2m_W/g}{f_\pi}$;\newline
* a very light scalar $x^0$ appears, with a mass of a few ten
$KeV$;\newline
* the ``quasi-standard'' Higgs boson is still very light, its mass being
$\simeq m_\pi$; this shows that the ``Higgs mass'' is not likely to be set
by $m_W$ but rather by the mass of pseudoscalar mesons, themselves in
relation with the masses of quarks. We shall confirm this fact in the case
of 2 generations;\newline
* that the Higgs mass may increase when more generations are added in
indeed  a good sign, but, then, (\ref{eq:lambda1gen}) may cause problems if
$m_\pi$ is replaced by $m_W/g$, since the quartic coupling of the Higgs
potential becomes then larger than $1$, such that perturbative techniques
can no longer be used.

\section{Quark masses determine the last two parameters}
\label{section:qmass1gen}

$\mu_X^3 \equiv \frac{<\bar u u+\bar d d>}{\sqrt{2}}$ is related to the
masses of $u$ and $d$ quarks and to $m_\pi^2$ by the GMOR relation
(\ref{eq:GMOR})
such that, combined with (\ref{eq:vx1gen}), the first equation of
(\ref{eq:params1gen}) yields $\rho_u +\rho_d= \frac{m_u+m_d}{\sqrt{2}f_\pi}$.

Because of the condition (\ref{eq:Lyukbos2}),
the second and third equations of (\ref{eq:params1gen}) yield
$\rho_u-\rho_d = -(\lambda_u+\lambda_d)\frac{\hat
v_X}{v_X}\frac{\mu_X^3}{\hat\mu_X^3}$, in which $\frac{\hat v_X}{v_X} =
\frac{2m_W/g}{f_\pi}=\sqrt{b_X}$ has been determined in (\ref{eq:bx1gen}),
 and  $\mu_X^3$ is known.

From the last of eqs.~(\ref{eq:params1gen}) and the condition
$\hat\delta_X=0$ (see footnote \ref{foot:hdelx}) one knows that
$\lambda_u=\lambda_d=\lambda_{u,d}$; then the third equation of
this same set yields
$\delta_{X\hat X}=\lambda_{u,d}\;\frac{2\mu_X^3}{v_X}$.

Last, 2 more relations are given by the $u$ and $d$ quark masses, which spring
from the VEV's of both Higgs bosons according to
\begin{equation}
m_u = \rho_u <[X^0]> + \lambda_u <[\hat X^3]> = \frac{v_X \rho_u + \hat v_X
\lambda_u}{\sqrt{2}}, \quad
m_d= \rho_d <[X^0]> + \lambda_d <[\hat X^3]> = \frac{v_X \rho_d+ \hat v_X
\lambda_d}{\sqrt{2}},
\label{eq:qmass1gen}
\end{equation}
which entail, using the results just  obtained above
\begin{equation}
\begin{split}
m_u+m_d &=
(\rho_u+\rho_d)\frac{v_X}{\sqrt{2}}+(\lambda_u+\lambda_d)\frac{\hat
v_X}{\sqrt{2}}= \frac{m_u+m_d}{2} +2\lambda_{u,d} \frac{\hat
v_X}{\sqrt{2}}\Rightarrow \lambda_{u,d}= \frac{m_u+m_d}{4\hat v_X/\sqrt{2}},\cr
m_u-m_d &= (\rho_u-\rho_d)\frac{v_X}{\sqrt{2}} +
\underbrace{(\lambda_u-\lambda_d)}_{0}\frac{\hat
v_X}{\sqrt{2}}=-2\lambda_{u,d}\frac{\hat
v_X}{\sqrt{2}}\frac{\mu_X^3}{\hat\mu_X^3}=-\frac{m_u+m_d}{2}\frac{\mu_X^3}{\hat\mu_X^3}.
\end{split}
\label{eq:params1}
\end{equation}
The last equation in (\ref{eq:params1}) yields
\begin{equation}
\hat r_X \equiv\frac{\hat\mu_X^3}{\mu_X^3} \equiv\frac{<\bar u u-\bar d d>}{<\bar u u+\bar
d d>} =-\frac12 \frac{m_u+m_d}{m_u-m_d},
\label{eq:hrx1gen}
\end{equation}
and the first
\begin{equation}
\delta_{X\hat X}= -\kappa_{\hat X X}=\frac{(m_u+m_d)\mu_X^3}{\sqrt{2} v_X
\hat v_X} \approx \frac{f_\pi m_\pi^2}{2m_W/g},
\label{eq:delxhx1gen}
\end{equation}
in which we have used the GMOR relation (\ref{eq:GMOR}) and the values of
$v_X$ and $\hat v_X$ that we have obtained in (\ref{eq:vx1gen}) and
(\ref{eq:hvx1gen0}).

\section{Summary and comments  for 1 generation}
\label{section:summary1gen}

We have now determined all parameters and the masses of the 2 Higgs bosons
in terms of physical quantities:
the mass of the $W$'s, the pions mass and their decay
constant $f_\pi$, the $SU(2)_L$ coupling constant $g$ and the 2 quark
masses $m_u$ and $m_d$. We summarize the results below.
\begin{equation}
\begin{split}
& \bullet\ \text{``bosonic'' VEV's :}\  <X^0>\equiv \frac{v_X}{\sqrt{2}}=
\frac{f_\pi}{\sqrt{2}},\quad
<\hat X^3>\equiv\frac{\hat v_X}{\sqrt{2}}=\frac{\sqrt{2}m_W}{g},\cr
&  \bullet\ \text{``fermionic'' VEV's :}\  <\bar u u+\bar d d>
\stackrel{(\ref{eq:GMOR})}{=}
 \frac{2f_\pi^2 m_\pi^2}{m_u+m_d},\quad \frac{<\bar u u-\bar d d>}{<\bar u
u+\bar d d>} = -\frac12 \frac{m_u+m_d}{m_u-m_d},\cr
&  \bullet\ \text{Higgs bosons masses :}\  m_{\hat x^3} = \sqrt{2}m_\pi,
\quad m_{x^0} = m_{\hat x^3}
\sqrt{\frac{<X^0>}{<\hat X^3>}}
=\sqrt{2}m_\pi\sqrt{\frac{f_\pi}{2m_W/g}},\cr
&  \bullet\ \text{Yukawa couplings :}\  \delta_X=m_\pi^2,\quad
\hat\delta_X=0,\quad
\delta_{X\hat X}=-\kappa_{\hat X X}\approx \frac{f_\pi m_\pi^2}{2m_W/g},\cr
& \bullet\ \text{scalar couplings :}\ m_H^2 =2\delta \approx
2\delta_X=2m_\pi^2,\quad \lambda_H = \frac{4\delta}{\hat v_X^2} \approx
\frac{g^2 m_\pi^2}{m_W^2}\ll1.
\end{split}
\label{eq:summ1gen}
\end{equation}
We already notice that the second equation in (\ref{eq:summ1gen}) points at
$m_d <0$
\footnote{otherwise one gets $\frac{<\bar u u-\bar d d>}{<\bar u u+\bar d
d>} \approx \frac 32$.} such that
\begin{equation}
\frac{<\bar u u-\bar d d>}{<\bar u u+\bar d d>} =\frac12
\frac{|m_d|-m_u}{|m_d|+m_u} \approx \frac16.
\label{eq:mdneg1gen}
\end{equation}
That $m_d <0$ is needed  will be confirmed in the case of 2 generations
\footnote{
In the GSW model, the sign of the fermion masses is irrelevant,
at least at the classical level since swapping the
sign involves a $\gamma_5$ transformation, which should be anomaly-free.
Indeed, by a transformation $d \to e^{i\beta\gamma_5}d$, $\bar d d$ becomes
$(\cos 2\beta\, \bar d d + i \sin 2\beta\, \bar d \gamma^5 d)$ such that, for
$\beta=\pi/2$, this is equivalent to swapping the sign of $m_d$.
One must however keep in mind that, in the present framework, this
transformation must also be operated on all quark bilinears:  for
example, $\bar u\gamma^5 d$ becomes $i\bar u d$  and the parity of bilinears
involving one $d$ quark gets swapped. In the 1-generation case,
the components of the $X$ quadruplet (\ref{eq:Xq1gen}) become 
proportional to $(\bar u u-\bar d d, \bar u\gamma^5 u + \bar d\gamma^5 d,
2i\,\bar u d, 2i\,\bar d u)$ such that, in particular, $<\bar u u+ \bar d
d>$ becomes $<\bar u u-\bar d d>$.}
.

The masses of the gauge bosons and of the pions have been used as inputs
and are therefore  suitably accounted for.
The mass of the quasi-standard Higgs boson is  still much too low,
but the situation will improve for 2 generations, showing that this mass
is connected with that of the heaviest pseudoscalar bound state.
The second Higgs boson is a very light scalar. A detailed study of such
light states and their couplings will be the subject of a forthcoming work
\cite{Machet5}, in which all couplings between the different fields
will be specially scrutinized because, there, lies non-standard physics.

The hierarchy between the ``bosonic'' VEV's of the 2 Higgs bosons is very
large, but  will be replaced by more numerous but smaller
hierarchies when 1 generation is added.

\section{The leptonic decay $\boldsymbol{\pi^+ \to \ell^+ \nu_\ell}$}
\label{section:1genlept}

It is generally believed than any tentative to build  a composite
``standard-like'' Higgs
boson with usual quarks is doomed to failure because of the factor
$f_\pi/m_W$. Either the mass of the Higgs boson that one gets is ${\cal
O}(f_\pi, m_\pi)$, or leptonic decays are off by the factor $f_\pi/m_W$
because the longitudinal component of the massive $W$ is also the
pseudo-goldstone pion (see \cite{Susskind}).

This argumentation becomes void in our framework, as already briefly
mentioned in the introduction, due to the following properties:\newline
*\ we are dealing with a model with several composite Higgs
multiplets;\newline
*\ to every such multiplet is attached two VEV's: a bosonic one, like the
$v/\sqrt{2}$ of the GSW model, and a fermionic one directly connected to
$<\bar q q>$ quark condensates: there are now enough scales to accommodate
for both chiral and weak physics;\newline
*\ low energy considerations fix the normalizations of a set of charged
pseudoscalar composite states, like the ones associated to $\pi^\pm, K^\pm,
\ldots$. That the corresponding mesons  get no longer normalized to
$1$ provides the ultimate mechanism to reconcile ``pion'' physics and weak
symmetry breaking.\newline
The argumentation that we develop below is independent of the relation
(\ref{eq:vx1gen}) and goes, as we shall see, beyond the trivial case of 1
generation.

$\pi^+ \to \ell^+ \nu_\ell$ decays are triggered by the crossed terms
\begin{equation}
\partial^\mu X^+ (ig)W_\mu^-\underbrace{T_L^+.X^-}_{-X^0-X^3}
\label{eq:pilept1}
\end{equation}
 which arise in the kinetic Lagrangian for the $X$ quadruplet.
Since $X^0$ gets a VEV $v_X/\sqrt{2}$, (\ref{eq:pilept1}) induces the
coupling
\begin{equation}
-ig \frac{v_X}{\sqrt{2}}\; \partial^\mu X^+ W_\mu^-.
\label{eq:pilept2}
\end{equation}
Using PCAC and the GMOR relation, we have already shown (see for example in
the introduction) that
\begin{equation}
X^+ = -i\frac{v_X}{f_\pi} \pi^+ = a_X \pi^+,\quad a_X=-i\frac{v_X}{f_\pi}.
\label{eq:Xpi1}
\end{equation}

From (\ref{eq:Xpi1}) we deduce that the matrix element ${\cal M}_\pi=
\langle \ell \nu_\ell | \pi^+\rangle$ is given by
\begin{equation}
{\cal M}_\pi
= \frac{1}{a_X} \underbrace{\langle \ell\nu_\ell | X^+\rangle}_{{\cal
M}_X}.
\label{eq:pilept3}
\end{equation}
${\cal M}_X$ can be easily calculated from the genuine Lagrangian, since,
in there, all fields are ``normalized to $1$''. From (\ref{eq:pilept3}) and
using the unitary gauge for the $W$, one gets
\footnote{In the two equations below, $v_\ell, \bar v_{\nu_\ell}$
 stand for Dirac spinors
and should not be confused with bosonic VEV's.}
\begin{equation}
{\cal M}_X = -ig \frac{v_X}{\sqrt{2}} (ip_\mu) \frac{-i
g_{\mu\nu}}{p^2-m_W^2}(ig)\;
v_\ell(k) \gamma^\mu\frac{1-\gamma_5}{2} \bar v_{\nu_\ell}(k'),
\end{equation}
and thus, by (\ref{eq:pilept3}) and at $p^2 = m_\pi^2 \ll m_W^2$
\begin{equation}
{\cal M}_\pi = ip_\mu \frac{g^2 f_\pi}{\sqrt{2}\,m_W^2} \;
v_\ell(k) \gamma^\mu\frac{1-\gamma_5}{2} \bar v_{\nu_\ell}(k'),
\end{equation}
which is the ``standard'' PCAC amplitude. It is controlled by the product
$v_X/a_X \simeq f_\pi$ {\em independent of} $v_X$
\footnote{such that $v_X$ can take any value, like $v_x \simeq \hat v_H$
for 2 generations.}.

\underline{\em Miscellaneous remarks}

$\bullet$\ In the usual S-matrix formalism
\footnote{See for example \cite{BjorkenDrell}, chapt.~16.},
$\langle 0\ |X^+(x)|\pi^+(p)\rangle =\sqrt{Z}\langle
0\ |X^+_{in}(x)|\pi^+(p)\rangle$ with $\langle 0 \
|X^+_{in}(x)|\pi^+(p)\rangle=
\int d^3q\;e^{-iqx}\langle 0\ | a_{in}(q)|\pi^+(p)\rangle= e^{-ipx}$,
$\sqrt{Z}$ represents the amplitude for creating a 1-pion state from the
vacuum with $X^+(x)$. In our case, $\sqrt{Z} \simeq v_X/f_\pi$. For 1
generation, it is equal to $1$ (see (\ref{eq:vx1gen}), but becomes much larger
for 2 generations.

$\bullet$\ While the components of $X$ are  interacting (composite)
bosonic fields entering the Lagrangian,
$\pi^\pm$ are the physical asymptotic states. So as to calculate the decay
rate, it is these physical states that we  normalize  according to
\begin{equation}
\langle p'\; in | p\; in\rangle = 2p^0 (2\pi)^3 \delta(\vec p-\vec p'),
\end{equation}
such that
\begin{equation}
|p\rangle= \int \underbrace{\frac{d^4p'}{(2\pi)^3}
\theta(p'_0)\delta({p'}^2-m^2)}_{d\mu(p')} |p'\rangle\langle p'|p\rangle.
\end{equation}
The phase space measure for outgoing particles is as usual
\begin{equation}
d\mu(k) = \delta(k^2 -m)\; d^4k \frac{1}{(2\pi)^3} \theta(k_0).
\end{equation}
and
\begin{equation}
d\Gamma = \frac{1}{2m_\pi}|{\cal M}_\pi|^2 (2\pi)^4 \delta^4(p-k-k')\;
d\mu(k)\;d\mu(k').
\end{equation}


\chapter{The case  $\boldsymbol{N=2}$ generations . General results}
\label{chapt:2gengen}

We give here  the basic formul{\ae} that will be used in
the rest of the paper: the explicit expressions of the quadruplets, the notations, the kinetic
terms and gauge bosons masses, the masses and orthogonality relations
of pseudoscalar mesons, and the
master formula (\ref{eq:cab0}) linking the mixing angles
$\theta_d$ and $\theta_u$ to the masses of charged pseudoscalars.
We also introduce the group $U(2)^g_L \times U(2)^g_R$
 that moves inside the space of quadruplets.

\section{The 8 Higgs quadruplets}
\label{section:quadruplets}

According to (\ref{eq:generquad}), we consider hereafter the $4$ following
quadruplets of the type
   $({\mathfrak s}^0,\vec {\mathfrak p})$
($\psi$ stands now for $(u,c,d,s)^t$)

\begin{equation}
\boxed{
\begin{split}
 X &=\frac{v_X}{\sqrt{2}\,\mu_X^3}\frac{1}{\sqrt{2}}\ \bar\psi\left(
\left(\begin{array}{rrrrr}
1 &  & \vline &  & \cr
 & 0 & \vline & & \cr
\hline
 & & \vline & 1 &  \cr
 & & \vline &  & 0 \end{array}\right),
\gamma^5\left(\begin{array}{rrrrr}
1 &  & \vline &  & \cr
 & 0 & \vline & & \cr
\hline
 & & \vline & -1 &  \cr
 & & \vline &  & 0 \end{array}\right),
2\gamma^5\left(\begin{array}{rrrrr}
 &  & \vline & 1 &  \cr
 &  & \vline &  & 0\cr
\hline
 & & \vline &  &  \cr
 & & \vline &  & \end{array}\right),
2\gamma^5\left(\begin{array}{rrrrr}
 &  & \vline &  &  \cr
 &  & \vline &  & \cr
\hline
1 &   & \vline &  &  \cr
 & 0  & \vline &  & \end{array}\right)
\right)\psi\cr
&= (X^0,X^3,X^+,X^-)
 \quad with\quad \mu_X^3 = \frac{<\bar u u+\bar d d>}{\sqrt{2}},
\end{split}
}
\label{eq:Xq2gen}
\end{equation}

\begin{equation}
\boxed{
\begin{split}
 H &=\frac{v_H}{\sqrt{2}\,\mu_H^3}\frac{1}{\sqrt{2}}\ \bar\psi\left(
\left(\begin{array}{ccccc}
0 &  & \vline &  & \cr
 & 1 & \vline & & \cr
\hline
 & & \vline & 0 &  \cr
 & & \vline &  & 1 \end{array}\right),
\gamma^5 \left(\begin{array}{ccccc}
0 &  & \vline &  & \cr
 & 1 & \vline & & \cr
\hline
 & & \vline & 0 &  \cr
 & & \vline &  & -1 \end{array}\right),
2\gamma^5 \left(\begin{array}{ccccc}
 &  & \vline & 0 &  \cr
 &  & \vline &  & 1\cr
\hline
 & & \vline &  &  \cr
 & & \vline &  & \end{array}\right),
2\gamma^5 \left(\begin{array}{ccccc}
 &  & \vline &  &  \cr
 &  & \vline &  & \cr
\hline
0 &   & \vline &  &  \cr
 & 1  & \vline &  & \end{array}\right)
\right)\psi\cr
&= (H^0,H^3,H^+,H^-)
 \quad with\quad \mu_H^3 = \frac{<\bar c c+\bar s s>}{\sqrt{2}},
\end{split}
}
\label{eq:Hq2gen}
\end{equation}

\begin{equation}
\boxed{
\begin{split}
 \Omega
&=\frac{v_\Omega}{\sqrt{2}\,\mu_\Omega^3}\frac12\ \bar\psi\left(
\left(\begin{array}{ccccc}
 & 1 & \vline &  & \cr
1 &  & \vline & & \cr
\hline
 & & \vline &  & 1 \cr
 & & \vline & 1 &  \end{array}\right),
\gamma^5\left(\begin{array}{ccccc}
 & 1 & \vline &  & \cr
1 &  & \vline & & \cr
\hline
 & & \vline &  & -1 \cr
 & & \vline & -1 &  \end{array}\right),
2\gamma^5\left(\begin{array}{ccccc}
 &  & \vline &  & 1 \cr
 &  & \vline & 1 & \cr
\hline
 & & \vline &  &  \cr
 & & \vline &  & \end{array}\right),
2\gamma^5\left(\begin{array}{ccccc}
 &  & \vline &  &  \cr
 &  & \vline &  & \cr
\hline
 & 1  & \vline &  &  \cr
1 &   & \vline &  & \end{array}\right)
\right)\psi\cr
&= (\Omega^0,\Omega^3,\Omega^+,\Omega^-)
 \quad with\quad \mu_\Omega^3 = \frac{<\bar u c+\bar c u +\bar d
s+\bar s d>}{2},
\end{split}
}
\label{eq:Omq2gen}
\end{equation}

\begin{equation}
\boxed{
\begin{split}
\Xi &=\frac{v_\Xi}{\sqrt{2}\,\mu_\Xi^3}\frac12\ \bar\psi\left(
\left(\begin{array}{rrrrr}
 & 1 & \vline &  & \cr
-1 &  & \vline & & \cr
\hline
 & & \vline &  & 1 \cr
 & & \vline & -1 &  \end{array}\right),
\gamma^5\left(\begin{array}{rrrrr}
 & 1 & \vline &  & \cr
-1 &  & \vline & & \cr
\hline
 & & \vline &  & -1 \cr
 & & \vline & 1 &  \end{array}\right),
2\gamma^5\left(\begin{array}{rrrrr}
 &  & \vline &  & 1 \cr
 &  & \vline & -1 & \cr
\hline
 & & \vline &  &  \cr
 & & \vline &  & \end{array}\right),
2\gamma^5\left(\begin{array}{rrrrr}
 &  & \vline &  &  \cr
 &  & \vline &  & \cr
\hline
 & 1  & \vline &  &  \cr
-1 &   & \vline &  & \end{array}\right)
\right)\psi\cr
&= (\Xi^0,\Xi^3,\Xi^+,\Xi^-)
 \quad with\quad \mu_\Xi^3 = \frac{<\bar u c -\bar c u + \bar d s  -\bar
 s d>}{2},
\end{split}
}
\label{eq:Xiq2gen}
\end{equation}
to which are added their $4$ parity-transformed alter egos of the type
$({\mathfrak p}^0, \vec {\mathfrak s})$, that we call
$\hat X, \hat H, \hat\Omega, \hat\Xi$. Since the algebraic aspect of the
last four is the same, but for the place of the $\gamma_5$ matrix, as that
of $X, H, \Omega, \Xi$, we omit writing them explicitly.
To  $\hat X, \hat H, \hat\Omega, \hat\Xi$ are associated the
bosonic VEV's
$\hat v_X, \hat v_H, \hat v_\Omega, \hat v_\Xi$, and the fermionic VEV's
\begin{equation}
\boxed{
\hat\mu_X^3 =\frac{<\bar u u-\bar d d>}{\sqrt{2}},
\hat\mu_H^3=\frac{<\bar c c-\bar s s>}{\sqrt{2}},
\hat\mu_\Omega^3 =\frac{<\bar u c+\bar c u - \bar d s-\bar s d>}{2},
\hat\mu_\Xi^3=\frac{<\bar u c-\bar c u-\bar d s+\bar s d>}{2}}
\end{equation}

The 8 Higgs bosons are the 8 scalars $[X^0], [H^0], [\Omega^0], [\Xi^0],
[\hat X^3], [\hat H^3], [\hat\Omega^3], [\hat\Xi^3]$, and
one has a total of $2 \times 2N^2 = 16$ VEV's to determine: 8 $v$'s and 8
$\mu^3$'s.

\subsection{Choosing the quasi-standard Higgs doublet}
\label{subsec:whichone}

We have now to choose  which quadruplet includes the 3 Goldstone
bosons of the spontaneously broken $SU(2)_L$, knowing that they will
disappear from the mesonic spectrum.

It is not suitable a priori that the charged Goldstones are pseudoscalars,
because charged pseudoscalar mesons are observed without ambiguity.
This excludes all $({\mathfrak s}, \vec{\mathfrak p})$-like (``unhatted'')
 quadruplets. Among $\hat X, \hat H, \hat\Omega, \hat\Xi$ it is easy to see
that the neutral pseudoscalar component of the last two are combinations of
neutral $K$ and $D$ mesons, that we  also want to keep in the mesonic
spectrum.
So, the only 2 left possibilities are $\hat X$ and $\hat H$. We choose
$\hat H$ because it involves the heavy quark $c$ and because it looks
more natural to eliminate from the spectrum $\bar c\gamma_5 c + \bar
s\gamma_5 s$ rather than $\bar u\gamma_5 u + \bar d\gamma_5 d$ which is
related to observed low mass pseudoscalar mesons. The reader can
argue that $\eta_c$ is also observed, but we have in mind that, for 3
generations, there is not so much problem in eliminating from the spectrum 
some pseudoscalar singlet including $\bar t\gamma_5 t$.

We accordingly postulate that, for 2 generations
\begin{equation}
\boxed{
W^+_\parallel \sim \hat H^+ \sim \bar c s\ (scalar),
W^-_\parallel \sim \hat H^- \sim \bar s c\ (scalar),
W^3_\parallel \sim \hat H^0 \sim \bar c\gamma_5 c + \bar s\gamma_5 s\
 (pseudoscalar)
}
\label{eq:Wpar}
\end{equation}

\subsection{Notations}\label{subsec:notations}

For a systematic and not too tedious
 treatment of the set of equations that will follow, we
introduce, as already mentioned in (\ref{eq:bdef}),
the following dimensionless ratios of bosonic VEV's, normalized
to the one of the ``quasi-standard'' Higgs multiplet $\hat H$
($\hat b_H$ is not present simply because it is identical to $1$ by
definition)
\begin{equation}
\boxed{
\begin{split}
& b_X\equiv \left(\frac{v_X}{\hat v_H}\right)^2,\quad
b_H\equiv \left(\frac{v_H}{\hat v_H}\right)^2,\quad
b_\Omega\equiv \left(\frac{v_\Omega}{\hat v_H}\right)^2,\quad
b_\Xi\equiv \left(\frac{v_\Xi}{\hat v_H}\right)^2,\cr
& \hat b_X\equiv \left(\frac{\hat v_X}{\hat v_H}\right)^2,\quad
\hat b_\Omega\equiv \left(\frac{\hat v_\Omega}{\hat v_H}\right)^2,\quad
\hat b_\Xi\equiv \left(\frac{\hat v_\Xi}{\hat v_H}\right)^2.
\end{split}
}
\label{eq:bdefs}
\end{equation}
Likewise we introduce dimensionless ratios of fermionic VEV's, normalized
this time to $\mu_X^3=\frac{<\bar u u+\bar d d>}{\sqrt{2}}$
  ($r_X=1$ by definition)
\begin{equation}
\boxed{
\begin{split}
r_H\equiv\frac{\mu_H^3}{\mu_X^2},\quad
r_\Omega\equiv\frac{\mu_\Omega^3}{\mu_X^2},\quad
r_\Xi\equiv\frac{\mu_\Xi^3}{\mu_X^3},\cr
\hat r_X \equiv \frac{\hat\mu_X^3}{\mu_X^3},\quad
\hat r_H \equiv \frac{\hat\mu_H^3}{\mu_X^3},\quad
\hat r_\Omega \equiv \frac{\hat\mu_\Omega^3}{\mu_X^3},\quad
\hat r_\Xi \equiv \frac{\hat\mu_\Xi^3}{\mu_X^3}.
\end{split}
}
\label{eq:rdef}
\end{equation}

To lighten forthcoming equations, we shall also introduce the following
ratios of bosonic to fermionic VEV's
\begin{equation}
\boxed{
\frac{1}{\nu_i^2}\equiv\frac{v_i}{\sqrt{2}\mu_i^3},\quad
\frac{1}{\bar\nu_i^4}\equiv\frac{1-b_i}{\nu_i^4},\quad
\frac{1}{\hat\nu_i^2}\equiv\frac{\hat v_i}{\sqrt{2}\hat \mu_i^3},\quad
\frac{1}{\overline{\hat\nu_i}^4}\equiv\frac{1-\hat b_i}{\hat \nu_i^4}
}
\label{eq:nudef}
\end{equation}

\section{The kinetic terms and the mass of the gauge bosons}
\label{section:gaugemass}

The kinetic terms are simply chosen to be a diagonal sum of the kinetic
terms for each of the 8 quadruplets
\begin{equation}
\sum_{i\in[X,\hat X, H,\hat H, \Omega,\hat\Omega, \Xi,\hat\Xi]}
(D_\mu \Delta_i)^\dagger (D^\mu\Delta_i),
\label{eq:kin2gen}
\end{equation}
in which $D_\mu$ is the covariant derivative with respect to the $SU(2)_L$
group.

From (\ref{eq:kin2gen}) one gets
\begin{equation}
\boxed{
m_W^2 = \frac{g^2}{4}\hat v_H^2(b_X+b_H+b_\Omega + b_\Xi
 + \hat b_X + \hat b_H
 +\hat b_\Omega + \hat b_\Xi)
}
\label{eq:mw2gen}
\end{equation}
We recall that $\hat b_H=1$ by definition. We shall also see in the next
chapter (eqs.~(\ref{eq:nulbxi}) and (\ref{eq:nulhbxi}))  that one can take
$b_\Xi=0=\hat b_\Xi$ as a consequence of $<\bar u c>=<\bar c u>$ and $<\bar d
s>=<\bar s d>$.

\section{Yukawa couplings and the Higgs potential}
\label{section:yukscal}

${\cal L}_{Yuk}$ is of the general form (\ref{eq:Lyuk0}) and must, by the choice
(\ref{eq:Wpar}), satisfy $\hat\delta_H=0$ (see (\ref{eq:truegol})).
In order to avoid classical
transitions between charged scalars and pseudoscalars, it must also
satisfy the 4 equations $\delta_{i\hat\imath}+\kappa_{\hat\imath i}=0, i\in
\{X,H,\Omega,\Xi\}$ in (\ref{eq:Lyukbos2}).
 This leaves 11 independent Yukawa couplings to determine.

All VEV's, the $b$'s, $\hat b$'s and the $\mu^3$'s, $\hat\mu^3$'s
 are a priori considered to be real. We shall see that this is well
supported by the solutions of our equations, but for a problem concerning
$\hat\mu_X^3 \equiv\frac{<\bar u u-\bar d d>}{\sqrt{2}}$.

The genuine scalar potential is given by (\ref{eq:Vgen}) in which $i$ spans the
set $\{X, H, \Omega,\Xi\}$. It only depends on 2 parameters, $m_H^2$ and
$\lambda_H$.
Subtracting from it the bosonised form of the Yukawa Lagrangian yields the
effective scalar potential $V_{eff}$ which determines the vacuum of the
theory.

Since $\hat \delta_H=0$ the effective potential $V_{eff}$
 for the Higgs multiplet
$\hat H$ is, by (\ref{eq:Lyukbos2}), identical to its genuine potential $V$.
It has accordingly its minimum at
$\hat v_H = v_0$ where $v_0$ has been defined in (\ref{eq:v0def}).

We also recall (see (\ref{eq:delrel1})) that the other equations
defining the minima of $V_{eff}$ entail
$\delta_i = \delta(1-b_i),\quad \hat\delta_i=\delta(1-\hat b_i).$

Note that both Yukawa couplings and the scalar potential can be written
as sums over pairs of parity-transformed quadruplets $(\Delta_i, \hat\Delta_i)$, 
the index $i$ spanning accordingly the set $\{X, H, \Omega,\Xi\}$.

\section{Charged pseudoscalar mesons}
\label{section:charged}

The Yukawa couplings induce, through the non-vanishing VEV's of $[X^0]$ and
$[\hat H^3]$ non-diagonal fermionic mass terms $\bar u c, \bar c u, \bar d
s, \bar s d$. As usual, the diagonalization of the two mass matrices leads
to the quark mass eigenstates $u_m, c_m, d_m, s_m$ which are connected to
flavor eigenstates by two rotations with respective angles $\theta_u$ and
$\theta_d$\  ($c_u$, $s_u$ mean respectively $\cos\theta_u$, $\sin\theta_u$
etc)
\begin{equation}
\left(\begin{array}{c} u \cr c\end{array}\right)
= \left(\begin{array}{cc} c_u & s_u \cr -s_u & c_u\end{array}\right)
\left(\begin{array}{c} u_m \cr c_m\end{array}\right),\quad
\left(\begin{array}{c} d \cr s\end{array}\right)
= \left(\begin{array}{cc} c_d & s_d \cr -s_d & c_d\end{array}\right)
\left(\begin{array}{c} d_m \cr s_m\end{array}\right),
\label{eq:rots}
\end{equation}
At this stage of our study, we shall not investigate
the link between Yukawa couplings and mixing angles, leaving this for 
section \ref{section:genferm}. All that we need for the present purpose
is to re-express the flavor quark bilinears in terms of bilinears
of quark mass eigenstates by using (\ref{eq:rots}). The corresponding
formul{\ae}, which are used throughout the paper,
have been gathered in Appendix \ref{section:flavmass}. 

From now onwards, the connexion between mesonic fields and bilinear quark
operators, that is, low energy relations (PCAC, GMOR),
is done through  $\bar q^i_m \gamma_5 q^j_m$
or $\bar q^i_m q^j_m$  involving quark mass eigenstates.

The main result of this section, the mixing formula (\ref{eq:cab0}),
 will be obtained from the  following very
simple statements concerning the charged $\pi^\pm, K^\pm, D^\pm, D_s^\pm$
pseudoscalar mesons only:\newline
* their  $mass^2$ are obtained from the appropriate ratios between the
 Yukawa couplings (as explained in subsection \ref{subsec:golds}) and the kinetic
 terms; for example, for charged pions, one selects in ${\cal
 L}_{Yuk}$ and in ${\cal L}_{kin}$ all terms proportional to $(\bar
u_m \gamma_5 d_m)(\bar d_m \gamma_5 u_m)$, eventually using the twofold
nature of the Higgs multiplets to express any
bosonic components  in terms of quark fields;\newline
* one demands that no non-diagonal transition occurs between any of them;
 for example one cancels in ${\cal L}_{Yuk}$
 the coefficient of the $ (\bar u_m \gamma_5 d_m) 
(\bar s_m\gamma_5 u_m)$ terms which would correspond to transitions between
$\pi^+$ and $K^+$.

All normalization factors that arise from low energy
theorems (PCAC, GMOR) cancel (in the ratios defining masses) or are
irrelevant (when setting some coefficient to zero). This entails in
particular that eq.~(\ref{eq:cab0}) does not depend on low energy
theorems but only on the much weaker hypothesis that charged mesonic
$\pi^+, K^+, D^+, D_s^+$ fields are respectively proportional to $\bar
u_m\gamma_5 d_m, \bar u_m\gamma_5 s_m, \bar c_m \gamma_5 d_m, \bar
c_m\gamma_5 s_m$.  It is therefore a robust result.

\subsection{Orthogonality}\label{subsec:orthog}

Charged pseudoscalar mesons can only be found in the ``non-hatted''
quadruplets $X, H, \Omega, \Xi$ and their quadratic terms can only be found in
the terms $-\delta_X X^\dagger X -\delta_H H^\dagger H -\delta_\Omega
\Omega^\dagger \Omega -\delta_\Xi \Xi^\dagger \Xi$ of ${\cal L}_{Yuk}$.

Using Appendix \ref{section:flavmass}, the 6 equations expressing the vanishing of non-diagonal transitions among them
are found to be the following:
\begin{equation}
\begin{split}
\pi^\pm\not\leftrightarrow K^\pm & \Leftrightarrow
\delta_X \frac{c_uc_dc_us_d}{\nu_X^4} - \delta_H
\frac{s_us_ds_uc_d}{\nu_H^4}
-\frac{\delta_\Omega}{2} \frac{s_{u+d}c_{u+d}}{\nu_\Omega^4}
+\frac{\delta_\Xi}{2} \frac{s_{u-d}c_{u-d}}{\nu_\Xi^4} =0,\cr
\pi^\pm \not\leftrightarrow D^\pm & \Leftrightarrow
\delta_X \frac{c_uc_ds_uc_d}{\nu_X^4} -\delta_H
\frac{s_us_dc_us_d}{\nu_H^4}
-\frac{\delta_\Omega}{2} \frac{s_{u+d}c_{u+d}}{\nu_\Omega^4} -
\frac{\delta_\Xi}{2} \frac{s_{u-d}c_{u-d}}{\nu_\Xi^4} =0,\cr
\pi^\pm \not\leftrightarrow D_s^\pm & \Leftrightarrow
\delta_X \frac{s_us_dc_uc_d}{\nu_X^4} +\delta_H
\frac{s_us_dc_uc_d}{\nu_H^4}
-\frac{\delta_\Omega }{2} \frac{s_{u+d}^2}{\nu_\Omega^4} +
\frac{\delta_\Xi}{2} \frac{s_{u-d}^2}{\nu_\Xi^4} =0,\cr
K^\pm \not\leftrightarrow D^\pm & \Leftrightarrow
\delta_X \frac{c_us_ds_uc_d}{\nu_X^4} +\delta_H
\frac{s_uc_dc_us_d}{\nu_H^4}
+\frac{\delta_\Omega}{2} \frac{c_{u+d}^2}{\nu_\Omega^4} -
\frac{\delta_\Xi}{2} \frac{c_{u-d}^2}{\nu_\Xi^4} =0,\cr
K^\pm \not\leftrightarrow D_s^\pm & \Leftrightarrow
\delta_X \frac{c_us_ds_us_d}{\nu_X^4} -\delta_H
\frac{s_uc_dc_uc_d}{\nu_H^4}
+\frac{\delta_\Omega }{2} \frac{s_{u+d}c_{u+d}}{\nu_\Omega^4} +
\frac{\delta_\Xi}{2} \frac{s_{u-d}c_{u-d}}{\nu_\Xi^4} =0,\cr
D^\pm \not\leftrightarrow D_s^\pm & \Leftrightarrow
\delta_X \frac{s_uc_ds_us_d}{\nu_X^4} -\delta_H
\frac{c_us_dc_uc_d}{\nu_H^4}
+\frac{\delta_\Omega}{2} \frac{s_{u+d}c_{u+d}}{\nu_\Omega^4} -
\frac{\delta_\Xi}{2} \frac{s_{u-d}c_{u-d}}{\nu_\Xi^4} =0.
\label{eq:ndiag+0}
\end{split}
\end{equation}
in which $c_{u-d}$ stands for $\cos (\theta_u-\theta_d)$ {\em etc} .

Using the relations (\ref{eq:delrel1}) allows to factor out $\delta$. Then
using the notations (\ref{eq:nudef}) transforms the 6 equations
of (\ref{eq:ndiag+0}) respectively into
\begin{equation}
\begin{split}
& (a) :
\frac{c_uc_dc_us_d}{\bar\nu_X^4} -\frac{s_us_ds_uc_d}{\bar\nu_H^4}
-\frac12 \frac{s_{u+d}\,c_{u+d}}{\bar\nu_\Omega^4} +\frac12
\frac{s_{u-d}\,c_{u-d}}{\bar\nu_\Xi^4}
=0,\cr
& (b) :
\frac{c_uc_ds_uc_d}{\bar\nu_X^4} -\frac{s_us_dc_us_d}{\bar\nu_H^4}
-\frac12 \frac{s_{u+d}\,c_{u+d}}{\bar\nu_\Omega^4} -\frac12
\frac{s_{u-d}\,c_{u-d}}{\bar\nu_\Xi^4}
=0,\cr
&  (c) :
\frac{s_us_dc_uc_d}{\bar\nu_X^4} +\frac{s_us_dc_uc_d}{\bar\nu_H^4}
-\frac12 \frac{s_{u+d}^2}{\bar\nu_\Omega^4} +\frac12
\frac{s_{u-d}^2}{\bar\nu_\Xi^4}
=0,\cr
& (d) :
\frac{c_us_ds_uc_d}{\bar\nu_X^4} +\frac{s_uc_dc_us_d}{\bar\nu_H^4}
+\frac12 \frac{c_{u+d}^2}{\bar\nu_\Omega^4} -\frac12
\frac{c_{u-d}^2}{\bar\nu_\Xi^4}
=0,\cr
& (e) :
\frac{c_us_ds_us_d}{\bar\nu_X^4} -\frac{s_uc_dc_uc_d}{\bar\nu_H^4}
+\frac12 \frac{s_{u+d}\,c_{u+d}}{\bar\nu_\Omega^4} +\frac12
\frac{s_{u-d}\,c_{u-d}}{\bar\nu_\Xi^4}
=0,\cr
& (f) :
\frac{s_uc_ds_us_d}{\bar\nu_X^4} -\frac{c_us_dc_uc_d}{\bar\nu_H^4}
+\frac12 \frac{s_{u+d}\,c_{u+d}}{\bar\nu_\Omega^4} -\frac12
\frac{s_{u-d}\,c_{u-d}}{\bar\nu_\Xi^4}
=0,
\label{eq:ndiag+1}
\end{split}
\end{equation}
or, equivalently, by recombining them,
\begin{equation}
\begin{split}
& (a)+(f) :
s_{2d}\left(\frac{1}{\bar\nu_X^4}-\frac{1}{\bar\nu_H^4}\right)=0,\cr
& (a)-(f) :
s_{2d}c_{2u}\left(\frac{1}{\bar\nu_X^4}+\frac{1}{\bar\nu_H^4}\right)-\frac{s_{2(u+d)}}{\bar\nu_\Omega^4}+\frac{s_{2(u-d)}}{\bar\nu_\Xi^4}=0,\cr
& (b)-(e) :
s_{2u}c_{2d}\left(\frac{1}{\bar\nu_X^4}+\frac{1}{\bar\nu_H^4}\right)-\frac{s_{2(u+d)}}{\bar\nu_\Omega^4}-\frac{s_{2(u-d)}}{\bar\nu_\Xi^4}=0,\cr
& (b)+(e) : s_{2u}\left(\frac{1}{\bar\nu_X^4}
-\frac{1}{\bar\nu_H^4}\right)=0,\cr
& (c)-(d) : \frac{1}{\bar\nu_\Omega^4}-\frac{1}{\bar\nu_\Xi^4}=0,\cr
& (c)+(d) : s_{2u}s_{2d}\left(\frac{1}{\bar\nu_X^4} +
\frac{1}{\bar\nu_H^4}\right)+\frac{c_{2(u+d)} }{\bar\nu_\Omega^4}
-\frac{c_{2(u-d)}}{\bar\nu_\Xi^4}=0.
\label{eq:ndiag+2}
\end{split}
\end{equation}
The solution of (\ref{eq:ndiag+2}) is
\begin{equation}
\frac{1}{\bar\nu_X^4} = \frac{1}{\bar\nu_H^4} =
\frac{1}{\bar\nu_\Omega^4}=\frac{1}{\bar\nu_\Xi^4}
\qquad \stackrel{(\ref{eq:nudef})}{\Leftrightarrow}
\qquad \frac{1-b_X}{\nu_X^4}=\frac{1-b_H}{\nu_H^4}
=\frac{1-b_\Omega}{\nu_\Omega^4}=\frac{1-b_\Xi}{\nu_\Xi^4}.
\label{eq:solnu1}
\end{equation}
Using the definitions of the $b$'s and $\nu^2$'s given in (\ref{eq:bdef})
and (\ref{eq:nudef}), (\ref{eq:solnu1}) also writes
\begin{equation}
\boxed{
\frac{b_X(1-b_X)}{\mu_X^6} =\frac{b_H(1-b_H)}{\mu_H^6}
=\frac{b_\Omega(1-b_\Omega)}{\mu_\Omega^6} =\frac{b_\Xi(1-b_\Xi)}{\mu_\Xi^6}}
\label{eq:solnu2}
\end{equation}
Note that we only took into account the non-diagonal quadratic terms that
occur in the Yukawa Lagrangian. Such terms are also present in the kinetic
Lagrangian and the conditions for their vanishing are not the same as
(\ref{eq:ndiag+0}). They however  vanish at small momentum such that the
solution (\ref{eq:solnu1}) (\ref{eq:solnu2})
 can only be considered to be valid at this limit.

\subsection{Masses}\label{subsec:masses}

From the ratios of the terms quadratic in the meson fields in the Yukawa
and kinetic terms, using (\ref{eq:delrel1}) and the notation
(\ref{eq:bdef}) one gets, with the help of Appendix \ref{section:flavmass}
\begin{equation}
\begin{split}
 m_{\pi^\pm}^2 &=\delta \frac{ \left(1-b_X\right)
(\frac{c_uc_d}{\nu_X^2})^2 +
 \left(1-b_H\right)(\frac{s_us_d}{\nu_H^2})^2 +
\left(1-b_\Omega\right) \frac12(\frac{s_{u+d}}{\nu_\Omega^2})^2
 + \left(1-b_\Xi\right)\frac12(\frac{s_{u-d}}{\nu_\Xi^2})^2 }
{(\frac{c_uc_d}{\nu_X^2})^2 + (\frac{s_us_d}{\nu_H^2})^2 +
\frac12(\frac{s_{u+d}}{\nu_\Omega^2})^2 +
\frac12(\frac{s_{u-d}}{\nu_\Xi^2})^2 },\cr
 m_{K^\pm}^2 &= \delta \frac{ \left(1-b_X\right)
(\frac{c_us_d}{\nu_X^2})^2 +
 \left(1-b_H\right)(\frac{s_uc_d}{ \nu_H^2})^2 +
\left(1-b_\Omega\right) \frac12(\frac{c_{u+d}}{\nu_\Omega^2})^2
+ \left(1-b_\Xi\right)\frac12(\frac{c_{u-d}}{\nu_\Xi^2})^2 }
{(\frac{c_us_d}{\nu_X^2})^2 + (\frac{s_uc_d}{ \nu_H^2})^2 +
\frac12(\frac{c_{u+d}}{\nu_\Omega^2})^2 +
\frac12(\frac{c_{u-d}}{\nu_\Xi^2})^2},\cr
 m_{D^\pm}^2 &= \delta \frac{ \left(1-b_X\right)
(\frac{s_uc_d}{\nu_X^2})^2 +
 \left(1-b_H\right)(\frac{c_us_d}{ \nu_H^2})^2 +
\left(1-b_\Omega\right) \frac12(\frac{c_{u+d}}{\nu_\Omega^2})^2
+ \left(1-b_\Xi\right)\frac12(\frac{c_{u-d}}{\nu_\Xi^2})^2 }
{(\frac{s_uc_d}{\nu_X^2})^2 + (\frac{c_us_d}{ \nu_H^2})^2 +
\frac12(\frac{c_{u+d}}{\nu_\Omega^2})^2 +
\frac12(\frac{c_{u-d}}{\nu_\Xi^2})^2 },\cr
 m_{D_s^\pm}^2 &= \delta \frac{ \left(1-b_X\right)
(\frac{s_us_d}{\nu_X^2})^2 +
 \left(1-b_H\right)(\frac{c_uc_d}{ \nu_H^2})^2 +
\left(1-b_\Omega\right) \frac12(\frac{s_{u+d}}{\nu_\Omega^2})^2
+ \left(1-b_\Xi\right)\frac12(\frac{s_{u-d}}{\nu_\Xi^2})^2 }
{(\frac{s_us_d}{\nu_X^2})^2 + (\frac{c_uc_d}{ \nu_H^2})^2 +
\frac12(\frac{s_{u+d}}{\nu_\Omega^2})^2 +
\frac12(\frac{s_{u-d}}{\nu_\Xi^2})^2 },
\label{eq:diag+0}
\end{split}
\end{equation}
which rewrites, using (\ref{eq:solnu1})
\begin{equation}
\begin{split}
& m_{\pi^\pm}^2 =\frac{\delta/\bar\nu_X^4}{(c_uc_d/\nu_X^2)^2 + (s_us_d/
\nu_H^2)^2 +
\frac12 (s_{u+d}/\nu_\Omega^2)^2 + \frac12 (s_{u-d}/\nu_\Xi^2)^2},\cr
& m_{K^\pm}^2 = \frac{\delta/\bar\nu_X^4}{(c_us_d/\nu_X^2)^2 + (s_uc_d/
\nu_H^2)^2 +
\frac12 (c_{u+d}/\nu_\Omega^2)^2 +
\frac12 (c_{u-d}/\nu_\Xi^2)^2},\cr
& m_{D^\pm}^2 =  \frac{\delta/\bar\nu_X^4}{(s_uc_d/\nu_X^2)^2 + (c_us_d/
\nu_H^2)^2 +
\frac12 (c_{u+d}/\nu_\Omega^2)^2 +
\frac12 (c_{u-d}/\nu_\Xi^2)^2},\cr
& m_{D_s^\pm}^2 =  \frac{\delta/\bar\nu_X^4}{(s_us_d/\nu_X^2)^2 + (c_uc_d/
\nu_H^2)^2 + \frac12 (s_{u+d}/\nu_\Omega^2)^2 +
\frac12 (s_{u-d}/\nu_\Xi^2)^2}.
\label{eq:diag+1}
\end{split}
\end{equation}
Recombining the 4 equations in (\ref{eq:diag+1}) yields
\begin{equation}
\begin{split}
& \delta \left( +\frac{1}{m_{\pi^\pm}^2}+\frac{1}{m_{K^\pm}^2} +
\frac{1}{m_{D^\pm}^2}+\frac{1}{m_{D_s^\pm}^2}\right)
=\frac{1}{1-b_X} +\frac{1}{1-b_H}+\frac{1}{1-b_\Omega}+
\frac{1}{1-b_\Xi},\cr
& \delta \left( +\frac{1}{m_{\pi^\pm}^2}-\frac{1}{m_{K^\pm}^2} +
\frac{1}{m_{D^\pm}^2}-\frac{1}{m_{D_s^\pm}^2}\right)
=c_{2d}\left(\frac{1}{1-b_X} - \frac{1}{1-b_H} \right),\cr
& \delta \left( +\frac{1}{m_{\pi^\pm}^2}+\frac{1}{m_{K^\pm}^2} -
\frac{1}{m_{D^\pm}^2}-\frac{1}{m_{D_s^\pm}^2}\right)
=c_{2u}\left(\frac{1}{1-b_X} - \frac{1}{1-b_H} \right),\cr
& \delta \left( +\frac{1}{m_{\pi^\pm}^2}-\frac{1}{m_{K^\pm}^2} -
\frac{1}{m_{D^\pm}^2}+\frac{1}{m_{D_s^\pm}^2}\right)
= c_{2u}c_{2d}\left(\frac{1}{1-b_X} + \frac{1}{1-b_H} \right)
-\frac{c_{2(u+d)}}{1-b_\Omega} -\frac{c_{2(u-d)}}{1-b_\Xi}.
\end{split}
\label{eq:diag+2}
\end{equation}

\section{Mixing : the master formula} \label{section:mix}

From the second and third equations of (\ref{eq:diag+2}) one gets,
independently of the scale $\delta$
\begin{equation}
\boxed{
\displaystyle\frac{c_{2u}-c_{2d}}{c_{2u}+c_{2d}}
\equiv\tan(\theta_d+\theta_u)\tan(\theta_d-\theta_u)
=\displaystyle\frac
{\displaystyle\frac{1}{m_{K^\pm}^2}-\displaystyle\frac{1}{m_{D^\pm}^2}}
{\displaystyle\frac{1}{m_{\pi^\pm}^2}-\displaystyle\frac{1}{m_{D_s^\pm}^2}}
}
\label{eq:cab1}
\end{equation}
which vanishes either at the chiral limit $m_\pi\to 0$ or when $m_K=m_D$.
(\ref{eq:cab1}) is independent of all the  VEV's. As we have already
mentioned, all normalization factors that occur in PCAC or GMOR relations
cancel out
\footnote{Eq.~(\ref{eq:cab1}) appears like a generalization of the result
by Oakes \cite{Oakes} which used the hypothesis that chiral $SU(2)\times
SU(2)$ breaking (by strong interactions) and the non-conservation of
strangeness (Cabibbo angle) in weak interactions have a common origin.}.
A major property of (\ref{eq:cab1}) is that $\theta_u$ and $\theta_d$
are not independent variables. In particular, the positivity of its r.h.s.
entails that the spectrum of charged pseudoscalar mesons is only compatible
with $\theta_u < \theta_d$.

\section{Neutral pseudoscalar mesons}\label{section:neutralgen}

We shall define hereafter the interpolating fields of $\pi^0$ and $\eta$ to
be respectively proportional to $\bar u\gamma_5 u - \bar d\gamma_5 d$ and
$\bar u\gamma_5 u + \bar d\gamma_5 d$, which does not exactly corresponds
to the divergences of the axial currents 
$\bar u\gamma^\mu\gamma_5 u - \bar d\gamma^\mu\gamma_5 d$ and
$\bar u\gamma^\mu\gamma_5 u + \bar d\gamma^\mu\gamma_5 d$. We make this
choice because of the negative sign that comes out for
the $d$ quark mass; we shall see that it leads to suitable orthogonality
relations and mass for the $\pi^0$, and that only $\eta$ looks somewhat
more problematic. Deeper investigations concerning this states are
postponed to further works, and we shall not use $\eta$ in the present one
to fit the parameters.

\subsection{Orthogonality}

The equations are the following
\begin{equation}
\boxed{
\begin{split}
& (a)\  D^0 \perp K^0 :
s_uc_u s_dc_d(-\frac{\delta_X}{\nu_X^4} +\frac{\hat\delta_X}{\hat\nu_X^4}
-\frac{\delta_H}{\nu_H^4}+\frac{\hat\delta_H}{\hat\nu_H^4})
+\frac12
c_{2u}c_{2d}(-\frac{\delta_\Omega}{\nu_\Omega^4}+\frac{\hat\delta_\Omega}{\hat\nu_\Omega^4})
+\frac12
(-\frac{\delta_\Xi}{\nu_\Xi^4}+\frac{\hat\delta_\Xi}{\hat\nu_\Xi^4})=0, \cr
& (b)\  \bar D^0 \perp \bar K^0 : idem,\cr
& (c)\  D^0 \perp \bar K^0 :
s_uc_u s_dc_d(-\frac{\delta_X}{\nu_X^4} +\frac{\hat\delta_X}{\hat\nu_X^4}
-\frac{\delta_H}{\nu_H^4}+\frac{\hat\delta_H}{\hat\nu_H^4})
+\frac12
c_{2u}c_{2d}(-\frac{\delta_\Omega}{\nu_\Omega^4}+\frac{\hat\delta_\Omega}{\hat\nu_\Omega^4})
-\frac12
(-\frac{\delta_\Xi}{\nu_\Xi^4}+\frac{\hat\delta_\Xi}{\hat\nu_\Xi^4})=0, \cr
& (d)\  \bar D^0 \perp K^0 : idem,\cr
& (e)\  \pi^0 \perp (D^0+\bar D^0) :
s_uc_u\frac{c_u^2+c_d^2}{2}\frac{\delta_X}{\nu_X^4}
+s_uc_u\frac{c_u^2-c_d^2}{2}\frac{\hat \delta_X}{\hat \nu_X^4}
-s_uc_u\frac{s_u^2+s_d^2}{2}\frac{\delta_H}{\nu_H^4}
-s_uc_u\frac{s_u^2-s_d^2}{2}\frac{\hat\delta_H}{\hat\nu_H^4}\cr
& \hskip 7cm -\frac12
c_{2u}\frac{s_{2u}+s_{2d}}{2}\frac{\delta_\Omega}{\nu_\Omega^4}-\frac12
c_{2u}\frac{s_{2u}-s_{2d}}{2}\frac{\hat\delta_\Omega}{\hat\nu_\Omega^4}
=0,\cr
& (f)\  \eta \perp (D^0+\bar D^0) :
s_uc_u\frac{c_u^2-c_d^2}{2}\frac{\delta_X}{\nu_X^4}
+s_uc_u\frac{c_u^2+c_d^2}{2}\frac{\hat \delta_X}{\hat \nu_X^4}
-s_uc_u\frac{s_u^2-s_d^2}{2}\frac{\delta_H}{\nu_H^4}
-s_uc_u\frac{s_u^2+s_d^2}{2}\frac{\hat\delta_H}{\hat\nu_H^4}\cr
& \hskip 7cm -\frac12
c_{2u}\frac{s_{2u}-s_{2d}}{2}\frac{\delta_\Omega}{\nu_\Omega^4}
-\frac12
c_{2u}\frac{s_{2u}+s_{2d}}{2}\frac{\hat\delta_\Omega}{\hat\nu_\Omega^4}
=0,\cr
& \pi^0 \perp (K^0-\bar K^0) : always\ true,\cr
& \eta \perp (K^0-\bar K^0) : always\ true,\cr
& \pi^0 \perp (D^0-\bar D^0) : always\ true,\cr
& \eta \perp (D^0-\bar D^0) : always\ true,\cr
& (K^0+\bar K^0) \perp (K^0-\bar K^0) : always\ true,\cr
& (D^0+\bar D^0) \perp (D^0-\bar D^0) : always\ true,\cr
& (g)\  \pi^0 \perp (K^0+\bar K^0) :
-s_dc_d\frac{c_u^2+c_d^2}{2}\frac{\delta_X}{\nu_X^4}
+s_dc_d\frac{c_u^2-c_d^2}{2}\frac{\hat \delta_X}{\hat \nu_X^4}
+s_dc_d\frac{s_u^2+s_d^2}{2}\frac{\delta_H}{\nu_H^4}
-s_dc_d\frac{s_u^2-s_d^2}{2}\frac{\hat\delta_H}{\hat\nu_H^4}\cr
& \hskip 7cm +\frac12
c_{2d}\frac{s_{2u}+s_{2d}}{2}\frac{\delta_\Omega}{\nu_\Omega^4}
-\frac12
c_{2d}\frac{s_{2u}-s_{2d}}{2}\frac{\hat\delta_\Omega}{\hat\nu_\Omega^4}
=0,\cr
& (h)\  \eta \perp (K^0+\bar K^0) :
-s_dc_d\frac{c_u^2-c_d^2}{2}\frac{\delta_X}{\nu_X^4}
+s_dc_d\frac{c_u^2+c_d^2}{2}\frac{\hat \delta_X}{\hat \nu_X^4}
+s_dc_d\frac{s_u^2-s_d^2}{2}\frac{\delta_H}{\nu_H^4}
-s_dc_d\frac{s_u^2+s_d^2}{2}\frac{\hat\delta_H}{\hat\nu_H^4}\cr
& \hskip 7cm +\frac12
c_{2d}\frac{s_{2u}-s_{2d}}{2}\frac{\delta_\Omega}{\nu_\Omega^4}
-\frac12
c_{2d}\frac{s_{2u}+s_{2d}}{2}\frac{\hat\delta_\Omega}{\hat\nu_\Omega^4}
=0,\cr
& (i)\  D^0 \perp \bar D^0 :
s_u^2c_u^2(\frac{\delta_X}{\nu_X^4} +\frac{\hat\delta_X}{\hat\nu_X^4}
+\frac{\delta_H}{\nu_H^4}+\frac{\hat\delta_H}{\hat\nu_H^4})
+\frac12
c_{2u}^2(\frac{\delta_\Omega}{\nu_\Omega^4}+\frac{\hat\delta_\Omega}{\hat\nu_\Omega^4})
-\frac12
(\frac{\delta_\Xi}{\nu_\Xi^4}+\frac{\hat\delta_\Xi}{\hat\nu_\Xi^4})=0,
\cr
& (j)\ K^0 \perp \bar K^0 :
s_d^2c_d^2(\frac{\delta_X}{\nu_X^4} +\frac{\hat\delta_X}{\hat\nu_X^4}
+\frac{\delta_H}{\nu_H^4}+\frac{\hat\delta_H}{\hat\nu_H^4})
+\frac12
c_{2d}^2(\frac{\delta_\Omega}{\nu_\Omega^4}+\frac{\hat\delta_\Omega}{\hat\nu_\Omega^4})
-\frac12
(\frac{\delta_\Xi}{\nu_\Xi^4}+\frac{\hat\delta_\Xi}{\hat\nu_\Xi^4})=0.
\end{split}
}
\label{eq:orthog1}
\end{equation}

\subsection{Masses of $\boldsymbol{\pi^0}$, $\boldsymbol{K^0}$ and
$\boldsymbol{D^0}$}

One gets
\begin{equation}
\hskip -2cm \boxed{ m_{\pi^0}^2=\displaystyle\frac
{
\left( \displaystyle\frac{c_u^2+c_d^2}{2}\right)^2
\displaystyle\frac{\delta_X}{\nu_X^4}
+
\left( \frac{c_u^2-c_d^2}{2}\right)^2
\displaystyle\frac{\hat\delta_X}{\hat\nu_X^4}
+
\left( \frac{s_u^2+s_d^2}{2}\right)^2
\displaystyle\frac{\delta_H}{\nu_H^4}
+
\left( \frac{s_u^2-s_d^2}{2}\right)^2
\displaystyle\frac{\hat\delta_H}{\hat\nu_H^4}
+\frac12
\left( \frac{s_{2u}+s_{2d}}{2}\right)^2
\displaystyle\frac{\delta_\Omega}{\nu_\Omega^4}
+ \frac12
\left( \frac{s_{2u}-s_{2d}}{2}\right)^2
\displaystyle\frac{\hat\delta_\Omega}{\hat\nu_\Omega^4}
}
{
\left( \displaystyle\frac{c_u^2+c_d^2}{2}\right)^2
\displaystyle\frac{1}{\nu_X^4}
+
\left( \frac{c_u^2-c_d^2}{2}\right)^2
\displaystyle\frac{1}{\hat\nu_X^4}
+
\left( \frac{s_u^2+s_d^2}{2}\right)^2
\displaystyle\frac{1}{\nu_H^4}
+
\left( \frac{s_u^2-s_d^2}{2}\right)^2
\displaystyle\frac{1}{\hat\nu_H^4}
+\frac12
\left( \frac{s_{2u}+s_{2d}}{2}\right)^2
\displaystyle\frac{1}{\nu_\Omega^4}
+ \frac12
\left( \frac{s_{2u}-s_{2d}}{2}\right)^2
\displaystyle\frac{1}{\hat\nu_\Omega^4}
}}
\label{eq:mpi02}
\end{equation}

\begin{equation}
\boxed{
m_{K^0}^2=\displaystyle\frac
{\displaystyle\frac{s_d^2 c_d^2}{2}\left(
\displaystyle\frac{\delta_X}{\nu_X^4}
+\displaystyle\frac{\hat\delta_X}{\hat\nu_X^4}
+\displaystyle\frac{\delta_H}{\nu_H^4}
+\displaystyle\frac{\hat\delta_H}{\hat\nu_H^4}
\right)
+\displaystyle\frac{c_{2d}^2}{4}\left(
\displaystyle\frac{\delta_\Omega}{\nu_\Omega^4}
+\displaystyle\frac{\hat\delta_\Omega}{\hat\nu_\Omega^4}
\right)
+\frac14\left(
\displaystyle\frac{\delta_\Xi}{\nu_\Xi^4}
+\displaystyle\frac{\hat\delta_\Xi}{\hat\nu_\Xi^4}
\right) }
{\displaystyle\frac{s_d^2 c_d^2}{2}\left(
\displaystyle\frac{1}{\nu_X^4}
+\displaystyle\frac{1}{\hat\nu_X^4}
+\displaystyle\frac{1}{\nu_H^4}
+\displaystyle\frac{1}{\hat\nu_H^4}
\right)
+\displaystyle\frac{c_{2d}^2}{4}\left(
\displaystyle\frac{1}{\nu_\Omega^4}
+\displaystyle\frac{1}{\hat\nu_\Omega^4}
\right)
+\frac14\left(
\displaystyle\frac{1}{\nu_\Xi^4}
+\displaystyle\frac{1}{\hat\nu_\Xi^4}
\right) }}
\label{eq:mk02}
\end{equation}

\medskip

\begin{equation}
\boxed{
m_{D^0}^2=\displaystyle\frac
{\displaystyle\frac{s_u^2 c_u^2}{2}\left(
\displaystyle\frac{\delta_X}{\nu_X^4}
+\displaystyle\frac{\hat\delta_X}{\hat\nu_X^4}
+\displaystyle\frac{\delta_H}{\nu_H^4}
+\displaystyle\frac{\hat\delta_H}{\hat\nu_H^4}
\right)
+\displaystyle\frac{c_{2u}^2}{4}\left(
\displaystyle\frac{\delta_\Omega}{\nu_\Omega^4}
+\displaystyle\frac{\hat\delta_\Omega}{\hat\nu_\Omega^4}
\right)
+\frac14\left(
\displaystyle\frac{\delta_\Xi}{\nu_\Xi^4}
+\displaystyle\frac{\hat\delta_\Xi}{\hat\nu_\Xi^4}
\right) }
{\displaystyle\frac{s_u^2 c_u^2}{2}\left(
\displaystyle\frac{1}{\nu_X^4}
+\displaystyle\frac{1}{\hat\nu_X^4}
+\displaystyle\frac{1}{\nu_H^4}
+\displaystyle\frac{1}{\hat\nu_H^4}
\right)
+\displaystyle\frac{c_{2u}^2}{4}\left(
\displaystyle\frac{1}{\nu_\Omega^4}
+\displaystyle\frac{1}{\hat\nu_\Omega^4}
\right)
+\frac14\left(
\displaystyle\frac{1}{\nu_\Xi^4}
+\displaystyle\frac{1}{\hat\nu_\Xi^4}
\right) }}
\label{eq:md02}
\end{equation}

\bigskip

We now have at our disposal all the tools to
successively study the case when one approximates $\theta_u$ with $0$, and
the more general one  when  this approximation is relaxed.

\section{Moving inside the space of Higgs multiplets}
\label{section:quadsym}

All quadruplets are  sets of 4 elements which are
 stable both by $SU(2)_L$ and $SU(2)_R$, according to the laws of
transformations (\ref{eq:ruleL}) and (\ref{eq:ruleR}).

There also exists a $U(2)^g_L \times U(2)^g_R$
 group of transformations orthogonal to the former, that moves
inside the 8-dimensional space of quadruplets. It includes
the group $U(1)_L \times U(1)_R$ the generators of which swap parity and
transform, for example, up to their normalizations and a sign,
 $X$ into $\hat X$.

In the case of 2 generations the corresponding 2 sets of 4 generators
are made of the $4\times 4$ identity matrix and of the 3 following ones
\begin{equation}
L^1=\frac12\,\left(\begin{array}{ccccc}
 & 1 & \vline & & \cr 1 & & \vline & & \cr
\hline
 & & \vline &  & 1\cr & & \vline & 1 &
\end{array}\right),\quad
L^2=-\frac{i}{2}\,\left(\begin{array}{ccccc}
 & 1 & \vline & & \cr -1 & & \vline & & \cr
\hline
 & & \vline & & 1\cr & & \vline & -1 & \end{array}\right),\quad
L^3=\frac12 \,\left(\begin{array}{ccccc}
1 &  & \vline & & \cr  & -1 & \vline & & \cr
\hline
 & & \vline & 1 &  \cr & & \vline &  & -1
\end{array}\right),
\label{eq:Ls}
\end{equation}
which act on quark bilinears according to (\ref{eq:group}) and
(\ref{eq:trans2}).
They satisfy the following commutation and anticommutation relations
\begin{equation}
[L^i,L^j]=i\epsilon_{ijk}L^k,\quad \{L_i,L^j\}=0, i\not=j.
\end{equation}
The generators $\vec L$ given in (\ref{eq:Ls}) commute with the generators
$\vec T$ of the gauge group given in (\ref{eq:Ts})
\begin{equation}
[\vec L, \vec T]=0,
\end{equation}
which makes them ``orthogonal'' groups,
while, for anticommutation
\begin{equation}
\{\vec L,\vec T\}=\vec L.
\end{equation}

This second chiral group of transformations cannot be a symmetry of the
theory as soon as the normalization factors $\frac{v}{\mu^3}$ are not
identical. It gets thus broken by $v_i\not= v_j \not= \hat v_i \not=\hat
v_j$ and $\mu_i^3 \not=\mu_j^3 \not= \hat\mu_i^3\not=\hat\mu_j^3$.

In the case of 1 generation, $L^1$ and $L^2$ collapse to $0$, such that
only $L^3$ is left, which becomes proportional to the unit matrix. The
chiral   group under consideration therefore shrinks down to
$U(1)_L \times U(1)_R$, in close connection, as we have seen in
(\ref{eq:par1gen}), with parity.  

More information concerning this group and its close relation to the
flipping of generations will be given in section \ref{section:symmetries}.


\chapter{$\boldsymbol{N=2}$ generations with $\boldsymbol{\theta_d\not= 0,\theta_u=0}$}
\label{chapt:2gentetaunul}

It is usual in the GSW model to perform a flavor rotation on the
$(u,c)$ quarks to align their flavor and mass eigenstates. This is
tantamount to setting $\theta_u=0$ such that the Cabibbo angle
$\theta_c=\theta_d-\theta_u$ becomes identical to $\theta_d$. In there,
it is allowed, as well, to keep instead $\theta_u \not=0$ and
to make a flavor rotation on $(d,s)$ quarks to tune $\theta_d$ to $0$.
In the extension that we propose, the situation is  different and  setting
$\theta_u = 0$ leads to problems  that will only
be cured at $\theta_u \not=0$ (chapter \ref{chapt:2gentetau}).

\section{Charged pseudoscalar mesons and the Cabibbo angle}\label{section:cab}

Flavor rotations that would
eventually tune a mixing angle to $0$ should be operated on the fermion
fields wherever they appear, which also includes all bilinear quark operators.
In this respect, they can no longer be considered as ``innocuous'',
all the more as  there is no reason why they should
correspond to an unbroken subgroup of $U(4)_L \times U(4)_R$.

We can thus only test the hypothesis that $\theta_u$ is small enough to be
neglected with respect to $\theta_d$, in which case
$\tan(\theta_d+\theta_u)\tan(\theta_d-\theta_u) \simeq \tan^2\theta_d
-\theta_u^2 \approx\tan^2\theta_d \approx \tan^2\theta_c$ and, from
(\ref{eq:cab1}) ``the'' mixing angle   is given approximately by
\begin{equation}
t^2\equiv\tan^2\theta_d\approx
\displaystyle\frac{\displaystyle\frac{1}{m_{K^\pm}^2}-\displaystyle\frac{1}{m_{D^\pm}^2}}
{\displaystyle\frac{1}{m_{\pi^\pm}^2}-\displaystyle\frac{1}{m_{D_s^\pm}^2}}
\approx\frac{m_{\pi^\pm}^2}{m_{K^\pm}^2}\left(1-\frac{m_{K^\pm}^2}{m_{D^\pm}^2}
+\frac{m_{\pi^\pm}^2}{m_{D_s^\pm}^2} + {\cal O}\Big(\frac{m_\pi^2,
m_K^2}{m_D^2, m_{D_s}^2}\Big)^2\right).
\label{eq:tantheta}
\end{equation}
Numerically, for the physical values of the charged pseudoscalar mesons
\cite{PDG}
\begin{equation}
m_{\pi^+}= 139.570\,MeV,\quad
m_{K^+}= 493.677\,MeV,\quad
m_{D^+}= 1.86962\,GeV,\quad
m_{D_s^+}=1.96849\,GeV.
\label{eq:mcharged}
\end{equation}
one gets $t^2 \approx .07473$ which corresponds to 
\begin{equation}
\theta_d \approx .26685,
\label{eq:tetadnum}
\end{equation}
to be compared with the experimental value (\ref{eq:numtetac}).
The value of the Cabibbo angle is not expected to be very sensitive to the
existence of heavier generations;
nevertheless, our result displays a $15\%$ discrepancy.
It can have two origins (if we forget about the absence of a
3rd generation).  The first is that our result is only
valid at the low momentum limit, where non-diagonal transitions coming
from kinetic terms vanish. While it is most probably an accurate limit for
pions, its reliability is  more questionable for heavier mesons.
The second points at a non-vanishing value of $\theta_u$, 
and, in chapter
\ref{chapt:2gentetau}, we will show that, indeed, $\theta_u$
plays a very important role.

\section{Determination of $\boldsymbol{b_X, b_H, b_\Omega, b_\Xi}$ in
terms of $\boldsymbol\delta$} \label{section:4bs}

The $b$ parameters have been defined in (\ref{eq:bdefs}).
First, we can take
\begin{equation}
\boxed{
b_\Xi=0
}
\label{eq:nulbxi}
\end{equation}
Indeed, $\mu_\Xi^3 \equiv \displaystyle\frac{<\bar u c-\bar c u+\bar d s-\bar s d>}{2}$
can be considered to be vanishing at the limit when the vacuum is
supposed to be invariant by $C$. The choice (\ref{eq:nulbxi}) corresponds
to a finite normalization factor $\frac{v_\Xi}{\sqrt{2}\mu_\Xi^3}$ for
the $\Xi$ quadruplet. Using the definition (\ref{eq:bdef}) of $b_\Xi$,
one can indeed write at this limit
$\frac{v_\Xi}{\sqrt{2}\mu_\Xi^3}\equiv \sqrt{\frac{b_\Xi\hat
v_H^2}{2\mu_\Xi^6}}\stackrel{b_\Xi \to 0}{\simeq}
 \sqrt{\frac{b_\Xi(1-b_\Xi)\hat v_H^2}{2\mu_\Xi^6}}$. One uses now the
solution (\ref{eq:solnu2}) to equate it to $\sqrt{\frac{b_X(1-b_X)\hat
v_H^2}{2\mu_X^6}}$ which is indeed  finite: the denominator is, up to
mixing, given by the GMOR relation in terms of quark and meson masses (see
(\ref{eq:GMORm}) below and the remark next to it) and
$\hat v_H^2$ is the VEV of the quasi-standard Higgs boson $\hat H^3$.
The other possible solution of (\ref{eq:solnu2}) would be $b_\Xi=1$, but it
does not correspond to a finite normalization for $\Xi$ and we reject it.

Then, $b_X, b_H, b_\Omega$ are determined by (\ref{eq:diag+2}) which yield,
using (\ref{eq:nulbxi}):
\begin{equation}
\begin{split}
&\frac{1}{1-b_\Omega}
=-1 + \delta \left(
-\frac{t^2}{1-t^2}\left(\frac{1}{m_{\pi^+}^2}+\frac{1}{m_{D_s^+}^2} \right)
+\frac{1}{1-t^2}\left( \frac{1}{m_{K^+}^2}+\frac{1}{m_{D^+}^2}\right)
\right),\cr
&\frac{1}{1-b_X}-\frac{1}{1-b_H}= 
\delta \left(
\frac{1}{m_{\pi^+}^2}+\frac{1}{m_{K^+}^2}
-\frac{1}{m_{D^+}^2} -\frac{1}{m_{D_s^+}^2} 
\right),\cr
&\frac{1}{1-b_X}+\frac{1}{1-b_H}= 
\delta \left(
\frac{1}{1-t^2}\left(\frac{1}{m_{\pi^+}^2}+\frac{1}{m_{D_s^+}^2} \right)
-\frac{t^2}{1-t^2}\left( \frac{1}{m_{K^+}^2}+\frac{1}{m_{D^+}^2}\right)
\right),
\end{split}
\label{eq:bxbhbom}
\end{equation}
in which $t^2 \equiv \tan^2\theta_c$ is given by (\ref{eq:tantheta}).

\begin{figure}
\centering
\includegraphics[width=10truecm, height=6truecm]{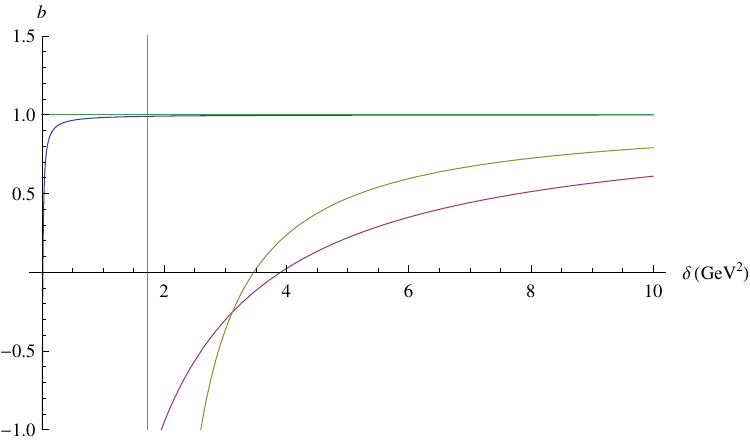}

\medskip
\caption{$b_X$ (blue), $b_H$ (purple) and $b_\Omega$ (yellow) as functions of
$\delta$ at $\theta_u=0$}
\label{figure:bxbhbom3}
\end{figure}

Calling
\begin{equation}
\begin{split}
& r_1\equiv \frac{1}{m_{\pi^+}^2} + \frac{1}{m_{K^+}^2} +\frac{1}{m_{D^+}^2}
+\frac{1}{m_{D_s^+}^2},\quad
r_2\equiv \frac{1}{m_{\pi^+}^2} - \frac{1}{m_{K^+}^2} +\frac{1}{m_{D^+}^2}
-\frac{1}{m_{D_s^+}^2},\cr
& r_3\equiv \frac{1}{m_{\pi^+}^2}+ \frac{1}{m_{K^+}^2}-\frac{1}{m_{D^+}^2}
-\frac{1}{m_{D_s^+}^2},\quad
r_4\equiv \frac{1}{m_{\pi^+}^2}- \frac{1}{m_{K^+}^2}-\frac{1}{m_{D^+}^2}
+\frac{1}{m_{D_s^+}^2},
\label{eq:4rdef}
\end{split}
\end{equation}
one gets
\begin{equation}
\begin{split}
b_X &= 1-\frac{2}
{\delta\left(r_3 +\displaystyle\frac{1}{1-t^2}\displaystyle\frac{r_1+r_4}{2}
-\displaystyle\frac{t^2}{1-t^2}\displaystyle\frac{r_1-r_4}{2}
\right)}
= 1-\frac{2}{\delta\left(r_3+\displaystyle\frac{r_1}{2}+\displaystyle\frac{r_4}{2\,c_{2d}}\right)},\cr
b_H &= 1-\frac{2}
{\delta\left(-r_3 +\displaystyle\frac{1}{1-t^2}\displaystyle\frac{r_1+r_4}{2}
-\displaystyle\frac{t^2}{1-t^2}\displaystyle\frac{r_1-r_4}{2}
\right)}
= 1-\frac{2}{\delta\left(-r_3+\displaystyle\frac{r_1}{2}+\displaystyle\frac{r_4}{2\,c_{2d}}\right)},\cr
b_\Omega &= 1-\frac{1}
{-1+\displaystyle\frac{\delta}{2}\left(r_1 -\displaystyle\frac{1+t^2}{1-t^2}\,r_4
\right)}
=1-\frac{1}{-1+\displaystyle\frac{\delta}{2}\left(r_1-\displaystyle\frac{r_4}{2\,c_{2d}}\right)}.
\end{split}
\end{equation}
We plot in Fig.\ref{figure:bxbhbom3} the 3 parameters $b_X, b_H, b_\Omega$
as functions of
$\delta$ for the physical values (\ref{eq:mcharged}) of the masses of the
charged pseudoscalar mesons.
%
%
Fig.\ref{figure:bxbhbom3} already provides information on the value of $\delta$ because the
$b$'s, being the squared of ratios of VEV's supposedly real should be
positive. We see that the strongest constraint is provided by the condition
$b_H \geq 0$, which entails
\begin{equation}
\begin{split}
b_H \geq 0\quad \Rightarrow\quad \delta &\geq \frac
{m_{D^+}^2 m_{D_s^+}^2 \Big(m_{K^+}^2 m_{\pi^+}^2(m_{D_s^+}^2 -m_{D^+}^2) +m_{D^+}^2
m_{D_s^+}^2(m_{K^+}^2-m_{\pi^+}^2)\Big)}
{m_{D^+}^2 m_{D_s^+}^2(m_{{K^+}^+}^2 m_{D^+}^2-m_{D_s^+}^2 m_{\pi^+}^2)+m_{K^+}^2
m_{\pi^+}^2(m_{D_s^+}^4-m_{D^+}^4)}\cr
& \cr
 &= m_{D_s^+}^2 + m_{\pi^+}^2 \frac{m_{D_s^+}^2 (m_{D_s^+}^2-m_{D^+}^2)
(m_{D^+}^2-m_{K^+}^2)}{m_{K^+}^2 m_{D^+}^4} +{\cal O}(m_{\pi^+}^4).
\end{split}
\label{eq:delbhnul}
\end{equation}
Numerically, the first line of (\ref{eq:delbhnul}) yields
\begin{equation}
\delta \geq 3.90923\,GeV^2,
\label{eq:delbh0num}
\end{equation}
to be compared with $m_{D_s}^2 \approx 3.87495\,GeV^2$.

By (\ref{eq:mhh3}) we accordingly know already that the quasi-standard
Higgs boson $\hat H^3$ has a mass
\begin{equation}
m_{\hat H^3} \geq \sqrt{2} m_{D_s^+},
\end{equation}
which is to be compared with the case of 1 generation, for which the
corresponding mass was $\sqrt{2}m_\pi$. As announced, this mass is
controlled by the heaviest pseudoscalar meson (or quark) mass.

\section{Neutral pseudoscalar mesons}\label{section:neutralpseudos}

\subsection{Orthogonality relations}\label{subsec:orthoneutral}

Eqs.~(\ref{eq:orthog1}) become at $\theta_u=0$
\begin{equation}
\begin{split}
& (a)\  D^0 \perp K^0 :
c_{2d}(-\frac{\delta_\Omega}{\nu_\Omega^4}+\frac{\hat\delta_\Omega}{\hat\nu_\Omega^4})
- (\frac{\delta_\Xi}{\nu_\Xi^4}-\frac{\hat\delta_\Xi}{\hat\nu_\Xi^4})=0, \cr
& (b)\  \bar D^0 \perp \bar K^0 : idem,\cr
& (c)\  D^0 \perp \bar K^0 :
c_{2d}(-\frac{\delta_\Omega}{\nu_\Omega^4}+\frac{\hat\delta_\Omega}{\hat\nu_\Omega^4})
+ (\frac{\delta_\Xi}{\nu_\Xi^4}-\frac{\hat\delta_\Xi}{\hat\nu_\Xi^4})=0, \cr
& (d)\  \bar D^0 \perp K^0 : idem,\cr
& (e)\  \pi^0 \perp (D^0+\bar D^0) :
 -s_{2d}\frac{\delta_\Omega}{\nu_\Omega^4}
+s_{2d}\frac{\hat\delta_\Omega}{\hat\nu_\Omega^4}
=0,\cr
& (f)\  \eta \perp (D^0+\bar D^0) :
s_{2d}\frac{\delta_\Omega}{\nu_\Omega^4}
-s_{2d}\frac{\hat\delta_\Omega}{\hat\nu_\Omega^4}
=0,\cr
& \pi^0 \perp (K^0-\bar K^0) : always\ true,\cr
& \eta \perp (K^0-\bar K^0) : always\ true,\cr
& \pi^0 \perp (D^0-\bar D^0) : always\ true,\cr
& \eta \perp (D^0-\bar D^0) : always\ true,\cr
& (K^0+\bar K^0) \perp (K^0-\bar K^0) : always\ true,\cr
& (D^0+\bar D^0) \perp (D^0-\bar D^0) : always\ true,\cr
& (g)\  \pi^0 \perp (K^0+\bar K^0) :
s_dc_d\left(-(1+c_d^2)\frac{\delta_X}{\nu_X^4}
+(1-c_d^2)\frac{\hat \delta_X}{\hat \nu_X^4}
+s_d^2\frac{\delta_H}{\nu_H^4}
+ s_d^2\frac{\hat\delta_H}{\hat\nu_H^4}\right)
+\frac{c_{2d} s_{2d}}{2} \left(
\frac{\delta_\Omega}{\nu_\Omega^4}
+\frac{\hat\delta_\Omega}{\hat\nu_\Omega^4}\right)
=0,\cr
& (h)\  \eta \perp (K^0+\bar K^0) :
s_dc_d\left(-(1-c_d^2)\frac{\delta_X}{\nu_X^4}
+(1+c_d^2)\frac{\hat \delta_X}{\hat \nu_X^4}
-s_d^2\frac{\delta_H}{\nu_H^4}
-s_d^2\frac{\hat\delta_H}{\hat\nu_H^4}\right)
 - \frac{c_{2d} s_{2d}}{2}\left(\frac{\delta_\Omega}{\nu_\Omega^4}
+\frac{\hat\delta_\Omega}{\hat\nu_\Omega^4}\right)
=0,\cr
& (i)\  D^0 \perp \bar D^0 : always\ true,\cr
& (j)\ K^0 \perp \bar K^0 :
s_d^2c_d^2(\frac{\delta_X}{\nu_X^4} +\frac{\hat\delta_X}{\hat\nu_X^4}
+\frac{\delta_H}{\nu_H^4}+\frac{\hat\delta_H}{\hat\nu_H^4})
+\frac12
c_{2d}^2(\frac{\delta_\Omega}{\nu_\Omega^4}+\frac{\hat\delta_\Omega}{\hat\nu_\Omega^4})
-\frac12
(\frac{\delta_\Xi}{\nu_\Xi^4}+\frac{\hat\delta_\Xi}{\hat\nu_\Xi^4})=0.
\end{split}
\label{eq:orthog2}
\end{equation}

$\bullet$\ (a) and (c) of (\ref{eq:orthog2}) yield
\begin{equation}
\frac{\delta_\Omega}{\nu_\Omega^4}=\frac{\hat\delta_\Omega}{\hat\nu_\Omega^4},\quad
\frac{\delta_\Xi}{\nu_\Xi^4}=\frac{\hat\delta_\Xi}{\hat\nu_\Xi^4},
\label{eq:solnu4}
\end{equation}
which also makes (e) and (f) of (\ref{eq:orthog2}) true.

Equations (\ref{eq:solnu1}), (\ref{eq:solnu4})
summarize into
\begin{equation}
\qquad \frac{1-b_X}{\nu_X^4}=\frac{1-b_H}{\nu_H^4}
=\frac{1-b_\Omega}{\nu_\Omega^4}=\frac{1-b_\Xi}{\nu_\Xi^4}
=\frac{1-\hat b_\Omega}{\hat \nu_\Omega^4}=\frac{1-\hat b_\Xi}{\hat
\nu_\Xi^4},
\label{eq:sumsolnu2}
\end{equation}
which is equivalent to
\begin{equation}
\frac{b_X(1-b_X)}{\mu_X^6}
=\frac{b_H(1-b_H)}{\mu_H^6}
=\frac{b_\Omega(1-b_\Omega)}{\mu_\Omega^6}
=\frac{b_\Xi(1-b_\Xi)}{\mu_\Xi^6}
=\frac{\hat b_\Omega(1-\hat b_\Omega)}{\hat \mu_\Omega^6}
=\frac{\hat b_\Xi(1-\hat b_\Xi)}{\hat \mu_\Xi^6}.
\label{eq:sumsolnu3}
\end{equation}
By the definition (\ref{eq:rdef}), this also translates into
\begin{equation}
\begin{split}
& r_H = \sqrt{\frac{b_H(1-b_H)}{b_X(1-b_X)}},\quad
r_\Omega = \sqrt{\frac{b_\Omega(1-b_\Omega)}{b_X(1-b_X)}},\quad
r_\Xi = \sqrt{\frac{b_\Xi(1-b_\Xi)}{b_X(1-b_X)}},\cr
& \hat r_\Omega = \sqrt{\frac{\hat b_\Omega(1-\hat
b_\Omega)}{b_X(1-b_X)}},\quad
\hat r_\Xi = \sqrt{\frac{\hat b_\Xi(1-\hat b_\Xi)}{b_X(1-b_X)}},
\end{split}
\label{eq:sumsolr1}
\end{equation}
to which should of course be added the definition / statements
\begin{equation}
\hat b_H = 1,\quad r_X=1,\quad \mu_\Xi^3=0=\hat\mu_\Xi^3
\quad\Rightarrow \quad r_\Xi=0=\hat r_\Xi.
\label{eq:defstat}
\end{equation}
Eqs.~(\ref{eq:sumsolnu3}) establish relations between bosonic and
fermionic VEV's, therefore between gauge and chiral symmetry breaking.

$\bullet$\ Mass-like  $K^0\bar K^0$ non-diagonal terms are
 proportional to
\begin{equation}
\frac{s_d^2c_d^2}{2}\left(
\frac{\delta_X}{\nu_X^4} +
\frac{\hat\delta_X}{\hat\nu_X^4}
+\frac{\delta_H}{\nu_H^4} +
\frac{\overbrace{\hat\delta_H}^{0}}{\hat\nu_H^4}
\right)
+\frac{c_{2d}^2}{4}\left(
\frac{\delta_\Omega}{\nu_\Omega^4} +
\frac{\hat\delta_\Omega}{\hat\nu_\Omega^4}\right)
-\frac14 \left(\frac{\delta_\Xi}{\nu_\Xi^4} +
\frac{\hat\delta_\Xi}{\hat\nu_\Xi^4}\right).
\end{equation}
From the conditions (\ref{eq:solnu4}) for the absence of neutral $K-D$
transitions the last two terms become
$\frac{c_{2d}^2}{2}\frac{\delta_\Omega}{\nu_\Omega^4} -\frac12
\frac{\delta_\Xi}{\nu_\Xi^4}$; then, from
 (\ref{eq:solnu1}) for charged mesons,
they become $-\frac{s_{2d}^2}{2}\frac{\delta_\Xi}{\nu_\Xi^4}$, which is
also  $-\frac{s_{2d}^2}{2}\frac{\delta_X}{\nu_X^4}$;
the first term transforms by (\ref{eq:solnu1}) into $\frac{s_d^2
c_d^2}{2}(2\frac{\delta_X}{\nu_X^4} + \frac{\hat \delta_X}{\hat \nu_X^4})$
such that the $K^0\bar K^0$ terms get finally proportional to
$s_d^2 c_d^2\left(\frac12
\frac{\hat\delta_X}{\hat\nu_X^4}-\frac{\delta_X}{\nu_X^4}\right)$.
That they vanish requires accordingly
\begin{equation}
\frac{\delta_X}{\nu_X^4}=\frac12 \frac{\hat\delta_X}{\hat \nu_X^4}.
\label{eq:KKperp}
\end{equation}
Using (\ref{eq:delrel1}) and the definitions (\ref{eq:nudef}), (\ref{eq:KKperp}) is
equivalent to
\begin{equation}
\frac{1-b_X}{\nu_X^4} = \frac12\frac{1-\hat b_X}{\hat\nu_X^4}
\Leftrightarrow
\frac{b_X(1-b_X)}{\mu_X^6} = \frac12 \frac{\hat b_X(1-\hat
b_X)}{\hat\mu_X^6}.
\label{eq:KKperp2}
\end{equation}

$\bullet$\ Using $\hat\delta_H=0$ (\ref{eq:truegol})
and (\ref{eq:solnu2}) for $X, H, \Omega,\hat\Omega$,
(7) and (8) of (\ref{eq:orthog2}) transform respectively into
\begin{equation}
\frac{\delta_X}{\nu_X^4}
=\frac12 \frac{\hat\delta_X}{\hat\nu_X^4},
\label{eq:solnu31}
\end{equation}
and
\begin{equation}
\frac{\delta_X}{\nu_X^4}
=\frac{1+c_d^2}{2c_d^2} \frac{\hat \delta_X}{\hat\nu_X^4}.
\label{eq:solnu32}
\end{equation}
(\ref{eq:solnu31}) and (\ref{eq:solnu32}) are incompatible.
(\ref{eq:solnu31}) is the same condition as (\ref{eq:KKperp}) that
 we found  by cancelling the non-diagonal $K^0-\bar K^0$ couplings.
So, while one can easily achieve the
orthogonality of $\pi^0$ to both $K^0 \pm \bar K^0$, this is not the case
for $\eta$.

$\bullet$\ For the same reasons as we chose $b_\Xi=0$ (see (\ref{eq:nulbxi})), 
(\ref{eq:sumsolnu2}) and the requirement that the normalization factor of
the $\hat\Xi$ quadruplet be finite allows to take
\begin{equation}
\boxed{
\hat b_\Xi=0}
\label{eq:nulhbxi}
\end{equation}

Relations (\ref{eq:nulbxi}) and (\ref{eq:nulhbxi}) largely simplify the
calculations
\footnote{They should be relaxed for 3 generations since,
at least at the perturbative (2-loops) level, $<\bar c u - \bar u c>$ and
$<\bar d s-\bar s d>$ get proportional to the $CP$-violating phase of the
Cabibbo-Kobayashi-Maskawa mixing matrix.}.

\subsection{Masses of $\boldsymbol{K^0}$ and $\boldsymbol{D^0}$}
\label{subsec:neutralmass}

One gets from (\ref{eq:mk02}) and (\ref{eq:md02})

\begin{equation}
m_{K^0}^2 = \frac
{\frac{s_d^2c_d^2}{2}\left(
\frac{\delta_X}{\nu_X^4} +
\frac{\hat\delta_X}{\hat\nu_X^4}
+\frac{\delta_H}{\nu_H^4} +
\frac{\hat\delta_H}{\hat\nu_H^4}
\right)
+\frac{c_{2d}^2}{4}\left(
\frac{\delta_\Omega}{\nu_\Omega^4} +
\frac{\hat\delta_\Omega}{\hat\nu_\Omega^4}
\right)
+\frac14 \left(\frac{\delta_\Xi}{\nu_\Xi^4} +
\frac{\hat\delta_\Xi}{\hat\nu_\Xi^4}
\right)}
{\frac{s_d^2 c_d^2}{2}\left(\frac{1}{\nu_X^4}+\frac{1}{\hat\nu_X^4}
+\frac{1}{\nu_H^4}+\frac{1}{\hat\nu_H^4}\right)
+\frac{c_{2d}^2}{4}\left(
\frac{1}{\nu_\Omega^4} + \frac{1}{\hat\nu_\Omega^4}\right)
+\frac14 \left(\frac{1}{\nu_\Xi^4} + \frac{1}{\hat\nu_\Xi^4}
\right)
},
\label{eq:mk0}
\end{equation}
and
\begin{equation}
m_{D^0}^2 =
\frac{\frac{\delta_\Omega}{\nu_\Omega^4}+\frac{\hat\delta_\Omega}{\hat\nu_\Omega^4}+\frac{\delta_\Xi}{\nu_\Xi^4}+\frac{\hat\delta_\Xi}{\hat\nu_\Xi^4}}
{\frac{1}{\nu_\Omega^4}+\frac{1}{\hat\nu_\Omega^4}
+\frac{1}{\nu_\Xi^4}+\frac{1}{\hat\nu_\Xi^4}}.
\label{eq:md0}
\end{equation}
Using (\ref{eq:delrel1}), (\ref{eq:sumsolnu2}) (which entails
$\hat\delta_H=0$), (\ref{eq:nulbxi}) and (\ref{eq:nulhbxi}) (which cancel
the terms  $\delta_\Xi$ and $\hat\delta_\Xi$ in their numerators),
 (\ref{eq:mk0}) and (\ref{eq:md0}) rewrite
\begin{equation}
m_{K^0}^2=\frac{4\delta}
{\displaystyle\frac12 s_{2d}^2\left(
\displaystyle\frac{1}{1-b_X} + \displaystyle\frac{2}{1-\hat b_X}
+\displaystyle\frac{1}{1-b_H}
+\displaystyle\frac{1}{b_X(1-b_X)}\displaystyle\frac{1}{\hat r_H^2}
\right)
+c_{2d}^2 \left(
\displaystyle\frac{1}{1-b_\Omega} + \displaystyle\frac{1}{1-\hat b_\Omega}
\right)
+2},
\label{eq:mk03}
\end{equation}
and
\begin{equation}
m_{D_0}^2=\frac{4\delta}{\displaystyle\frac{1}{1-b_\Omega}
+\displaystyle\frac{1}{1-\hat
b_\Omega} +2}.
\label{eq:md03}
\end{equation}
 (\ref{eq:mk03}) and (\ref{eq:md03}) combine into
\begin{equation}
\displaystyle\frac{m_{K_0}^2}{m_{D_0}^2}=
\displaystyle\frac
{\displaystyle\frac{1}{1-b_\Omega}+\displaystyle\frac{1}{1-\hat b_\Omega} +
2}
{\displaystyle\frac12 s_{2d}^2\left(
\displaystyle\frac{1}{1-b_X} + \displaystyle\frac{2}{1-\hat b_X}
+\displaystyle\frac{1}{1-b_H}
+\displaystyle\frac{1}{b_X(1-b_X)}\displaystyle\frac{1}{\hat r_H^2}
\right)
+c_{2d}^2 \left(
\displaystyle\frac{1}{1-b_\Omega} + \displaystyle\frac{1}{1-\hat b_\Omega}
\right)
+2}.
\label{eq:mdmk1}
\end{equation}
Since $b_\Omega$ is known by (\ref{eq:bxbhbom}) as a function of $\delta$,
(\ref{eq:md02}) defines $\hat b_\Omega$ as a function of $\delta$, too.

Eq.~(\ref{eq:mdmk1}) involves $\hat r_H^2
\equiv\hat \mu_H^6/\mu_X^6$;  $\sin^2 2\theta_d \equiv
4t^2/(1+t^2)^2$ and $\cos^2 2\theta_d \equiv
\left(\frac{1-t^2}{1+t^2}\right)^2$ are known by (\ref{eq:tantheta}),
 $b_X, b_H, b_\Omega$ and $\hat b_\Omega$ are known as functions
of $\delta$; therefore, (\ref{eq:mdmk1})
 defines $\hat b_X$ as a function of $\delta$ and $\hat r_H$.

We  avoid at the moment to use
$\pi^0, \eta \ldots$ as inputs. This is because
these mesons are known to mix and, accordingly, there is uncertainty
concerning  their interpolating fields in terms of quark bilinears.
We shall see later that the $\pi^0$ as we defined it, proportional to $\bar
u\gamma_5 u -\bar d\gamma_5 d$, gets a suitable mass
(together with appropriate orthogonality relations as we already
mentioned in subsection \ref{subsec:orthoneutral}).

\subsection{First hints at a problem concerning $\boldsymbol{\hat b_X}$}
\label{subsec:hbxpel}

One can already get from (\ref{eq:mdmk1}) some valuable information concerning
$\hat b_X \equiv \left(\frac{\hat v_X}{\hat v_H}\right)^2=
\left( \frac{<\hat X^3>}{<\hat H^3>}\right)^2$.
One has to make a reasonable estimate of $\hat r_H^2$. This is fairly easy
from its definition since
$\hat r_H\equiv\frac{\hat\mu_H^3}{\mu_X^3}
= \frac{<\bar c c-\bar s s>}{<\bar u u+\bar d d>}$: its modulus
is presumably $\leq 1$ because heavy quarks being ``more classical'' that light
quarks should undergo less condensation in the vacuum. We shall see later
from fermionic considerations that, indeed $\hat r_H \approx .6$.
Numerical evaluations then show that, for $\delta \geq m_{D_s}^2$,
\begin{equation}
\hat b_X > 1.
\end{equation}
This is confirmed by a formal expansion at the chiral limit $m_\pi \to 0$
\footnote{Such expansions cannot always be trusted. However, numerical checks
show that (\ref{eq:hbxexp}) is reasonably accurate. In particular, the pole
at $\hat r_H^2=1$ of (\ref{eq:hbxexp}) only gets moved in the exact
formula to $\hat r_H^2 \approx 1.08$.\label{foot:polehbx}}
\begin{equation}
\hat b_X \stackrel{m_\pi \to 0}{\simeq} 1+ \frac{m_\pi^2}{\delta}\frac{2
\hat
r_H^2}{1-\hat r_H^2} +{\cal O}(m_\pi^4) > 1.
\label{eq:hbxexp}
\end{equation}
A similar expansion for $b_X$ is
\begin{equation}
b_X \stackrel{m_\pi \to 0}{\simeq} 1- \frac{m_\pi^2}{\delta} +{\cal
O}(m_\pi^4) < 1,
\label{eq:bxexp}
\end{equation}
which is in fair agreement with the curve on Fig.\ref{figure:bxbhbom3}.

Eq.~(\ref{eq:solnu31}), which controls the orthogonality of  $\pi^0$ 
to $K^0 + \bar K^0$, identical to (\ref{eq:KKperp}) which controls the orthogonality
of $K^0$ to $\bar K^0$ are 
incompatible with (\ref{eq:hbxexp}) and (\ref{eq:bxexp}): indeed, $b_X <1$ and
$\hat b_X>1$ lead, by (\ref{eq:solnu31}), to $\hat \nu_X^4/\nu_X^4 \equiv
(b_X/\hat b_X)(\hat\mu_X^3/\mu_X^3)^2 <0$. Since $b_X$ and $\hat b_X$ are
real, this could only occur for $\hat\mu_X^3/\mu_X^3
\equiv <\bar u u-\bar d d>/<\bar u u+\bar d d>$ imaginary. $\mu_X^3$
being real by the GMOR relation (see (\ref{eq:GMORm}) below),
 $\hat\mu_X^3\equiv <\bar u u-\bar d
d>/\sqrt{2}$
should be imaginary. This is manifestly impossible since $(\bar q_i
q_i)^\dagger = \bar q_i q_i$.

\section{Charged scalars}\label{section:chargedscal}

The same investigation that we did in section \ref{section:cab} for charged
pseudoscalars we now do for charged scalars.
It is not our goal here to extensively study the spectrum of charged scalar
mesons. We want only to show that they are expected to align with flavor
eigenstates.
Like for charged pseudoscalars, we shall consider a priori that the
interpolating field of the charged scalar pion $\pi^{s +}$
is proportional to $\bar u_m d_m$, that of the charged scalar $D_s$ meson,
that we note $D_s^{s +}$, is proportional to $\bar c_m s_m$ {\em etc}.

\subsection{Orthogonality relations}\label{subsec:orthoscal}

The set of equations equivalent to (\ref{eq:ndiag+0}) is (remember
that we work at $\theta_u=0$)
\begin{equation}
\begin{split}
\frac{s_dc_d}{\overline{\hat\nu_X}^4}
&
-\frac12\frac{s_dc_d}{\overline{\hat\nu_\Omega}^4}-\frac12\frac{s_d
c_d}{\overline{\hat\nu_\Xi}^4}=0 \qquad (\pi^{s +} \not \leftrightarrow
K^{s +}),\cr
&
-\frac12\frac{s_dc_d}{\overline{\hat\nu_\Omega}^4}+\frac12\frac{s_d
c_d}{\overline{\hat\nu_\Xi}^4}=0 \qquad(\pi^{s +} \not \leftrightarrow
D^{s +}),\cr
&
-\frac12\frac{s_d^2}{\overline{\hat\nu_\Omega}^4}+\frac12\frac{s_d^2}{\overline{\hat\nu_\Xi}^4}=0
\qquad (\pi^{s +} \not \leftrightarrow  D_s^{s +}),\cr
&
+\frac12\frac{c_d^2}{\overline{\hat\nu_\Omega}^4}-\frac12\frac{c_d^2}{\overline{\hat\nu_\Xi}^4}=0
\qquad (K^{s +} \not \leftrightarrow  D^{s +}),\cr
&
+\frac12\frac{s_dc_d}{\overline{\hat\nu_\Omega}^4}-\frac12\frac{s_d
c_d}{\overline{\hat\nu_\Xi}^4}=0 \qquad (K^{s +} \not \leftrightarrow  D_s^{s
+}),\cr
\frac{s_dc_d}{\overline{\hat\nu_H}^4}
&
+\frac12\frac{s_dc_d}{\overline{\hat\nu_\Omega}^4}+\frac12\frac{s_d
c_d}{\overline{\hat\nu_\Xi}^4}=0 \qquad (D^{s +} \not \leftrightarrow  D_s^{s
+}).
\end{split}
\label{eq:chascal}
\end{equation}
The main difference between (\ref{eq:chascal}) and (\ref{eq:ndiag+0}) is
that $\hat \delta_H = \delta(1-\hat b_H) = 0$ while $\delta_H$ is
different from $0$. One term is consequently ``missing'' in equation 6 of
(\ref{eq:chascal}).

Equations 2 to 5 of (\ref{eq:chascal}) lead to $\frac{1-\hat b_\Omega}{\hat
\nu_\Omega^4}= \frac{1-\hat b_\Xi}{\hat\nu_\Xi^4}$, which has already been
obtained  from pseudoscalar mesons in (\ref{eq:sumsolnu2}).
Then, equation 1 leads to 
\begin{equation}
\frac{1-\hat b_X}{\hat\nu_X^4}=\frac{1-\hat
b_\Omega}{\hat\nu_\Omega^4},
\label{eq:solnu8}
\end{equation}
 while, because $\hat b_H=1$, the same equation 6 entails 
\begin{equation}
\frac{1-\hat b_X}{\hat\nu_X^4}=-\frac{1-\hat
b_\Omega}{\hat\nu_\Omega^4}.
\label{eq:solnu9}
\end{equation}
Both (\ref{eq:solnu8}) and (\ref{eq:solnu9}) are in contradiction with
(\ref{eq:solnu31}). (\ref{eq:solnu8}) is incompatible
 because it does not exhibit the factor
$1/2$ present in (\ref{eq:solnu31}) (we recall that by (\ref{eq:sumsolnu2})
$\frac{1-\hat b_\Omega}{\hat\nu_\Omega^4}=\frac{1-b_X}{\nu_X^4}$);
(\ref{eq:solnu9}) is also manifestly incompatible with 
(\ref{eq:solnu31}). 

The only way to reconcile these orthogonality relations with the results
that we have obtained for pseudoscalar mesons is to turn to $0$ the mixing
angle for scalars, that is to set $\theta_d=0$ in (\ref{eq:chascal}). Then,
scalar mesons are bound states of quark flavor eigenstates.

\subsection{Masses of charged scalars}\label{subsec:scalmass}

Let us confirm the proposition above by focusing on
the $(\pi^{s +},K^{s +})$ and $(D^{s +}, D_s^{s +})$ systems.
They indeed correspond to the problematic equations 1 and 6 in the system
(\ref{eq:chascal}).

As far as the second pair is concerned, one finds that its mass matrix is
proportional to
\begin{equation}
\delta\left(\begin{array}{cc} c_d^2 & s_d c_d \cr s_d c_d & s_d^2
\end{array}\right),
\end{equation}
which displays, as expected, a vanishing eigenvalue : the model has been
indeed built such that the {\em flavor} $D_s^{s \pm}$ are true
Goldstones of the broken $SU(2)_L$.

We then focus on the $(\pi^{s +}, K^{s +})$ system. Its mass matrix is found
to be proportional to
\begin{equation}
\delta\left(\begin{array}{cc}
\displaystyle\frac{c_d^2}{\hat\nu_X^4}(1-\hat b_X)
 +\displaystyle\frac{s_d^2}{\hat\nu_\Omega^4}(1-\hat b_\Omega) &
-\displaystyle\frac{s_dc_d}{\hat\nu_\Omega^4}(1-\hat b_\Omega)
 \cr
-\displaystyle\frac{s_dc_d}{\hat\nu_\Omega^4}(1-\hat b_\Omega)
& \displaystyle\frac{s_d^2}{\hat\nu_X^4}(1-\hat b_X)
+\displaystyle\frac{c_d^2}{\hat\nu_\Omega^4}(1-\hat
b_\Omega)\end{array}\right)
\propto \delta\left(\begin{array}{cc}
s_d^2 -\epsilon & -s_d c_d \cr
-s_d c_d & c_d^2 -\epsilon,
\end{array}\right),
\label{eq:scalpikmass}
\end{equation}
in which $\epsilon$ is a small positive number.
(\ref{eq:scalpikmass}) displays a small negative eigenvalue $-\epsilon$.
Since the corresponding kinetic terms, proportional to
\begin{equation}
\left(\begin{array}{cc}
\frac{c_d^2}{\hat\nu_X^4}
 +\frac{s_d^2}{\hat\nu_\Omega^4} &
-\frac12\frac{s_dc_d}{\hat\nu_\Omega^4}-\frac12\frac{s_d
c_d}{\hat\nu_\Xi^4} \cr
-\frac12\frac{s_dc_d}{\hat\nu_\Omega^4}-\frac12\frac{s_d
c_d}{\hat\nu_\Xi^4}
& \frac{s_d^2}{\hat\nu_X^4}
+\frac{c_d^2}{\hat\nu_\Omega^4}\end{array}\right),
\end{equation}
have 2 positive eigenvalues, we face the issue that, for 2 generations,
the binary system of charged scalar pions and kaons, if aligned with quark
mass eigenstates, involves a tachyonic state.

This problem fades away when setting, like in the previous subsection,
 $\theta_d=0$, that is aligning scalar mesons with flavor eigenstates. The
masses of the scalar charged pion and kaon become, then,  proportional to
$\frac{1-\hat b_X}{\hat\nu_X^4}$ and $\frac{1-\hat
b_\Omega}{\hat\nu_\Omega^4}$ that is, by (\ref{eq:solnu31}) and
(\ref{eq:sumsolnu3}), to
$2\frac{1-b_X}{\nu_X^4}=2\frac{b_X(1-b_X)}{\mu_X^6}$
and $\frac{1-b_X}{\nu_X^4}=\frac{b_X(1-b_X)}{\mu_X^6}$.

We therefore conclude that the alignment of the 3 Goldstone bosons of the
broken gauge $SU(2)_L$ symmetry with flavor eigenstates triggers the same
alignment for charged scalar  mesons.

\section{Summary of bosonic constraints}\label{section:sumbos}

We summarize below the results that we have obtained
from the sole bosonic constraints and list  what remains to be done to
determine all the parameters of the theory.

We need the reference points for the $b$'s and the $r$'s, that is $\hat
v_H$ and $\mu_X^3$. $\mu_X^3$ is a  condensate of quark flavor
eigenstates while the low energy theorems (PCAC, GMOR relation) involve
mass eigenstates:
\begin{equation}
i(m_u+m_d)\bar u_m \gamma_5 d_m =\sqrt{2} f_\pi m_{\pi^+}^2 \pi^+,
\label{eq:PCACm}
\end{equation}
\begin{equation}
(m_u+m_d)<\bar u_m u_m + \bar d_m d_m> = 2f_\pi^2 m_{\pi^+}^2.
\label{eq:GMORm}
\end{equation}
 Since the mixing angle $\theta_c$ is small, we shall  make the approximation that
 $\mu_X^3$ is close to the quark condensate of mass eigenstates
$\mu_X^3 \approx \frac{<\bar u_m u_m + \bar d_m d_m>}{\sqrt{2}}$ which
is given by the GMOR relation (\ref{eq:GMORm}). We shall
accordingly approximate
\begin{equation}
\mu_X^3 \equiv\frac{<\bar u u+\bar d d>}{\sqrt{2}} \approx \frac{<\bar u_m
u_m + \bar d_m d_m>}{\sqrt{2}} = \frac{\sqrt{2} f_\pi^2 m_\pi^2}{m_u+m_d}.
\label{eq:GMOR2}
\end{equation}

By (\ref{eq:bxbhbom}) and (\ref{eq:md0}),
 $b_X, b_H, b_\Omega, \hat b_\Omega$ are known functions of $\delta$;
therefore, by (\ref{eq:sumsolr1}), $r_H, r_\Omega, \hat r_\Omega$ are also
 known functions of $\delta$.
By (\ref{eq:mdmk1}), $\hat b_X$ is a known function of $\delta$ and $\hat r_H$.

Numerically,  one gets ($\delta$ being expressed in $GeV^2$)
\begin{equation}
\begin{split}
& b_X\equiv\left(\frac{v_X}{\hat v_H}\right)^2 \approx 1 -
\frac{0.0181324}{\delta},\quad
b_H\equiv\left(\frac{v_H}{\hat v_H}\right)^2  \approx 1 -
\frac{3.90923}{\delta},\quad
b_\Omega\equiv\left(\frac{v_\Omega}{\hat v_H}\right)^2  \approx 1 -
\frac{1}{-1 + 0.576693\; \delta},\cr
&\hskip -1cm
\hat b_X\equiv\left(\frac{\hat v_X}{\hat v_H}\right)^2  \approx
\frac{0.0017312\; \hat r_H^2 - 0.113608\; \delta\; \hat r_H^2 +
 \delta^2 (-0.87758 +  \hat r_H^2)}
{0.00115414\; \hat r_H^2 -0.0817829\;\delta\; \hat r_H^2 + \delta^2
(-0.87758 +  \hat r_H^2)},\quad
\hat b_\Omega\equiv\left(\frac{\hat v_\Omega}{\hat v_H}\right)^2  \approx 1
- \frac{1}{-1 + 0.573491\;\delta},
\end{split}
\end{equation}
to which should be added $\hat b_H=1$ by definition and $b_\Xi=0=\hat
b_\Xi$ (\ref{eq:nulbxi}) and (\ref{eq:nulhbxi}).

As will be confirmed later, $\delta$ stands very close to $m_{D_s}^2$, such
that we already know that $b_X$ is very close to and smaller that $1$,
$b_H$ is very small, $b_\Omega \approx \hat b_\Omega \approx .2$. Since
$\hat r_H^2$ is of order $1$, $\hat b_X$ is very close to $1$ and slightly
larger. This gives already interesting results concerning the Higgs
spectrum which varies like $\sqrt{b}$ (see (\ref{eq:mhiggs})):
 $X^0, \hat X^3, \hat H^3$ are quasi
degenerate, $H^0$ is very light, $\Omega^0$ and $\hat\Omega^3$ have
intermediate mass scales. As far as $\Xi^0$ and $\hat \Xi^3$ are concerned,
they are at the moment massless but they cannot be true Goldstones bosons
and  we shall show in subsection \ref{subsec:lights}
 that they are expected to get small masses by quantum corrections. 

Hierarchies between bosonic VEV's are accordingly ${\cal O}(1)$ except for
$\hat v_H/v_H \equiv 1/\sqrt{b_H}$ which is still large, but much smaller
than for 1 generation (see also subsection \ref{subsec:hier}). This is the same
type of ``see-saw'' mechanism that we witnessed for one generation between
the quasi-standard Higgs doublet and its parity transformed, and which
gives birth to a very light Higgs bosons $H^0$.

Equation (\ref{eq:mw2gen}) determines the value of $\hat v_H$ as a
function of $\delta$ and $\hat r_H$
\begin{equation}
\boxed{
\hat v_H =\frac{2m_W}{g} \frac{1}{\sqrt{b_X+b_H+b_\Omega + 1 + \hat
b_X+\hat b_\Omega}}
}
\label{eq:hvh}
\end{equation}
Since, as we shall confirm, $\delta \approx m_{D_s}^2$, $b_X \approx
1\approx \hat b_X$, $b_\Omega \approx \hat b_\Omega \approx .2$, one can
already state
\begin{equation}
\hat v_H \approx 143\;GeV.
\label{eq:hvh0}
\end{equation}
The value of $\hat v_H$ comes out smaller than in the GSW model because the
$W$ mass receives contributions from the VEV's of several Higgs bosons.
Each of them has therefore less to contribute.

Once the $b$'s are determined, (\ref{eq:delrel1}) gives the values of the
$\delta_i$'s and (\ref{eq:sumsolr1}) provides the ratios of fermionic
VEV's (except $\hat r_X$ which is problematic, see section
\ref{section:neutmass} below).
We shall come back later to their numerical values.


Accordingly, at this point, $\delta$ and $\hat r_H$ are still to be
determined, together with the $\delta_{i \hat\imath}=-\kappa_{\hat\imath i}$'s.
This makes a total of 10 parameters still to be determined to have full
control of the theory. For what concerns us here, mainly the spectrum of
Higgs bosons, we mainly need $\delta$ and $\hat v_H$. The other parameters
will only be needed to determine the couplings of the various fields to each
other.

In  section \ref{section:genferm}, we shall use the fermionic
sector of the theory to determine $\delta$ and $\hat v_H$.

\section{General fermionic constraints}\label{section:genferm}

Yukawa couplings provide quark mass terms as functions of the various
parameters and VEV's. Setting  $\theta_u$ to $0$ constrains in particular
 the non-diagonal
$\mu_{uc}$ and $\mu_{cu}$ mass terms to vanish.

\subsection{Quark mass terms}\label{subsec:qmass}

From the Yukawa Lagrangian (\ref{eq:Lyuk0}) and $i\in [X,H,\Omega,\Xi]$, one gets
the following diagonal quark mass terms
\begin{equation}
\begin{split}
\mu_u &=\frac{\hat v_H^2}{2\sqrt{2}}\left[
\frac{\delta(1-b_X)b_X}{\mu_X^3}
+\frac{\delta(1-\hat b_X)\hat b_X}{\hat\mu_X^3}
+\delta_{X\hat X}\,\sqrt{b_X \hat b_X}\left(
\frac{1}{\mu_X^3}-\frac{1}{\hat\mu_X^3}
\right) \right],\cr
\mu_d &=\frac{\hat v_H^2}{2\sqrt{2}}\left[
\frac{\delta(1-b_X)b_X}{\mu_X^3}
-\frac{\delta(1-\hat b_X)\hat b_X}{\hat\mu_X^3}
+\delta_{X\hat X}\,\sqrt{b_X \hat b_X}\left(
\frac{1}{\mu_X^3}+\frac{1}{\hat\mu_X^3}
\right) \right],\cr
\mu_c &=\frac{\hat v_H^2}{2\sqrt{2}}\left[
\frac{\delta(1-b_H)b_H}{\mu_H^3}
+\frac{\delta(\overbrace{1-\hat b_H}^{0})\hat b_H}{\hat\mu_H^3}
+\delta_{H\hat H}\,\sqrt{b_H \hat b_H}\left(
\frac{1}{\mu_H^3}-\frac{1}{\hat\mu_H^3}
\right) \right]\cr
&=\frac{\hat v_H^2}{2\sqrt{2}}\left[
\frac{\delta(1-b_H)b_H}{\mu_H^3}
+\delta_{H\hat H}\,\sqrt{b_H} \left(
\frac{1}{\mu_H^3}-\frac{1}{\hat\mu_H^3}
\right) \right] ,\cr
\mu_s &=\frac{\hat v_H^2}{2\sqrt{2}}\left[
\frac{\delta(1-b_H)b_H}{\mu_H^3}
-\frac{\delta(\overbrace{1-\hat b_H}^{0})\hat b_H}{\hat\mu_H^3}
+\delta_{H\hat H}\,\sqrt{b_H \hat b_H}\left(
\frac{1}{\mu_H^3}+\frac{1}{\hat\mu_H^3}
\right) \right]\cr
&= \frac{\hat v_H^2}{2\sqrt{2}}\left[
\frac{\delta(1-b_H)b_H}{\mu_H^3}
+\delta_{H\hat H}\,\sqrt{b_H} \left(
\frac{1}{\mu_H^3}+\frac{1}{\hat\mu_H^3}
\right) \right],
\end{split}
\label{eq:diagqmass}
\end{equation}

and the following non-diagonal quark mass terms

\begin{equation}
\begin{split}
\mu_{uc}&= \frac{\hat v_H^2}{4}\left[
\frac{\delta(1-b_\Omega)b_\Omega}{\mu_\Omega^3}
+\frac{\delta(1-\hat b_\Omega)\hat b_\Omega}{\hat\mu_\Omega^3}
+\delta_{\Omega\hat \Omega}\,\sqrt{b_\Omega \hat b_\Omega}\left(
\frac{1}{\mu_\Omega^3}-\frac{1}{\hat\mu_\Omega^3}
\right)\right. \cr
&\hskip 2cm\left.+ \frac{\delta(1-b_\Xi)b_\Xi}{\mu_\Xi^3}
+\frac{\delta(1-\hat b_\Xi)\hat b_\Xi}{\hat\mu_\Xi^3}
+\delta_{\Xi\hat \Xi}\,\sqrt{b_\Xi \hat b_\Xi}\left(
\frac{1}{\mu_\Xi^3}-\frac{1}{\hat\mu_\Xi^3}
\right)
\right]\ \leftarrow\ =0\cr
&= \frac{\hat v_H^2}{4}\left[
\frac{\delta(1-b_\Omega)b_\Omega}{\mu_\Omega^3}
+\frac{\delta(1-\hat b_\Omega)\hat b_\Omega}{\hat\mu_\Omega^3}
+\delta_{\Omega\hat \Omega}\,\sqrt{b_\Omega \hat b_\Omega}\left(
\frac{1}{\mu_\Omega^3}-\frac{1}{\hat\mu_\Omega^3}
\right)\right],\cr
\mu_{cu}&= \frac{\hat v_H^2}{4}\left[
\frac{\delta(1-b_\Omega)b_\Omega}{\mu_\Omega^3}
+\frac{\delta(1-\hat b_\Omega)\hat b_\Omega}{\hat\mu_\Omega^3}
+\delta_{\Omega\hat \Omega}\,\sqrt{b_\Omega \hat b_\Omega}\left(
\frac{1}{\mu_\Omega^3}-\frac{1}{\hat\mu_\Omega^3}
\right)\right.\cr
&\hskip 2cm \left.- \frac{\delta(1-b_\Xi)b_\Xi}{\mu_\Xi^3}
-\frac{\delta(1-\hat b_\Xi)\hat b_\Xi}{\hat\mu_\Xi^3}
-\delta_{\Xi\hat \Xi}\,\sqrt{b_\Xi \hat b_\Xi}\left(
\frac{1}{\mu_\Xi^3}-\frac{1}{\hat\mu_\Xi^3}
\right)
\right]\ \leftarrow\ =0\cr
&= \frac{\hat v_H^2}{4}\left[
\frac{\delta(1-b_\Omega)b_\Omega}{\mu_\Omega^3}
+\frac{\delta(1-\hat b_\Omega)\hat b_\Omega}{\hat\mu_\Omega^3}
+\delta_{\Omega\hat \Omega}\,\sqrt{b_\Omega \hat b_\Omega}\left(
\frac{1}{\mu_\Omega^3}-\frac{1}{\hat\mu_\Omega^3}
\right)\right],
\end{split}
\label{eq:ndiagqmass1}
\end{equation}
\begin{equation}
\begin{split}
\mu_{ds}&= \frac{\hat v_H^2}{4}\left[
\frac{\delta(1-b_\Omega)b_\Omega}{\mu_\Omega^3}
-\frac{\delta(1-\hat b_\Omega)\hat b_\Omega}{\hat\mu_\Omega^3}
+\delta_{\Omega\hat \Omega}\,\sqrt{b_\Omega \hat b_\Omega}\left(
\frac{1}{\mu_\Omega^3}+\frac{1}{\hat\mu_\Omega^3}
\right)\right.\cr
&\hskip 2cm \left.+ \frac{\delta(1-b_\Xi)b_\Xi}{\mu_\Xi^3}
-\frac{\delta(1-\hat b_\Xi)\hat b_\Xi}{\hat\mu_\Xi^3}
+\delta_{\Xi\hat \Xi}\,\sqrt{b_\Xi \hat b_\Xi}\left(
\frac{1}{\mu_\Xi^3}+\frac{1}{\hat\mu_\Xi^3}
\right)
\right]\ \leftarrow\ =0\cr
&=   \frac{\hat v_H^2}{4}\left[
\frac{\delta(1-b_\Omega)b_\Omega}{\mu_\Omega^3}
-\frac{\delta(1-\hat b_\Omega)\hat b_\Omega}{\hat\mu_\Omega^3}
+\delta_{\Omega\hat \Omega}\,\sqrt{b_\Omega \hat b_\Omega}\left(
\frac{1}{\mu_\Omega^3}+\frac{1}{\hat\mu_\Omega^3}
\right)\right],\cr
\mu_{sd}&= \frac{\hat v_H^2}{4}\left[
\frac{\delta(1-b_\Omega)b_\Omega}{\mu_\Omega^3}
-\frac{\delta(1-\hat b_\Omega)\hat b_\Omega}{\hat\mu_\Omega^3}
+\delta_{\Omega\hat \Omega}\,\sqrt{b_\Omega \hat b_\Omega}\left(
\frac{1}{\mu_\Omega^3}+\frac{1}{\hat\mu_\Omega^3}
\right)\right.\cr
&\hskip 2cm \left.- \frac{\delta(1-b_\Xi)b_\Xi}{\mu_\Xi^3}
+\frac{\delta(1-\hat b_\Xi)\hat b_\Xi}{\hat\mu_\Xi^3}
-\delta_{\Xi\hat \Xi}\,\sqrt{b_\Xi \hat b_\Xi}\left(
\frac{1}{\mu_\Xi^3}+\frac{1}{\hat\mu_\Xi^3}
\right)
\right]\ \leftarrow\ =0\cr
&=  \frac{\hat v_H^2}{4}\left[
\frac{\delta(1-b_\Omega)b_\Omega}{\mu_\Omega^3}
-\frac{\delta(1-\hat b_\Omega)\hat b_\Omega}{\hat\mu_\Omega^3}
+\delta_{\Omega\hat \Omega}\,\sqrt{b_\Omega \hat b_\Omega}\left(
\frac{1}{\mu_\Omega^3}+\frac{1}{\hat\mu_\Omega^3}
\right)\right].
\end{split}
\label{eq:ndiagqmass2}
\end{equation}
To write eqs. (\ref{eq:diagqmass}), (\ref{eq:ndiagqmass1}) and
(\ref{eq:ndiagqmass2}), all bosonic and fermionic VEV's have been
supposed to be real. This is in particular legitimate for all diagonal
$<\bar q_i q_i>$ fermionic condensates of hermitian operators.
We have no reasons a priori to
consider that bosonic VEV's could become complex, and no sign either that
$<\bar u c>=<\bar c u>$  and $<\bar d s>=<\bar s d>$
could get some imaginary parts. 

Some explanations are due concerning the vanishing of the lines marked with
arrows in (\ref{eq:ndiagqmass1}), (\ref{eq:ndiagqmass2}) above, which
ensures in particular the symmetry relations $\mu_{uc}=\mu_{cu},
\mu_{ds}=\mu_{sd}$.\newline
Among the terms with arrows stand, for example,
$\frac{\delta(1-b_\Xi)b_\Xi}{\mu_\Xi^3}$. By (\ref{eq:sumsolnu2}), 
calling the constant ratio $(1-b_X)/\nu_X^4 = \beta=(1-b_\Xi)/\nu_\Xi^4$,
 this term rewrites
$\delta\frac{b_\Xi}{\mu_\Xi^3} \beta \nu_\Xi^4 = \delta
\frac{b_\Xi}{\mu_\Xi^3}\beta \frac{2 \mu_\Xi^6}{\hat v_\Xi^2}=\frac{2\delta
b_\Xi\beta \mu_\Xi^3}{b_\Xi \hat v_H^2} = \frac{2\delta \beta
\mu_\Xi^3}{\hat
v_H^2}$; since $\mu_\Xi^3=0$, this term vanishes.
So, $\frac{\delta b_\Xi(1-b_\Xi)}{\mu_\Xi^3} \stackrel{\mu_\Xi^3
=0}{=}0$ and, likewise,  $\frac{\delta \hat b_\Xi(1-\hat
b_\Xi)}{\hat \mu_\Xi^3} \stackrel{\hat \mu_\Xi^3 =0}{=}0$. Next,
consider  $\delta_{\Xi\hat\Xi}\sqrt{b_\Xi \hat b_\Xi}
\left(\frac{1}{\mu_\Xi^3}\pm \frac{1}{\hat\mu_\Xi^3}\right)$.
By the same argumentation one gets 
$\delta_{\Xi\hat\Xi}\sqrt{b_\Xi \hat b_\Xi}
\left(\frac{1}{\mu_\Xi^3}\pm
\frac{1}{\hat\mu_\Xi^3}\right)=\frac{\sqrt{2}\delta_{\Xi\hat\Xi}}{\hat v_H
\sqrt{\beta}}\left(\sqrt{(1-b_\Xi)\hat b_\Xi} \pm \sqrt{(1-\hat
b_\Xi)b_\Xi} \right)$, which vanishes for $b_\Xi=0=\hat b_\Xi$. These
properties ensure in particular that, as soon as their diagonal elements
are real, the mass matrices of $(u,c)$ and $(d,s)$ quarks are hermitian
\footnote{Then (see for example \cite{MachetPetcov}), no bi-unitary
transformation is needed to diagonalize them such that the quarks masses
become identical to the eigenvalues of the mass matrices.}.

\subsection{Tuning $\boldsymbol{\theta_u}$ to zero}
\label{subsec:nulthetau}

Since $\tan 2\theta_u = -\frac{2\mu_{uc}}{\mu_u-\mu_c}$, tuning
$\theta_u$ to $0$ goes along with constraining $\mu_{uc}$ to vanish.
 Using the dimensionless $b$ and
$r$ variables defined in (\ref{eq:bdef}) and (\ref{eq:rdef}), this is
equivalent, by (\ref{eq:ndiagqmass1}), to
\begin{equation}
\theta_u=0 \quad\Leftrightarrow \quad
\delta_{\Omega\hat\Omega}=-\frac{\delta}{\sqrt{b_\Omega\hat b_\Omega}}
\frac{1}{\hat r_\Omega-r_\Omega}\left(
\hat r_\Omega b_\Omega(1-b_\Omega)
+ r_\Omega \hat b_\Omega(1-\hat b_\Omega)
\right),
\label{eq:delomhom1}
\end{equation}
which determines $\delta_{\Omega\hat\Omega}$ as a function of $\delta$.

Tuning $\theta_u$ to $0$ also entails that the mass  eigenvalues
are $\mu_u=m_u$ and $\mu_c=m_c$.
By (\ref{eq:diagqmass}), this is equivalent to
\begin{equation}
\delta_{X\hat X}=\frac{\delta}{\sqrt{b_X \hat b_X}(\hat r_X
-\underbrace{r_X}_{1})}
\left(
\frac{2\sqrt{2}}{\rho}\underbrace{r_X}_{1}\hat r_X \frac{m_u}{m_u+m_d}
- \hat r_X b_X(1-b_X) -\underbrace{r_X}_{1} \hat b_X(1-\hat b_X)
\right),
\label{eq:delxhx1}
\end{equation}
and
\begin{equation}
\delta_{H\hat H}=\frac{\delta}{\sqrt{b_H \underbrace{\hat b_H}_{1}}(\hat r_H
-r_H)}\left(
\frac{2\sqrt{2}}{\rho} r_H\hat r_H \frac{m_c}{m_u+m_d}
- \hat r_H b_H(1-b_H) -r_H \hat b_H (\underbrace{1-\hat b_H}_{0})
\right),
\label{eq:delhhh1}
\end{equation}
in which $m_u, m_d, m_c$ are now considered as physical inputs and in which
 we have introduced the dimensionless variable $\rho$
\begin{equation}
\rho= \frac{\delta \hat v_H^2}{(m_u+m_d)\mu_X^3}
\stackrel{(\ref{eq:mw2gen})}{=} \frac{4\delta m_W^2}{g^2(m_u+m_d)\mu_X^3}
\;\frac{1}{b_X + b_H + b_\Omega + 1 + \hat b_X + \hat b_\Omega}.
\label{eq:rhodef}
\end{equation}
$\rho$ is a function of $\delta, \hat r_H$ (and $\mu_X^3$ but for this we
can take its approximation (\ref{eq:GMOR2})).
Eqs.~(\ref{eq:delxhx1}) and (\ref{eq:delhhh1}) therefore determine
 $\delta_{X\hat X}$ and $\delta_{H\hat H}$ as
functions of $\delta$ and $\hat r_H$.

Taking (\ref{eq:delomhom1}), (\ref{eq:delxhx1}) and (\ref{eq:delhhh1}) into
account, one gets now from (\ref{eq:diagqmass}) and (\ref{eq:ndiagqmass2})
\begin{equation}
\mu_d
=-\frac{(m_u+m_d)\rho}{\sqrt{2}}
\frac {b_X(1-b_X)+\hat b_X(1-\hat b_X)}{\hat r_X-1}
+m_u\frac{\hat r_X+1}{\hat r_X-1},
\label{eq:mud}
\end{equation}

\begin{equation}
\mu_s
=-\frac{(m_u+m_d)\rho}{\sqrt{2}} \frac{b_H(1-b_H)}{\hat r_H
-r_H}
+ m_c\frac{\hat r_H + r_H}{\hat r_H -r_H},
\label{eq:mus}
\end{equation}

\begin{equation}
\mu_{ds}
=-\frac{(m_u+m_d)\rho}{2} \frac{b_\Omega(1-b_\Omega)+\hat
b_\Omega(1-\hat b_\Omega)}{\hat r_\Omega-r_\Omega}.
\label{eq:muds}
\end{equation}
The masses of the $d$ and $s$ quarks are given by
\begin{equation}
\begin{split}
m_d &=\frac12\left(
(\mu_d+\mu_s)-\sqrt{(\mu_d-\mu_s)^2+4\mu_{ds}^2}\right),\cr
m_s &=\frac12\left(
(\mu_d+\mu_s)+\sqrt{(\mu_d-\mu_s)^2+4\mu_{ds}^2}\right),
\end{split}
\label{eq:mdms}
\end{equation}
and the $\theta_d =\theta_c$ mixing angle by
\begin{equation}
\tan 2\theta_d =\frac{-2\mu_{ds}}{\mu_d-\mu_s}.
\label{eq:tan2theta}
\end{equation}

$m_d, m_s$ and $\theta_c$ are accordingly functions of $\delta$ and $\hat
r_H$.

\section{Constraints of reality}\label{section:reality}

The Yukawa Lagrangian (\ref{eq:Lyuk0}) being hermitian, fermionic mass
terms, in particular $\mu_d, \mu_s, \mu_{ds}$
cannot be but real. For $\theta_u=0$, which translates into
(\ref{eq:delomhom1}, \ref{eq:delxhx1}, \ref{eq:delhhh1}),
their expressions are given in (\ref{eq:mud}, \ref{eq:mus}, \ref{eq:muds}).
These formul{\ae}  assume that all bosonic $b,\hat b$ and fermionic
$\mu^3,\hat\mu^3$ VEV's are real, which a priori does not pose any problem,
except for $\hat\mu_X^3$, because the value of $\hat b_X$ determined from
the spectrum of pseudoscalar mesons is larger than $1$: then, $K^0$ can
only be orthogonal to $\bar K^0$ and $\pi^0$ to $K^0 + \bar K^0$ if
$\hat\mu_X^3$ becomes imaginary.

Therefore, we have, at least formally, to allow $\hat\mu_X^3$ to become
complex, which has consequences on $\mu_u$ and $\mu_d$ in (\ref{eq:diagqmass}).

\subsection{Reality of $\boldsymbol{\mu_d}$}
\label{subsec:mudreal}

Because $\mu_d+\mu_s=m_d+m_s$ and since $\mu_s$, as we shall see in
subsection \ref{subsec:realmus}), has no problem of reality,
$\mu_d$ should stay real whatever happens to $\hat\mu_X^3$. We accordingly
constrain its imaginary part to vanish.

All desired orthogonality conditions are supposed to be satisfied.
In particular (\ref{eq:solnu31})  is equivalent to
 $\hat b_X(1-\hat b_X) =2\hat r_X^2 b_X(1-b_X)$; therefore, 
 $\hat r_X^2$ is real, and $\hat r_X^2 -1$ too. Thanks to this, one easily
gets from (\ref{eq:mud})
\begin{equation}
\Im (\mu_d) = \frac{1}{\hat r_X^2-1}\Bigg[
 -\frac{(m_u+m_d)\rho}{\sqrt{2}} b_X(1-b_X)(1+2\hat r_X^2) \Im \hat r_X +
2m_u \Im \hat r_X
\Bigg],
\end{equation}
such that the reality condition of $\mu_d$ reads
\begin{equation}
\mu_d\ real\ \Leftrightarrow 2m_u =  \frac{(m_u+m_d)\rho}{\sqrt{2}}\Big(
b_X(1-b_X) + \hat b_X(1-\hat b_X)\Big)=\frac{\delta \hat
v_H^2}{\sqrt{2}\mu_X^3} \Big(
b_X(1-b_X) + \hat b_X(1-\hat b_X)\Big).
\label{eq:realmud}
\end{equation}
Note that this entails in particular that $b_X(1-b_X) + \hat b_X(1-\hat
b_X)$ is a very small positive number (we suppose that $m_u>0$).
When (\ref{eq:realmud}) is realized, one also gets from (\ref{eq:mud})
\begin{equation}
\mu_d =\frac{1}{\hat r_X^2-1}\Bigg(
\underbrace{-\frac{m_u+m_d}{\sqrt{2}}\rho\; b_X(1-b_X)(1+2\hat
r_X^2)}_{-2m_u}(1+\Re \hat r_X) +
m_u(1+\hat r_X^2 +2\Re \hat r_X)\Bigg)
= m_u\frac{-1+\hat r_X^2}{\hat r_X^2-1}.
\end{equation}
So,
\begin{equation}
\mu_d\ real\ \Rightarrow\mu_d = m_u,
\label{eq:mud=mu}
\end{equation}
which goes accordingly to $0$ at the chiral limit.
When (\ref{eq:realmud}) is realized, one also gets from (\ref{eq:delxhx1})
\begin{equation}
\delta_{X\hat X} = \delta\frac{\hat
b_X(1-\hat b_X)}{\sqrt{b_X\hat b_X}}.
\label{eq:delxhx2}
\end{equation}

\subsection{First consequence: determination
of $\boldsymbol{|\hat r_H|}$}\label{subsec:hvh2}

\begin{figure}
\centering
\includegraphics[width=10truecm, height=6truecm]{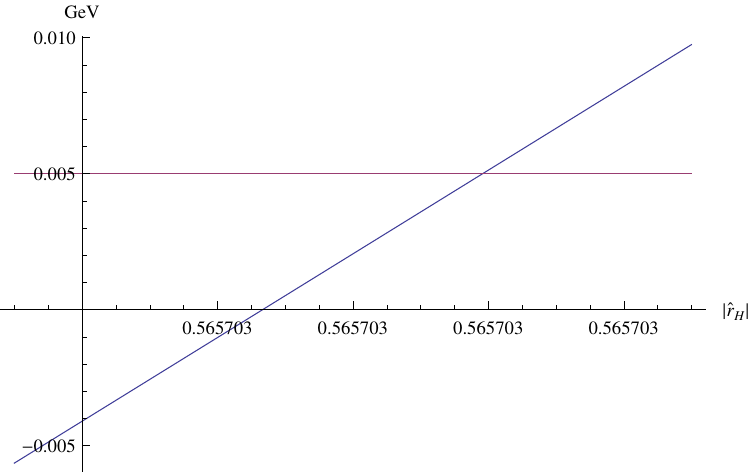}

\caption{The r.h.s. of (\ref{eq:realmud}) is plotted as a function of $|\hat
r_H|$ for $\delta \approx m_{D_s}^2$ and $\hat v_H \approx 143\,GeV$; the
horizontal line is the value of the l.h.s. $2m_u \approx 5\,MeV$}

\label{figure:hrh3}
\end{figure}

The determination of $\hat r_H$ is done through (\ref{eq:realmud}).
We plot in Fig.~\ref{figure:hrh3} its r.h.s. in which
we have inserted the estimate (\ref{eq:hvh0}) for $\hat v_H$ and $\delta
\approx m_{D_s}^2$, as a function of $|\hat r_H|$. The horizontal red line
is the value of $2m_u \approx 5\,MeV$.

In practice, due to the smallness of $m_u$ and the large value of
$\delta \hat v_H^2/\sqrt{2}\mu_X^3$, the solution of this equation is
practically the same as that of $b_X(1-b_X) + \hat b_X(1-\hat b_X)=0$.
It is very precise, with furthermore an extremely small sensitivity to
variations of $\delta$ and of $\hat v_H$.

One gets
\begin{equation}
|\hat r_H| \approx .565703.
\end{equation}

A reasonable approximate value can be obtained from the expansions of $b_X$
and $\hat b_X$ (\ref{eq:bxexp}) and (\ref{eq:hbxexp}) at the chiral limit.
They yield $b_X(1-b_X) + \hat b_X(1-\hat b_X)\approx 0$ at $\hat r_H^2
=1/3=(.58)^2$.

\subsection{Second consequence: another determination of
 the mixing angle}\label{subsec:mixingagain}

Once (\ref{eq:mud=mu}) has been implemented, (\ref{eq:mdms})
and (\ref{eq:tan2theta}) yield
\begin{equation}
m_d = m_u\frac{1+\sqrt{1+\tan^2 2\theta_c}}{2} +
\mu_s\frac{1-\sqrt{1+\tan^2 2\theta_c}}{2}.
\label{eq:mdbis}
\end{equation}
If we believe that $m_d >0$ and that, at least, $m_d \geq 1.5\,m_u$, we are
having troubles with (\ref{eq:mdbis}). One indeed needs the second term in
its r.h.s. to be positive; but, the numerator of the fraction being
negative, this requires $\mu_s<0$.
However, $m_d+m_s=\mu_d+\mu_s$ becomes, due to (\ref{eq:realmud}),
$m_d+m_s=m_u+\mu_s$ such that 
\begin{equation}
\mu_s= m_d+m_s-m_u
\label{eq:musbis}
\end{equation}
 has to be positive.
The only solution is therefore  
\begin{equation}
m_d <0
\label{eq:mdneg}
\end{equation}
like we have found for 1 generation.
Using (\ref{eq:mdbis}) and (\ref{eq:musbis}) leads straightforwardly to the
result
\begin{equation}
\tan^2\theta_d = \frac{m_u-m_d}{m_s-m_u}=\frac{m_u+|m_d|}{m_s-m_u},
\label{eq:thetabis}
\end{equation}
which is, like (\ref{eq:tantheta}) a fairly accurate formula if one takes
\cite{PDG} $m_u\approx 2.5\,MeV, |m_d|\approx 5\,MeV, m_s\approx 100\,MeV$.
It is certainly not a new output since similar estimates have already been
obtained (see for example \cite{Oakes} \cite{Weinberg} \cite{Fritzsch}).
 However, the way it has been obtained is new.

Note that a consequence of (\ref{eq:musbis}) is that, at the chiral limit
$m_u,m_d \to 0$, $\mu_s \to m_s$. Furthermore, since we know that $m_u,
|m_d| \ll m_s$, at a good approximation $\mu_s \approx m_s$.

\subsection{Reality of $\boldsymbol{\mu_s}$}\label{subsec:realmus}

Since $\hat r_H$ has been shown to be non-pathological (we will determine
its sign later), the reality of $\mu_s$ as given by (\ref{eq:mus}) can
only be put in jeopardy if $r_H$ becomes complex, which means, by
(\ref{eq:sumsolr1}), if $b_H<0$. This would correspond to a
physically unacceptable imaginary $v_H$ (once $\hat v_H$ is indeed real).
Despite this reservation, we may play the same game as for $\mu_d$ and
require that, when $b_H$ formally becomes negative, the imaginary part of
$\mu_s$ should vanish.
 
\begin{equation}
\mu_s=\frac{1}{\hat r_H^2-r_H^2}\Bigg[
-(\hat r_H + r_H)b_X(1-b_X)r_H^2 \frac{(m_u+m_d)\rho}{\sqrt{2}} + m_c(\hat
r_H^2 + r_H^2 +2 r_H \hat r_H)
\Bigg],
\end{equation}
such that
\begin{equation}
\Im \mu_s = \frac{\Im r_H}{\hat r_H^2-r_H^2}\Bigg[
-\frac{(m_u+m_d)\rho}{\sqrt{2}}b_H(1-b_H) + 2m_c\hat r_H
\Bigg],
\end{equation}
and
\begin{equation}
\Im \mu_s =0 \Leftrightarrow
2m_c=\frac{(m_u+m_d)\rho}{\sqrt{2}}\frac{b_H(1-b_H)}{\hat r_H}
 =\frac{(m_u+m_d)\rho}{\sqrt{2}}\frac{r_H^2 b_X(1-b_X)}{\hat r_H}
=\frac{\delta \hat v_H^2}{\sqrt{2}\mu_X^3}\frac{r_H^2 b_X(1-b_X)}{\hat
r_H}.
\label{eq:musreal}
\end{equation}
When this is realized
\begin{equation}
\mu_s =\Re\mu_s = \frac{1}{\hat r_H^2-r_H^2}\Bigg[
-(\hat r_H + \Re r_H)\underbrace{b_X(1-b_X)r_H^2
\frac{(m_u+m_d)\rho}{\sqrt{2}}}_{2m_c \hat r_H}
+m_c(
\hat r_H^2 + r_H^2 +2\hat r_H \Re r_H)
\Bigg]=-m_c.
\end{equation}
This result being in flagrant disagreement with what we have obtained
before, in particular at the chiral limit at which $\mu_s=m_s$, we conclude
that a negative $b_H$ is totally excluded, which requires, as we have
already written in (\ref{eq:delbh0num}), $\delta \geq 3.90923\,GeV^2$. It
will be satisfied by our result in (\ref{eq:deltanum}).

\section{The fermionic mixing angle : a paradox}\label{section:fermix}

Using (\ref{eq:realmud}) and (\ref{eq:musbis}), (\ref{eq:tan2theta})
rewrites (we recall $\theta_d=\theta_c$)
\begin{equation}
\tan 2\theta_d= \frac{2\mu_{ds}}{m_d+m_s-2m_u},
\label{eq:tan2thetabis}
\end{equation}
or, equivalently 
\begin{equation}
\mu_{ds}= (m_d+m_s-2m_u)\frac{\tan\theta_c}{1-\tan^2\theta_c}.
\label{eq:muds3}
\end{equation}
On the other side, $\mu_{ds}$ is given by (\ref{eq:muds}) and
(\ref{eq:rhodef})
\begin{equation}
\mu_{ds}= \frac{\delta \hat
v_H^2}{2\mu_X^3}\;\frac{b_\Omega(1-b_\Omega)+\hat b_\Omega(1-\hat
b_\Omega)}{r_\Omega-\hat r_\Omega}.
\label{eq:muds4}
\end{equation}
Neither in (\ref{eq:muds3}) nor in (\ref{eq:muds4}) did we make use of any
bosonic relation, except those in (\ref{eq:delrel1})
 of the type $\delta_i=\delta(1-b_i)$ which
result from the minimization of the effective potential.
At this point (\ref{eq:sumsolr1}), relating the $r$'s to the $b$'s, has not
been used, nor any of the bosonic relations connecting the $b$'s to
$\delta$ and to the masses of pseudoscalar mesons.
In this respect the mixing angle occurring in (\ref{eq:muds4}) appears as
the ``fermionic mixing angle''.

$\delta \hat v_H^2/2\mu_X^3$ which occurs in the r.h.s. of (\ref{eq:muds4})
 is a very large scale while, a
priori, at least from bosonic considerations and measurements,
 we expect a small (bosonic) $\theta_c$, which, by (\ref{eq:muds3}),
requires a small $\mu_{ds}$.

Then, to reconcile (\ref{eq:muds3}) and (\ref{eq:muds4}) one needs:\newline
* either a very small value for $\frac{b_\Omega(1-b_\Omega) + \hat
b_\Omega(1-\hat b_\Omega)}{r_\Omega-\hat r_\Omega}$;\newline
* or a ``fermionic mixing angle'' which is different from the bosonic
mixing angle and which is close to maximal ($\pi/4$) to enlarge the
r.h.s. of (\ref{eq:muds3}).

Even if the $b$'s are probably subject to uncertainties, it is very
unlikely that $b_\Omega$ and $\hat b_\Omega$ are in reality close to $0$ or
$1$. If they stay ${\cal O}(1)$ as we determined from bosonic
considerations, then the only solution concerning (\ref{eq:muds4}) is that
$r_\Omega - \hat r_\Omega \equiv \sqrt{2}\frac{<\bar d s+ \bar s d>}{<\bar
u u+\bar d d>}$ becomes very large.\newline
Let us now consider the bosonic evaluation of this quantity. By
(\ref{eq:sumsolnu3}), $r_\Omega-\hat r_\Omega=
\frac{\sqrt{b_\Omega(1-b_\Omega)}
-\sqrt{\hat b_\Omega(1-\hat b_\Omega)} }{\sqrt{b_X(1-b_X)}}$ and we have
found that $b_\Omega$ is not very different from $\hat b_\Omega$
\footnote{All other masses and parameters being left untouched, 
moving the mass of the neutral $D^0$ meson from  $m_{D^0}=1864.86\,MeV$ \cite{PDG}
 down to $m_{D^0}=1862.29\,MeV$ is enough to match $b_\Omega=\hat
b_\Omega$, that is, to reach the pole of $\mu_{ds}$.}
. It is
thus very unlikely that $r_\Omega-\hat r_\Omega$ becomes very large unless
$b_X$ comes extremely close to $1$, much closer than what we found  from bosonic
considerations.\newline
In this case,  sticking to a small mixing angle, the r.h.s. of
(\ref{eq:muds4}) can only match the one of (\ref{eq:muds3}) if very little
confidence can be attached to the bosonic determinations of the $b$'s, in
particular that of $b_X$. One knows that  $b_X \to 1$ at the chiral limit $(m_u, m_d,
m_\pi \to 0, \mu_X^3\ fixed)$, and we may wonder whether  our calculations
could be, for some unknown reason, only valid at this  limit.
This is far from satisfying, because then, since the Cabibbo angle also
 vanishes at this limit (see (\ref{eq:tantheta})),
 no credit whatsoever should be
then granted to our calculations (\ref{eq:tan2theta}) and (\ref{eq:thetabis}).
 This is why we   look for other possibilities:\newline
* it may happen that adding the 3rd generation of quarks
yield a value of $b_X$ naturally much closer to $1$;\newline
* the r.h.s. of (\ref{eq:muds4}) is instead doomed to stay naturally large,
such that its matching with the r.h.s. of (\ref{eq:muds3}) calls for a quasi-maximal
fermionic mixing angle $\tan^2 \theta_d \approx 1$. This reminds of leptons,
where large fermionic mixing angle(s) seem to be ``natural''. There, 
they  are directly measured from the corresponding asymptotic states,
 which  is not the case for mesons.

At this stage, the situation for 2 generations
 can only be summarized as follows: {\em the
``bosonic'' evaluation of $\mu_{ds}$ (we mean by this the evaluation of
the r.h.s. of (\ref{eq:muds4}) in which the $b$'s and $r$'s are calculated
 from bosonic considerations) corresponds to a quasi-pole of its fermionic
expression (\ref{eq:muds3}) and, accordingly,
to $\theta_d^{fermionic}\approx \pi/4$}.

We shall see in chapter \ref{chapt:2gentetau} that, in reality, the
situation already largely improves when one switches on $\theta_u \not=0$,
which in particular brings $b_\Omega$ (and presumably also $\hat b_\Omega$)
very close to $0$. But this cannot  be guessed at this stage of the
study.

\section{The chiral limit and the $\boldsymbol s$ quark mass; the sign of
$\boldsymbol{\hat r_H}$}
\label{section:chilim}

By going to the chiral limit $m_u, m_d, m_\pi \to 0$, with
 $<\bar u_m u_m + \bar d_m d_m>$ fixed, we shall show that
 $\hat r_H\equiv\frac{<\bar c c-\bar s
s>}{<\bar u u+\bar d d>}$ is positive and get an estimate of how close
$\delta$ stays to $m_{D_s}^2$.

We know by (\ref{eq:tantheta}) that the Cabibbo angle goes to $0$ at the chiral
limit $m_\pi\to 0$. This also means by (\ref{eq:tan2theta})
 that, at the fermionic level,  $\mu_{ds} \to 0$, such that
 in particular, $\mu_s \simeq m_s$ as given by (\ref{eq:mus}). 
According to (\ref{eq:rhodef}), $(m_u+m_d)\rho \equiv \frac{4\delta
m_W^2}{g^2\mu_X^3}\frac{1}{b_X+b_H+b_\Omega +1 + \hat b_X +\hat b_\Omega}$
is a very large mass scale. Indeed, we already know that $\delta \geq
m_{D_s}^2$, $g={\cal O}(1)$, $\mu_X^3$ is supposed to stay constant at the
chiral limit and keep a value close to its physical value as given by the
GMOR relation, and the $b$'s are of order $1$ or smaller. So, the role
of $(m_u+m_d)\rho$
in (\ref{eq:mus}) should be damped by $b_H$ becoming very small, which, as
we saw on (\ref{eq:delbhnul}), can only happen for $\delta \to m_{D_s}^2$.

Let us be more precise.  When $m_\pi\to 0$,
\begin{equation}
\begin{split}
b_X(1-b_X) &\simeq \frac{m_\pi^2}{\delta} +
\frac{m_\pi^4}{\sigma^4}+\ldots,\cr
with\quad \frac{1}{\sigma^4}
&=-\frac{1}{\delta}\left(\frac{1}{\delta}-\frac{1}{m_{D^+}^2}+\frac{1}{m_{K^+}^2}
\right);
\end{split}
\label{eq:bxexp2}
\end{equation}
\begin{equation}
\begin{split}
b_H(1-b_H) &\simeq \frac{m_{D_s}^2(\delta-m_{D_s}^2)}{\delta^2}
-\frac{m_\pi^2}{\tau^2}+\ldots,\cr
with\quad \tau^2 &=\frac{\delta^2 m_D^4
m_K^2}{m_{D_s}^2(\delta -2 m_{D_s}^2)(m_D^2-m_{D_s}^2)(m_D^2-m_K^2)}
\stackrel{\delta \approx m_{D_s}^2}{\approx}\frac{m_D^4
m_K^2}{(m_{D_s}^2-m_D^2)(m_D^2-m_K^2)}+\ldots.
\end{split}
\label{eq:bhexp2}
\end{equation}
(\ref{eq:sumsolr1}), (\ref{eq:bxexp2}) and (\ref{eq:bhexp2}) yield
\begin{equation}
r_H^2 = \frac{b_H(1-b_H)}{b_X(1-b_X)}
 \approx
\frac{\frac{m_{D_s}^2(\delta-m_{D_s}^2)}{\delta^2}-\frac{m_\pi^2}{\tau^2}}
{\frac{m_\pi^2}{\delta}+\underbrace{\frac{m_\pi^4}{\sigma^4}}_{can\ be\
neglected}}
\approx \frac{m_{D_s}^2(\delta-m_{D_s}^2)
-\delta^2\frac{m_\pi^2}{\tau^2}}{\delta m_\pi^2}
\approx \frac{m_{D_s}^2(\delta-m_{D_s}^2)}{\delta
m_\pi^2}-\frac{\delta}{\tau^2}.
\end{equation}
Then, from (\ref{eq:mus}) one gets for $\delta$ close to $m_{D_s}^2$
\begin{equation}
m_s \simeq\mu_s \approx\frac
{-\frac{(m_u+m_d)\rho}{\sqrt{2}}\left(\frac{m_{D_s}^2(\delta-m_{D_s}^2)}{\delta^2}
-\frac{m_\pi^2}{\tau^2} \right)
+m_c\left(\hat r_H + \sqrt{\frac{m_{D_s}^2(\delta-m_{D_s}^2)}{\delta
m_\pi^2}-\frac{\delta}{\tau^2}} \right)}
{\hat r_H -\sqrt{\frac{m_{D_s}^2(\delta-m_{D_s}^2)}{\delta
m_\pi^2}-\frac{\delta}{\tau^2}}}.
\label{eq:muschir}
\end{equation}
Inside the $\sqrt{\ }$'s an in-determination arises when both
$m_\pi \to 0$ and $\delta \to m_{D_s}^2$. To lift it, the following
limit has to be taken
\begin{equation}
\delta \stackrel{m_\pi\to 0}{\sim}m_{D_s}^2 + \omega^2 m_\pi^2.
\label{eq:deltachir}
\end{equation}
Plugging (\ref{eq:deltachir}) into (\ref{eq:muschir}) yields
\begin{equation}
m_s \approx \mu_s \stackrel{m_\pi\to 0}{\approx}
\frac{
-\frac{(m_u+m_d)\rho}{\sqrt{2}}
\left(\frac{\omega}{m_{D_s}^2}-\frac{1}{\tau^2}
 \right) m_\pi^2
+m_c\left(\hat r_H +\sqrt{\omega-\frac{\delta}{\tau^2}} \right)}
{\hat r_H -\sqrt{\omega-\frac{\delta}{\tau^2}}}.
\label{eq:muschir2}
\end{equation}

\begin{figure}
\centering
\includegraphics[width=10truecm, height=6truecm]{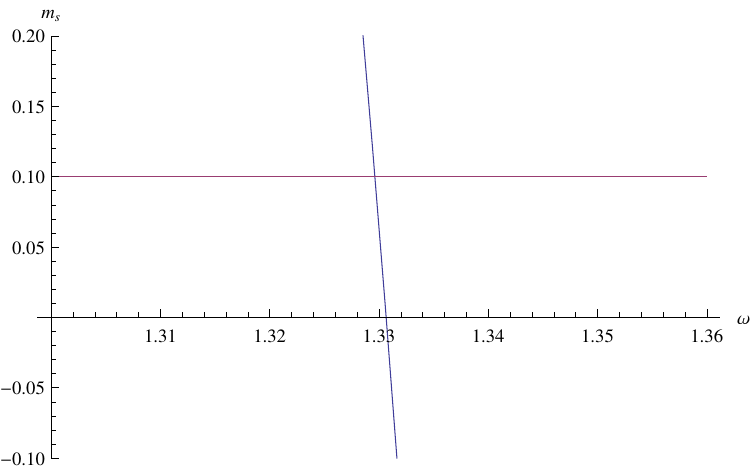}

\caption{The chiral limit of $m_s$ as a function of $\omega$ for $\hat
r_H= + 59215$ and $m_c=1.2755\,GeV$}

\label{figure:mschir}
\end{figure}

We display in Fig.\ref{figure:mschir} the curve giving $m_s$ as a function of $\omega$
 for $\hat r_H = +.59215 >0$, which is the only sign that
can fit the value of $m_s$. We truncated the range of $\omega$  so as to avoid
the pole that appears in (\ref{eq:muschir2}) in this case. The $c$ quark
mass has been taken as $m_c = 1.2755\,GeV$ \cite{PDG}.

This determines $\omega \approx 1.33$, together with
\begin{equation}
\hat r_H\equiv \frac{<\bar c c-\bar s s>}{<\bar u u+\bar d d>} \approx +
.565703,
\label{eq:hrhnum}
\end{equation}
and, by (\ref{eq:deltachir}),
\begin{equation}
\delta = m_{D_s}^2 + (1.33\, m_\pi)^2 \approx 3.9094\,GeV^2.
\label{eq:deltanum0}
\end{equation}

The value (\ref{eq:deltanum0})
 is very close to the lower bound (\ref{eq:delbh0num}) below which $b_H$
becomes negative.

Strictly speaking, we have not proved that, for physical
pions, $\delta$ stays close to this value and does not become much larger.
However, the parameters of all formal expansions that we have been using
($\frac{m_\pi^2}{m_K^2}, \frac{m_\pi^2}{m_D^2},
\frac{m_\pi^2}{m_{D_s}^2}, \frac{m_\pi^2}{\delta}$) are presumably small enough 
for a good convergence and we think reasonable, at this stage,  not to
 expect large deviations from the value (\ref{eq:deltanum0}).

\section{Summarizing the solutions of the equations}\label{section:solution}

The set of solutions includes (\ref{eq:hrhnum}) and (\ref{eq:deltanum0}).
 We also recall (\ref{eq:tantheta})
\begin{equation}
t^2\equiv\tan^2\theta_d=
\displaystyle\frac{\displaystyle\frac{1}{m_{K^\pm}^2}-\displaystyle\frac{1}{m_{D^\pm}^2}}
{\displaystyle\frac{1}{m_{\pi^\pm}^2}-\displaystyle\frac{1}{m_{D_s^\pm}^2}}
\approx .07473 \Rightarrow \theta_c \approx .2668.
\end{equation}

Plugging (\ref{eq:hrhnum}) and (\ref{eq:deltanum0}) in (\ref{eq:bxbhbom}),
(\ref{eq:sumsolr1}),  (\ref{eq:md0}), (\ref{eq:mdmk1}), (\ref{eq:hvh})
yield
\begin{equation}
\begin{split}
& b_X \approx .99536,\quad b_H\approx 4.64\;10^{-5},\quad
b_\Omega\approx .20289,\quad b_\Xi=0,\cr
& \hat b_X\approx 1.0046,\quad \hat b_H=1,\quad \hat b_\Omega \approx
.19486,\quad \hat b_\Xi=0.
\end{split}
\label{eq:bnum}
\end{equation}

\begin{equation}
\hat v_H \approx 142.973\; GeV.
\end{equation}

Together with (\ref{eq:bnum}), this yields
\begin{equation}
v_X \approx 142.641\,GeV,\quad  \hat v_X \approx 143.301\,GeV,\quad
 v_H \approx 973\,MeV,\quad
v_\Omega \approx 64.40\,GeV,\quad \hat v_\Omega \approx 63.11\,GeV.
\label{eq:vnum}
\end{equation}
One has also found
\begin{equation}
\begin{split}
& r_X=1,\quad r_H \approx .1002,\quad r_\Omega\approx 5.9187,\quad
r_\Xi=0,\cr
& \hat r_X = ?,\quad \hat r_H \approx .59215,\quad \hat r_\Omega\approx
5.8295,
\quad \hat r_\Xi=0.
\end{split}
\label{eq:rnum}
\end{equation}

The following relations have been obtained from fermionic considerations.
$m_d$
has been determined to be negative, in particular from the expression of
the mixing angle (\ref{eq:thetabis})
\begin{equation}
\tan^2\theta_d = \frac{m_u-m_d}{m_s-m_u}=\frac{m_u+|m_d|}{m_s-m_u} \approx
.076923 \Rightarrow \theta_c \approx .2705,
\end{equation}
in good agreement with the value extracted from (\ref{eq:tan2theta}) (both
leading however to a slightly too large value for the Cabibbo angle
(\ref{eq:numtetac})).
The quark masses
 $m_u\approx 2.5\,MeV, m_d \approx -5\,MeV, m_s \approx
100\,MeV, m_c \approx 1.2755\,GeV$ have been used as inputs.
In particular, from the values of $m_u,m_d$, the GMOR relation yielded the
value of $\mu_X^3$
\begin{equation}
m_u\approx 2.5\;MeV,\  m_d \approx -5\;MeV \quad \Rightarrow\quad
\frac{<\bar u u+\bar d d>}{\sqrt{2}}=\mu_X^3
\stackrel{(\ref{eq:GMOR2})}{\approx} \frac{\sqrt{2}f_\pi^2
m_{\pi^+}^2}{m_u+m_d} \approx
-0.09370\;GeV^3.
\end{equation}

From the values (\ref{eq:bnum}) of the $b$'s we deduce by (\ref{eq:mhh3})
and
(\ref{eq:mhiggs}) the masses of the Higgs bosons
\begin{equation}
\begin{array}{ccc}
m_{X^0}\approx 2.7897\,GeV& m_{\hat H^3}\approx 2.7962\,GeV &
m_{\hat
X^3}\approx 2.8026\,GeV\cr
&& \cr
 & \hskip -3mmm_{\Omega^0}\approx 1.2595\,GeV\hskip 15mm
m_{\hat\Omega^3}\approx 1.2343\,GeV &\cr
&& \cr
  & m_{H^0}\approx 19\,MeV &  \cr
&& \cr
 & \hskip -5mm m_{\Xi^0}\ small  \hskip 17mm  m_{\hat \Xi^3} \ small
\end{array}
\label{eq:higgsmasses}
\end{equation}

From the values (\ref{eq:rnum}) of the $r$'s and their definition
(\ref{eq:rdef})
 we deduce the fermionic VEV's
\begin{equation}
\begin{split}
<\bar c c> &= \frac{r_H + \hat r_H}{\sqrt{2}}\, \mu_X^3 \approx .47\,
\mu_X^3 <0,\cr
<\bar s s> &= \frac{r_H - \hat r_H}{\sqrt{2}}\, \mu_X^3 \approx
-.33\,\mu_X^3>0,\cr
<\bar u c>=<\bar c u>  &= \frac{r_\Omega + \hat r_\Omega}{2}\, \mu_X^3
\approx 5.87\, \mu_X^3 <0,\cr
<\bar d s>=<\bar s d>  &= \frac{r_\Omega - \hat r_\Omega}{2}\, \mu_X^3
\approx .045\,\mu_X^3 <0.
\end{split}
\label{eq:vevnum}
\end{equation}
We notice a  large non-diagonal $<\bar u c>=<\bar c u>$ condensation,
which could certainly not be predicted on perturbative grounds since it
starts occurring only at 2-loops. This result will get modified when
$\theta_u$ is no longer approximated by $0$.
\newline

As far as $\hat\mu_X^3=\frac{<\bar u u-\bar d d>}{\sqrt{2}}$ is
concerned, we cannot  find any acceptable (real) value compatible
with the orthogonality of $K^0$ to $\bar K^0$, and of $\pi^0$ to
$K^0+\bar K^0$ since these 2 conditions are associated with
(\ref{eq:KKperp2}) while one has found $\hat b_X >1$ from the ratio
$m_{K^0}^2/m_{D^0}^2$.

\subsection{The small value of $\boldsymbol{b_H}$ versus
$\boldsymbol{b_X\approx \hat b_X \approx \hat b_H=1}$}
\label{subsec:smallbh}

It would be desirable to have an analytic expression for $b_H$ which reflects its
small value. It unfortunately turns out that its complete analytical
expression in terms of $\delta$ and charged pseudoscalar masses,
despite its relative simplicity,
 has no trustable expansion in powers of $m_\pi$ at the chiral limit $m_\pi
\to 0$. The most meaningful expansion that we could get is by
writing $\delta = \delta_{(b_H=0)} +\epsilon\, m_\pi^2$, in which
$\delta_{(b_H=0)}$ is the value of $\delta$ at which $b_H$ vanishes, given
by the r.h.s. of the first line of (\ref{eq:delbhnul}) and $\epsilon = \frac{\delta
-\delta_{(b_H=0)}}{m_\pi^2}\stackrel{(\ref{eq:deltachir},\ref{eq:deltanum0})}{=}\omega^2 + \frac{m_{D_s}^2
-\delta_{(b_H=0)}}{m_\pi^2} \approx .0093$. One gets then
\begin{equation}
b_H \equiv \frac{v_H}{\hat v_H} \approx \epsilon \frac{m_\pi^2}{m_{D_s}^2},
\label{eq:bhexp}
\end{equation}
which exhibits a sort of ``see-saw'' mechanism between the two scales
$m_\pi$ and $m_{D_s}$, strengthened by the small factor $\epsilon$. Though
$\omega = {\cal O}(1)$,
$\epsilon$ gets $\ll 1$ because of two near cancellations: the first occurs
 between $m_{D_s}^2$ and $\delta_{(b_H=0)}$, and the second between $\omega^2$ and
$\frac{m_{D_s}^2 -\delta_{(b_H=0)}}{m_\pi^2}$. Eq.~(\ref{eq:bhexp})
yields $b_H \to 0$ at the chiral limit, which is compatible with the exact
result $b_H \stackrel{m_\pi\to 0}{\longrightarrow}
1-\frac{m_{D_s}^2}{\delta}$ since, by (\ref{eq:deltachir}), $\delta
\stackrel{m_\pi\to 0}{\longrightarrow} m_{D_s}^2$.

\begin{figure}
\centering

\includegraphics[width=6 cm, height=6 cm]{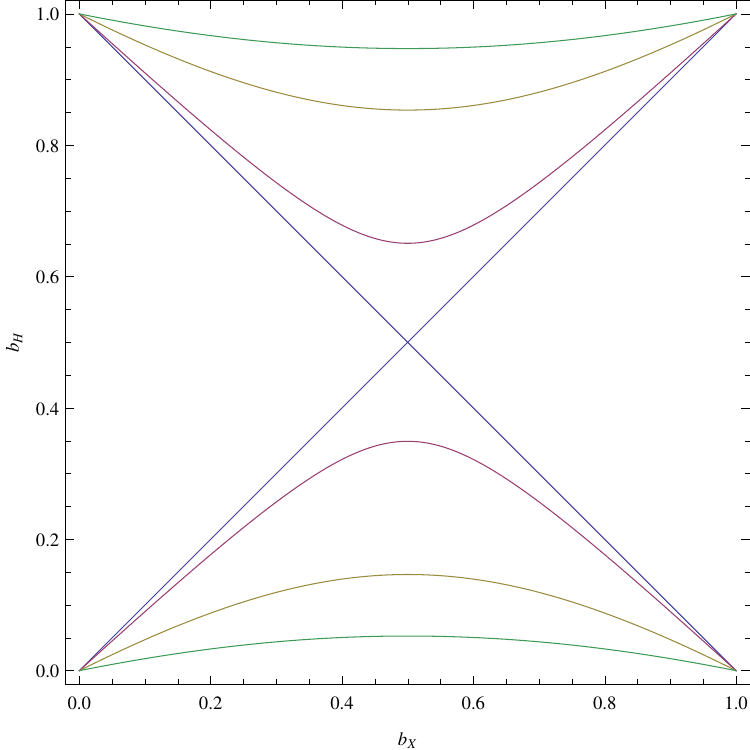}

\caption{$b_H$ at $\theta_u=0$ as a function of $b_X$ for
$1/r_H^2=1$ (blue), $1.1$ (purple), $2$ (yellow) and $5$ (green)}

\label{figure:seesaw}
\end{figure}

A ``see-saw'' mechanism between $b_H$ and $b_X$ can be traced back to the
orthogonality relations between charged pseudoscalars (\ref{eq:solnu2}),
which entail in particular 
 (see (\ref{eq:sumsolr1})).
\begin{equation}
b_X(1-b_X) = \frac{b_H(1-b_H)}{r_H^2},\quad r_H=\frac{<\bar
c c+\bar s s>}{<\bar u u+\bar d d>}.
\label{eq:see-saw1}
\end{equation}
Since $b_X
\approx 1 \Rightarrow v_X \approx \hat v_H$, the same phenomenon occurs
between $b_H$ and $\hat b_H$. It is due to the property $|r_H| <1$ which
one can understand by the fact that heavy quarks being more ``classical''
that light ones, they are expected to condense less. One can indeed easily
check that
\begin{equation}
|r_H| \ll 1 \Rightarrow b_H \ll b_X\ or\ b_H \gg b_X.
\label{eq:see-saw2}
\end{equation}
It is easily visualized on Fig.\ref{figure:seesaw} in which we plot $b_H$ as a function of
$b_H$ for $1/r_H^2=1$ (blue), $1/r_H^2=1.1$ (purple), $1/r_H^2=2$ (yellow) and
$1/r_H^2 = 5$ (green). Since we found $1/r_H^2 \approx 100$ (see
(\ref{eq:rnum})), the see-saw mechanism is  very effective. 

\subsection{Hierarchies}\label{subsec:hier}

The largest hierarchy among bosonic VEV's is $1/\sqrt{b_H}\approx 151$. It is
already much smaller that the one occurring for 1 generation, that we
recall (see (\ref{eq:hier1gen})) to be $\approx 2858$. The other ones,
which are given by $1/\sqrt{b}$'s are all  ${\cal O}(1)$.

Fermionic hierarchies are given by the $r$'s and do not exceed $r_\Omega
\approx 5-6$. Since it is hard in the 2-generation case to have a reliable
calculation of $<\bar d d>/<\bar u u>$, one cannot draw, yet, definitive
conclusions.

Seemingly, when more generations are added, the general trend is a
decrease of the hierarchies among VEV's. This can be easily understood
because large masses, like that of gauge bosons, get ``shared'' by several
VEV's, and so are the masses of heavy quarks. The number of Higgs bosons
growing like $2N^2$, one can naively expect that, for example, bosonic
 hierarchies are 8 times smaller for 2 generations than for 1 generation.
This is of course just an estimate,
and we have seen that the decrease is even stronger.

If this trend goes on, we can expect still smaller hierarchies for 3
generations, because large mass scales will be shared among 18 quadruplets.
The ``$\tan \beta$'' which was disquietingly huge for 1 generation could then
become replaced by a series of much smaller and more natural numbers.

\subsection{The light Higgs bosons $\boldsymbol{H^0}$,
  $\boldsymbol{\Xi^0}$ and $\boldsymbol{\hat\Xi^3}$}\label{subsec:lights}

Like for 1 generation,  a light Higgs $H^0$ arises in the
Higgs quadruplet that is  parity transformed of the one that
contains the 3 Goldstones of
the broken $SU(2)_L$. Its mass lies well below that of the lightest
pseudoscalars such that, in particular, it cannot decay into 2 pions.

As far as $\Xi^0$ and $\hat\Xi^3$ are concerned, we could not calculate
explicitly their masses, but there are reasonable argument that they should
not vanish nor become large. They should not vanish because they
do not correspond to any true Goldstone boson. Also,
because the hypothesis $<\bar c u>=<\bar uc>$ which led to their classical
vanishing for 2 generations should no longer be true with 3 generations. To
have an intuitive idea of this, at least at the perturbative level, it is
enough to realize that such condensates occur at 2-loops
with an intermediate gauge
boson line, and will accordingly be sensitive to the $CP$ violating
phase(s) that cannot be avoided in this case. Their effects are expected to
be small.

Next, even for 2 generations, one expects quantum corrections
to their classical masses through fermion loops as follows.
Let us note $(\bar q_i q_j)^*$ the fermion loop with quarks $\bar q_i$ and
$q_j$. From the Yukawa Lagrangian, one gets, at second order,
 the couplings
$\delta_{\Xi\hat\Xi}^2 \Big(
\Xi^0 \frac{\hat v_\Xi}{\sqrt{2}\hat\mu_\Xi^3}\frac14
(\bar u c-\bar c u-\bar d s+\bar s d)^\ast 
\frac{\hat v_\Xi}{\sqrt{2}\hat\mu_\Xi^3} \Xi^0$
$+\hat\Xi^3 \frac{v_\Xi}{\sqrt{2}\mu_\Xi^3}\frac14
(\bar u c-\bar c u+\bar d s-\bar s d)^\ast
 \frac{ v_\Xi}{\sqrt{2}\mu_\Xi^3} \hat\Xi^3 \Big)$.
In section \ref{section:4bs}, we have emphasized that, at $b_\Xi=0$,
 the normalization $v_\Xi/\sqrt{2}\mu_\Xi^3$ of the $\Xi$ quadruplet is a
constant, and so is the one of the $\Hat\Xi$ quadruplet. They are the same
as for the $X$ quadruplet, 
$1/\nu_\Xi^2 = 1/\hat\nu_\Xi^2 =1/\nu_X^2 = \hat v_H
\sqrt{b_X(1-b_X)}/\sqrt{2}\mu_X^3$.
One gets accordingly the couplings
$\frac{\delta_{\Xi\hat\Xi}^2}{4\nu_X^4}\Big(\Xi^0
(\bar uc-\bar c u-\bar d s+\bar s d)^\ast \Xi^0$
$+ \hat\Xi^3 (\bar u c-\bar c u+\bar d s-\bar s d)^\ast \hat\Xi^3\Big)$.
The same $(uc)$ and $(ds)$ fermion loops 
occur for $\Xi^0$ and $\hat\Xi^3$; they are accordingly expected to yield
the same mass terms for $\Xi^0$ and $\hat\Xi^3$.
This is of course only an intuitive perturbative argument. In particular
the fermion loops are quadratically divergent and need to be regularized.
One can also notice that the argument is in some way iterative because the
fermion loops can themselves be ``saturated'' by virtual $\hat\Xi^3$ and
$\Xi^0$ lines.

\section{The masses of $\boldsymbol{\pi^0}$, $\boldsymbol{K^0}$ and
$\boldsymbol{D^0}$; tracing why $\boldsymbol{\hat b_X = (\hat v_X/\hat
v_H)^2 >1}$ is needed}
\label{section:neutmass}

The physical value  $\hat b_X>1$ that we determined
from the spectrum of pseudoscalar mesons is problematic because it is in
contradiction with the 2 identical relations (\ref{eq:solnu31}) and
(\ref{eq:KKperp})
that control the orthogonality of $\pi^0$ to $K^0 + \bar K^0$ and the
absence of non-diagonal $K^0-\bar K^0$ terms. We now look deeper for the
origin of this problem.

In addition to (\ref{eq:mk02}) and (\ref{eq:md02}) for the masses of
neutral kaon and $D$ mesons, one gets for the neutral pion (we recall that
its interpolating field we chose proportional to $(\bar u\gamma_5 u -
\bar d \gamma_5 d$)
\begin{equation}
m_{\pi^0}^2= \displaystyle\frac{2\delta (1+s_d^2)}
{ \displaystyle\frac{(1+c_d^2)^2}{1-b_X} +
2\displaystyle\frac{(1-c_d^2)^2}{1-\hat b_X}
+\displaystyle\frac{s_d^4}{1-b_H}
+\displaystyle\frac{s_d^4}{b_X(1-b_X)}\displaystyle\frac{1}{\hat r_H^2}
+2s_d^2 c_d^2
\left(\displaystyle\frac{1}{1-b_\Omega}+\displaystyle\frac{1}{1-\hat
b_\Omega}\right)
}.
\end{equation}

At the values of the parameters that we determined (see section
\ref{section:solution}), one gets
\begin{equation}
m_{D^0} \approx 1.865\,GeV, \quad m_{K^0}\approx 497.61\,MeV,\quad
m_{\pi^0}\approx 139.38\,MeV,
\end{equation}
which is quite satisfying and shows in particular that these
values of the parameters can not only nicely fit $m_{K^0}^2/m_{D^0}^2$ as
we did in subsection \ref{subsec:neutralmass},
 but also the absolute masses of the neutral mesons.

\begin{figure}
\centering

\includegraphics[width=10 cm, height=6 cm]{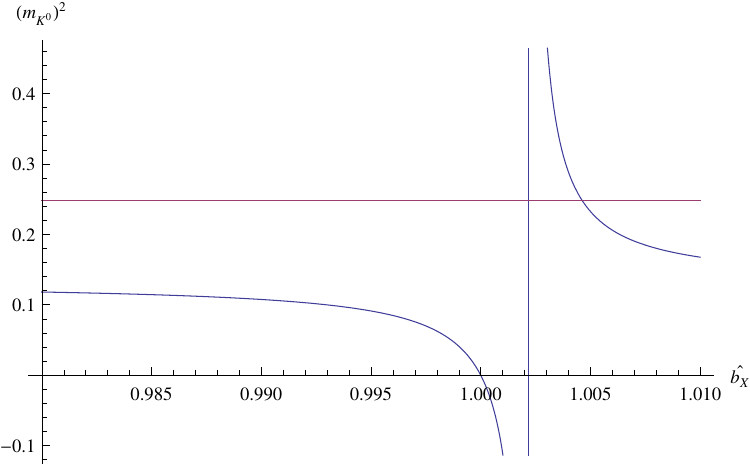}

\caption{$m_{K^0}^2$ as a function of $\hat b_X$ at $\theta_u=0$.
The other parameters are
fixed to their determined values. The horizontal line is at the physical
value of $m_{K^0}^2$}

\label{figure:mk0hbx}
\end{figure}

Then the question arises: why are we induced to the troublesome $\hat b_X
>1$?
The numerical analysis shows that, while the value of $m_{D^0}$ is quite
stable for $\hat b_X$ close to $1$, it is not the case for $m_{K^0}$. For
$\hat b_X<1$ one gets too small a value of $m_{K^0}$, which, however, turns
out to have a pole at $\hat b_X$ slightly above $1$ (see
Fig.\ref{figure:mk0hbx}).
 The suitable mass can
only be recovered above the pole, which is of course a very artificial and
ad-hoc solution.

%
To give an idea of by how much $m_{K^0}$ is found too small, we give
the values of the same masses at $\hat b_X = .95$ : $m_{D^0} \approx
1.865\,
GeV, \quad m_{K^0}\approx 354\,MeV,\quad m_{\pi^0}\approx 139.18\,MeV$.

The conclusion is therefore that the masses of the charged mesons $\pi^\pm,
K^\pm, D^\pm, D_s^\pm$, and those of $\pi^0$ and $D^0$ can
be accounted for very ``naturally'', but, then, the mass of the $K^0$ meson
is
short by $140\,MeV$ (nearly the pion mass), unless one goes to $\hat b_X
>1$. This unnatural solution raises then another issue connected to the
orthogonality of $\pi^0$ to $K^0 + \bar K^0$. Indeed, (\ref{eq:KKperp})
yields
then a complex unrealistic value for $\hat r_X \equiv <\bar u u-\bar d
d>/<\bar u
u+\bar d d>$.

\subsection{Can one restore $\boldsymbol{\hat b_X<1}$ ?}
\label{subsec:need3gen}

One may  question the
way we have defined the $\pi^0$ interpolating field and/or the necessity
to cancel $K^0\bar K^0$  non-diagonal mass terms (this
is akin to giving up  the neutral kaon mass eigenstates as
 $K^0 \pm \bar K^0$). The mixing among neutral mesons may also be more
subtle than usually thought of, and the whole set of interpolating fields
that we have chosen be much too naive.

There is most probably no need to go to such extremes because, as we shall
see in chapter \ref{chapt:2gentetau},
the situation largely improves  when one allows
for $\theta_u \not=0$.
This goes however with a value $\hat r_H \geq .945$ larger than the one
that we found here, which means that
$<\bar c c-\bar s s> \simeq <\bar u u-\bar d d>$.
So, large condensates of heavy quarks are required.
A hint in this direction appears in the expansions (\ref{eq:hbxexp}) and
(\ref{eq:bxexp}). Since (\ref{eq:bxexp}) depends on the sole parameter
$\delta$,  $b_X \leq 1$ looks a robust property. This is however not
quite the case for $\hat b_X$ because its expansion (\ref{eq:hbxexp}) also
depends on the ratio $\hat r_H$ of fermionic VEV's and as soon as $|\hat
r_H| \geq 1$ (see also footnote \ref{foot:polehbx})
 $\hat b_X$ becomes $\leq 1$ as desired.
$|\hat r_H| \geq 1$ means, for 2 generations, $|<\bar c c-\bar s s>| \geq
|<\bar u u+\bar d d>|$, which reasonably cannot be realized.
However, for more
generations, it may translate into the necessity for the condensate of
the heaviest quark to become larger than that of light quarks,
for example $|<\bar t t>| \geq |<\bar u u+ \bar d d>|$.
One must accordingly keep in mind that, to cure the (small) remaining
discrepancy that may subsist for the masses of neutral pseudoscalars, one
may have to invoke large condensates for heavy quarks, which can only
eventually be achieved for more than 2 generations.

The mechanism likely to generate such a large condensation remains of course to
be uncovered. However, as we will show in the subsequent paper \cite{Machet5},
some scalar(s)  turn out to be strongly coupled to
quarks, which  can  induce  the formation of a massive
 bound state (see subsection \ref{subsec:normalization}). The resulting
 breaking of the chiral symmetry  can then go along with quark
condensation in the appropriate channel. At this point, these are of course
only conjectures.

\section{Problems with the leptonic decays $\boldsymbol{\pi^+ \to \ell^+ \nu_\ell}$ and
$\boldsymbol{K^+ \to \ell^+ \nu_\ell}$ at $\boldsymbol{\theta_u=0}$}
\label{section:piKlept1}

At this point, the situation looks globally satisfying, but for a few
points that need to be clarified,  concerning  the orthogonality of a
few mesons, the mass of the $K^0$  and the resulting $\hat b_X >1$,
and a paradoxical  fermionic mixing angle. 
However this is illusory and we show below that the approximation
$\theta_u=0$  is completely at a loss to explain the leptonic decays of
$\pi^+$ and $K^+$, which makes  the situation  worse than for 1
generation. Leptonic decays are the smoking gun of something very serious
that we have missed.

\subsection{No mixing at all : $\boldsymbol{\theta_d=0=\theta_u}$}

\subsubsection{$\bullet\ \boldsymbol{\pi^+ \to \ell^+\, \nu_\ell}$
 decay}

$\pi^+$ occurs only in the $X$ quadruplet, and one falls back on the same
situation as in the case of 1 generation only (see section
\ref{section:1genlept}):
the leptonic decays of charged pions are suitably described (of course up
to the Cabibbo factor $\cos\theta_c$).

\subsubsection{$\bullet\ \boldsymbol{K^+ \to \ell^+\, \nu_\ell}$
 decay}

We refer the reader to eq.~(\ref{eq:amKlept}) below for $\theta_d=0$. While the
amplitude is expected to vanish at $\theta_d=0$, one finds it proportional
to $\sqrt{b_\Omega(1-b_\Omega)}$. It is therefore unsuitably described,
unless $b_\Omega \to 0$. We shall see that this only happens when both
$\theta_d$ and $\theta_u$ are taken not vanishing.

\subsection{$\boldsymbol{\theta_d \not=0}$ and  setting
$\boldsymbol{\theta_u=0}$ : a lack of
cancellation in leptonic decay amplitudes}

\subsubsection{$\bullet\ \boldsymbol{\pi^+ \to \ell^+\, \nu_\ell}$
decay}

As soon as there is mixing, $\pi^+ \propto \bar u_m\gamma_5 d_m$ occurs in
the 4 quadruplets $X,H, \Omega, \Xi$ such that
\begin{equation}
\pi^+ = \beta_X X^+ + \beta_H H^+ + \beta_\Omega \Omega^+ + \beta_\Xi
\Xi^+.
\label{eq:expi}
\end{equation}
One has accordingly for the looked for matrix element ${\cal M}_\pi$
\begin{equation}
\begin{split}
{\cal M}_{\pi} &\equiv\langle \ell\nu_\ell | \pi^+\rangle =
\beta_X \langle \ell\nu_\ell | X^+\rangle
+ \beta_H \langle \ell\nu_\ell | H^+\rangle
+ \beta_\Omega \langle \ell\nu_\ell | \Omega^+\rangle
+ \beta_\Xi \langle \ell\nu_\ell | \Xi^+\rangle.
\end{split}
\label{eq:pilept4}
\end{equation}
The leptonic decay of charged pions receive now a priori contributions from
the 4 quadruplets $X, H, \Omega, \Xi$. In the case of no mixing we noticed
that the amplitude was controlled by $v_X/a_X \equiv v_X \beta_X$, and it is
now controlled by the sum 
\begin{equation}
v_X\beta_X + v_H\beta_H + v_\Omega \beta_\Omega +
v_\Xi \beta_\Xi.
\label{eq:pilept5}
\end{equation}
One needs therefore to know $\beta_X, \beta_H, \beta_\Omega, \beta_\Xi$. To
that purpose we write
\begin{equation}
\begin{split}
X^+ &= a_X \pi^+ + f_X K^+ + c_X D^+ + d_X D_s^+,\cr
H^+ &= a_H \pi^+ + f_H K^+ + c_H D^+ + d_H D_s^+, \cr
\Omega^+ &= a_\Omega \pi^+ + f_\Omega K^+ + c_\Omega D^+ + d_\Omega D_s^+,
\cr
\Xi^+ &= a_\Xi \pi^+ + f_\Xi K^+ + c_\Xi D^+ + d_\Xi D_s^+.
\end{split}
\end{equation}
Using PCAC and GMOR for $\pi^+$, PCAC for $K^+, D^+, D_s^+$,  one gets
\begin{equation}
\begin{split}
& a_X= c_d\frac{v_X}{f_\pi}, \quad
a_H = 0,\quad
a_\Omega= -a_X
\frac{s_{d}}{c_d}\sqrt{\frac{b_\Omega}{b_X}}\frac{1}{r_\Omega}
\frac{1}{\sqrt{2}},\quad
a_\Xi= -a_X  \frac{s_d}{c_d}\sqrt{\frac{b_\Xi}{b_X}}\frac{1}{r_\Xi}
\frac{1}{\sqrt{2}},\cr
& f_X= a_X \frac{s_d}{c_d}
\underbrace{\frac{f_K}{f_\pi}\frac{m_K^2}{m_\pi^2}\frac{m_u+m_d}{m_u+m_s}}_{F_K},\quad
 f_H= 0,\cr
 & \hskip 1cm
f_\Omega= a_X
\sqrt{\frac{b_\Omega}{b_X}}\frac{1}{r_\Omega}
\frac{1}{\sqrt{2}}
\frac{f_K}{f_\pi}\frac{m_K^2}{m_\pi^2}\frac{m_u+m_d}{m_u+m_s},\quad
f_\Xi= a_X \sqrt{\frac{b_\Xi}{b_X}}\frac{1}{r_\Xi}
\frac{1}{\sqrt{2}}
\frac{f_K}{f_\pi}\frac{m_K^2}{m_\pi^2}\frac{m_u+m_d}{m_u+m_s},\cr
& c_X= 0,\quad
 c_H= -a_X \frac{s_d}{c_d}\sqrt{\frac{b_H}{b_X}}\frac{1}{r_H}
\frac{f_D}{f_\pi}\frac{m_D^2}{m_\pi^2}\frac{m_u+m_d}{m_c+m_d},\cr
 & \hskip 1cm
c_\Omega= a_X
\sqrt{\frac{b_\Omega}{b_X}}\frac{1}{r_\Omega}
\frac{1}{\sqrt{2}}
\frac{f_D}{f_\pi}\frac{m_D^2}{m_\pi^2}\frac{m_u+m_d}{m_c+m_d},\quad
c_\Xi= -a_X \sqrt{\frac{b_\Xi}{b_X}}\frac{1}{r_\Xi}
\frac{1}{\sqrt{2}}
\frac{f_D}{f_\pi}\frac{m_D^2}{m_\pi^2}\frac{m_u+m_d}{m_c+m_d},\cr
& d_X= 0,\quad
 d_H= a_X \sqrt{\frac{b_H}{b_X}}\frac{1}{r_H}
\frac{f_{D_s}}{f_\pi}\frac{m_{D_s}^2}{m_\pi^2}\frac{m_u+m_d}{m_c+m_s},\cr
 & \hskip 1cm
d_\Omega= a_X
\frac{s_d}{c_d}\sqrt{\frac{b_\Omega}{b_X}}\frac{1}{r_\Omega}
\frac{1}{\sqrt{2}}
\frac{f_{D_s}}{f_\pi}\frac{m_{D_s}^2}{m_\pi^2}\frac{m_u+m_d}{m_c+m_s},\quad
d_\Xi= -a_X \frac{s_d}{c_d}\sqrt{\frac{b_\Xi}{b_X}}\frac{1}{r_\Xi}
\frac{1}{\sqrt{2}}
\frac{f_{D_s}}{f_\pi}\frac{m_{D_s}^2}{m_\pi^2}\frac{m_u+m_d}{m_c+m_s}.
\end{split}
\label{eq:afcd1}
\end{equation}
Because of (\ref{eq:sumsolr1}),
$\sqrt{\frac{b_H}{b_X}}\frac{1}{r_H}= \sqrt{\frac{1-b_H}{1-b_X}}$,\quad
$\sqrt{\frac{b_\Omega}{b_X}}\frac{1}{r_\Omega}=
\sqrt{\frac{1-b_\Omega}{1-b_X}}$,\quad
$\sqrt{\frac{b_\Xi}{b_X}}\frac{1}{r_\Xi}= \sqrt{\frac{1-b_\Xi}{1-b_X}}
\stackrel{b_\Xi=0}{=}\sqrt{\frac{1}{1-b_X}}$.

Because of (\ref{eq:expi}) one must have
\begin{equation}
\begin{split}
& \beta_X a_X + \beta_H a_H + \beta_\Omega a_\Omega + \beta_\Xi a_\Xi=1,\cr
& \beta_X f_X + \beta_H f_H + \beta_\Omega f_\Omega + \beta_\Xi f_\Xi=0,\cr
& \beta_X c_X + \beta_H c_H + \beta_\Omega c_\Omega + \beta_\Xi c_\Xi=0,\cr
& \beta_X d_X + \beta_H d_H + \beta_\Omega d_\Omega + \beta_\Xi d_\Xi=0.
\end{split}
\label{eq:betaeqs}
\end{equation}
The solution of (\ref{eq:betaeqs}) and (\ref{eq:afcd1}) is
\begin{equation}
\begin{split}
& \beta_X=\frac{c_d^2}{a_X}=c_d\frac{f_\pi}{v_X},\quad
\beta_H=0,\cr
& \beta_\Omega=-\frac{1}{\sqrt{2}}c_ds_d
\sqrt{\frac{1-b_\Omega}{1-b_X}}\frac{1}{a_X}=
-\frac{1}{\sqrt{2}}s_d
\sqrt{\frac{1-b_\Omega}{1-b_X}}\frac{f_\pi}{v_X}, \cr
& \beta_\Xi =
-\frac{1}{\sqrt{2}}c_ds_d\sqrt{\frac{1-b_\Xi}{1-b_X}}\frac{1}{a_X}
=-\frac{1}{\sqrt{2}}s_d
\sqrt{\frac{1-b_\Xi}{1-b_X}}\frac{f_\pi}{v_X}.
\end{split}
\label{eq:betasol1}
\end{equation}
According to (\ref{eq:pilept5}) 
 the leptonic decay amplitude of $\pi^+$ is accordingly proportional to
\begin{equation}
f_\pi \; c_d\left(1 -\underbrace{\frac{s_d}{\sqrt{2}}
\sqrt{\frac{b_\Omega(1-b_\Omega)}{b_X(1-b_X)}}}_{\approx 1}\right).
\label{eq:ampilept}
\end{equation}
Eq.~(\ref{eq:nulbxi}) has been used to get rid of the  contribution of $\Xi$.
The numerical estimates uses the value of $b_\Omega$ which we had found in the
case $\theta_u=0$ (see section \ref{section:solution}).

The nice agreement that took place with no
mixing at all disappears when we turn on the $d-s$ mixing only.
Another contribution is needed to cancel the one of $\Omega$, which does
not exist presently.

\subsubsection{$\bullet\ \boldsymbol{K^+ \to \ell^+\, \nu_\ell}$ decay}

The equivalent of (\ref{eq:expi}) is now
\begin{equation}
K^+ = \zeta_X X^+ + \zeta_H H^+ + \zeta_\Omega \Omega^+ + \zeta_\Xi
\Xi^+;
\label{eq:exK}
\end{equation}
that of (\ref{eq:pilept4}) is
\begin{equation}
\begin{split}
{\cal M}_{K} &\equiv\langle \ell\nu_\ell | K^+\rangle =
\zeta_X \langle \ell\nu_\ell | X^+\rangle
+ \zeta_H \langle \ell\nu_\ell | H^+\rangle
+ \zeta_\Omega \langle \ell\nu_\ell | \Omega^+\rangle
+ \zeta_\Xi \langle \ell\nu_\ell | \Xi^+\rangle,
\end{split}
\label{eq:Klept4}
\end{equation}
which is now proportional to (the other terms are trivial factors)
\begin{equation}
v_X \zeta_X + v_H \zeta_H + v_\Omega \zeta_\Omega + v_\Xi \zeta_\Xi.
\label{eq:Klept5}
\end{equation}
The system (\ref{eq:betaeqs}) is replaced with
\begin{equation}
\begin{split}
& \zeta_X a_X + \zeta_H a_H + \zeta_\Omega a_\Omega + \zeta_\Xi a_\Xi=0,\cr
& \zeta_X f_X + \zeta_H f_H + \zeta_\Omega f_\Omega + \zeta_\Xi f_\Xi=1,\cr
& \zeta_X c_X + \zeta_H c_H + \zeta_\Omega c_\Omega + \zeta_\Xi c_\Xi=0,\cr
& \zeta_X d_X + \zeta_H d_H + \zeta_\Omega d_\Omega + \zeta_\Xi d_\Xi=0.
\end{split}
\label{eq:zetaeqs}
\end{equation}
Combined with (\ref{eq:afcd1}), it leads to
\begin{equation}
\begin{split}
& \zeta_X=s_dc_d\frac{1}{a_X}\frac{1}{F_K}
=s_d\frac{1}{F_K}\frac{f_\pi}{v_X},\quad
\zeta_H=0,\cr
& \zeta_\Omega=\frac{1}{\sqrt{2}}c_d^2
\sqrt{\frac{1-b_\Omega}{1-b_X}}\frac{1}{a_X}\frac{1}{F_K}
=\frac{1}{\sqrt{2}}c_d\sqrt{\frac{1-b_\Omega}{1-b_X}}
\frac{1}{F_K}\frac{f_\pi}{v_X},\cr
& \zeta_\Xi=\frac{1}{\sqrt{2}}c_d^2
\sqrt{\frac{1-b_\Xi}{1-b_X}}\frac{1}{a_X}\frac{1}{F_K}
=\frac{1}{\sqrt{2}}c_d\sqrt{\frac{1-b_\Xi}{1-b_X}}
\frac{1}{F_K}\frac{f_\pi}{v_X},\cr
& with\ F_K =\frac{f_K}{f_\pi}\frac{m_K^2}{m_\pi^2}\frac{m_u+m_d}{m_u+m_s}.
\end{split}
\label{eq:zetasol1}
\end{equation}
Like for pion leptonic decays, $b_\Xi=0$ ensures the vanishing of the $\Xi$
contribution, such that, by (\ref{eq:Klept5}), the amplitude gets controlled
by
\begin{equation}
v_X \zeta_X + v_\Omega \zeta_\Omega =
\frac{f_\pi}{F_K}\; s_d \left(
1 +\frac{1}{\sqrt{2}} \frac{c_d}{s_d} \sqrt{\frac{b_\Omega(1-b_\Omega)}{b_X(1-b_X)}}
\right).
\label{eq:amKlept}
\end{equation}
Like for charged pions, the fair agreement that would be obtained from the
contribution of $X^+$ alone gets totally spoiled by that from $\Omega^+$.

Note: the parameter $f_\pi/F_K$ that occurs in (\ref{eq:amKlept}) also
writes
\begin{equation}
\frac{f_\pi}{F_K} \approx f_K \frac{<\bar u u+ \bar d d>}{<\bar u u+ \bar s
s>}.
\label{eq:fpiFK}
\end{equation}

\section{Conclusion for the case $\boldsymbol{\theta_d\not=0, \theta_u =0}$}
\label{section:endtetaunul}

The approximation of setting $\theta_u=0$ has given a fairly good estimate
of $\theta_d$, and a matching with the physics of pseudoscalar mesons much
better than could have been anticipated. However,  dark
points subsist than we enumerate below.

* we find too large  values of $b_\Omega, \hat b_\Omega$ and of the non-diagonal
quark condensates $<\bar u c>, <\bar c u>$;

* the mass of the neutral kaon is found too small by $140\,MeV$ unless one
goes to $\hat b_X >1$; however, this value conflicts with strongly
desired orthogonality relations; 

* the value of the $d-s$ mixing angle that we find from bosonic
argumentation practically coincides with a pole of the ratio of fermionic mass
terms $\frac{\mu_{ds}}{\mu_d-\mu_s}$, which would corresponds to a maximal
``fermionic'' mixing; this paradox could be avoided if $b_\Omega$ and $\hat b_\Omega$
were very small, which does not occur at $\theta_u=0$;

* leptonic decays of $\pi^+$ and $K^+$ are totally off, unless one also
switches off $\theta_d$ or if $b_\Omega, \hat b_\Omega \to 0$.
%


\chapter{$\boldsymbol{N=2}$ generations with $\boldsymbol{\theta_d \not= 0, \theta_u\not =0}$}
\label{chapt:2gentetau}

We show in this chapter that the situation largely improves by taking $\theta_u
\not=0$ and proceed as follows.\newline
* $\theta_d-\theta_u=\theta_c$ will be taken to its experimental value
(\ref{eq:numtetac});\newline
* $\theta_u$ will be taken to satisfy (\ref{eq:cab1});\newline
* $b_X, b_H, b_\Omega$ are expressed according to (\ref{eq:diag+2}) as functions
of $\theta_u, \theta_d$ and $\delta$;\newline
* that leptonic decays of $\pi^+$ and $K^+$ are suitably
described provides a relation between $\theta_u$, $b_H$ and $b_\Omega$ which
determines the value of $\delta$;\newline
* the  masses of $\pi^0$, $K^0$ and $D^0$ are then used to determine
$\hat b_X$, $\hat b_\Omega$ and $\hat r_H$.

 The masses of neutral pseudoscalar mesons are now
fairly well described with $\hat b_X <1$. $b_\Omega$
(and probably $\hat b_\Omega$, too) become very small, which presumably solves
the paradox of bosonic versus fermionic mixing. At the opposite, $b_H$ departs
from its previously very small value. Leptonic decays of charged pions and
kaons are correctly accounted for.

\section{An estimate of $\boldsymbol{\theta_u}$}
\label{section:valtetau}

\begin{figure}
\centering
\medskip
\includegraphics[width=10 cm, height=6 cm]{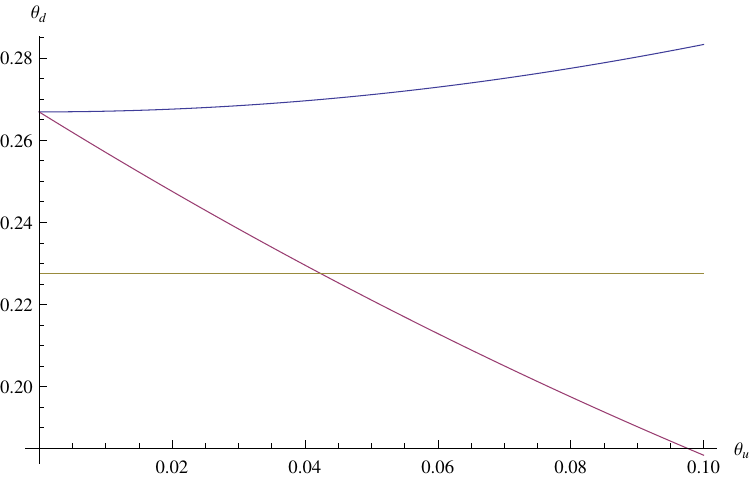}

\caption{$\theta_d$ (blue) and $\theta_d-\theta_u$ (purple) as
functions of $\theta_u$, compared with the experimental
 Cabibbo value $\theta_c \approx .22759$ (yellow)}

\label{figure:thetadthetau}
\end{figure}

On Fig.\ref{figure:thetadthetau} we plot in blue $\theta_d$ as a function of $\theta_u$ as given by
(\ref{eq:cab1}),
and we also plot in purple the corresponding value of
$\theta_d-\theta_u$ that we identify with the ``Cabibbo angle''
of the GSW model.
This last curve crosses the physical value (\ref{eq:numtetac}) drawn
in yellow at 
\begin{equation}
\boxed{\theta_u \approx .04225 \ll \theta_d}
\label{eq:tetaunum}
\end{equation}

\begin{figure}
\centering
\medskip
\includegraphics[width=10 cm, height=6 cm]{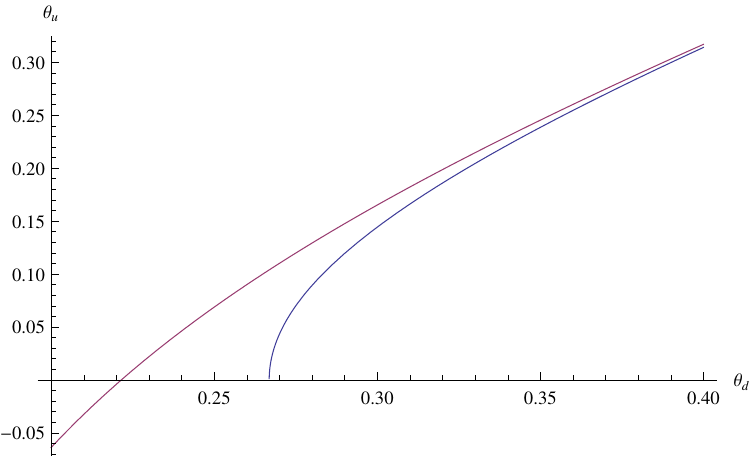}

\caption{$\theta_u$ as a function of $\theta_d$ (blue curve); its
expansion at 2nd order in $a$, parameter given in (\ref{eq:tetauan})
(purple curve)}

\label{figure:exptetau}
\end{figure}

A peculiarity of $\theta_u$ as given by (\ref{eq:cab1}) as a function of
$\theta_d$
\begin{equation}
\theta_u =\arccos\left(\frac{1}{\sqrt{2}}\sqrt{
 1 + \frac{1+a}{1-a} \cos{2\theta_d}}\right), \quad
a=\frac
 {\frac{1}{m_{K^\pm}^2}-\frac{1}{m_{D^\pm}^2}}
 {\frac{1}{m_{\pi^\pm}^2}-\frac{1}{m_{D_s^\pm}^2}}.
\label{eq:tetauan}
\end{equation}
is that:\newline
*\ it is, as shown on Fig.~\ref{figure:exptetau} (blue curve),
 a rather rapidly varying function of $\theta_d$;\newline
*\  it has no reliable expansion at $\theta_d \to 0$ (since (\ref{eq:cab1}) has no
solution);
if one brutally perform a formal expansion of $\theta_u$ in powers of
$\theta_d$ one gets
\begin{equation}
 \theta_u \stackrel{\theta_d\to 0}{\longrightarrow}
\arccos{\frac{1}{\sqrt{1-a}}} + \frac{1 + a}{2 \sqrt{1 - a}
\sqrt{-\displaystyle\frac{a}{1 - a}}}\;\theta_d^2 +\ldots
\end{equation}
which is clearly meaningless  since $a<1$ yields a second term that is
imaginary;\newline
*\ its expansion in powers of $a \simeq \frac{m_\pi^2}{m_K^2}\ll1$
 starts with
\begin{equation}
\begin{split}
& \theta_u \stackrel{a\to 0}{\longrightarrow}
\theta_d -\frac{a}{\tan 2\theta_d} -\frac{a^2}{\tan 2\theta_d}\frac{1}{\sin^2
2\theta_d} + \ldots
\end{split}
\label{eq:exptetau}
\end{equation}
in which the limit $\theta_d\to 0$ must coincide with $a\to0$, that is
$m_\pi\to 0$.

The purple curve on Fig.\ref{figure:exptetau} corresponds to the
first 2 terms of the expansion (\ref{eq:exptetau}). It shows that there are
quantities, in particular the ones that depend on $\theta_u$,
 for which $\theta_d$ cannot be straightforwardly
considered as a small number.

\section{Leptonic decays of pions and kaons}
\label{section:newpiKlept}

Since $\theta_u \to 0$ has been seen in the previous chapter to
lead to erroneous leptonic decays of charged pions and kaons, we start our
new investigations with these decays.

For $\theta_u\not=0$ and $\theta_d \not=0$, eqs.~(\ref{eq:afcd1}) are
replaced with
\begin{equation}
\begin{split}
& a_X= c_uc_d\frac{v_X}{f_\pi}, \quad
a_H = a_X \frac{s_us_d}{c_uc_d} \sqrt{\frac{b_H}{b_X}}\frac{1}{r_H},\quad
a_\Omega= a_X (-)
\frac{s_{u+d}}{c_uc_d}\sqrt{\frac{b_\Omega}{b_X}}\frac{1}{r_\Omega}
\frac{1}{\sqrt{2}},\quad
a_\Xi= a_X  \frac{s_{u-d}}{c_uc_d}\sqrt{\frac{b_\Xi}{b_X}}\frac{1}{r_\Xi}
\frac{1}{\sqrt{2}},\cr
& f_X= a_X \frac{s_d}{c_d}
\underbrace{\frac{f_K}{f_\pi}\frac{m_K^2}{m_\pi^2}\frac{m_u+m_d}{m_u+m_s}}_{F_K},\quad
 f_H= a_X (-)\frac{s_u}{c_u}\sqrt{\frac{b_H}{b_X}}\frac{1}{r_H}
\frac{f_K}{f_\pi}\frac{m_K^2}{m_\pi^2}\frac{m_u+m_d}{m_u+m_s},\cr
 & \hskip 1cm
f_\Omega= a_X
\frac{c_{u+d}}{c_uc_d}\sqrt{\frac{b_\Omega}{b_X}}\frac{1}{r_\Omega}
\frac{1}{\sqrt{2}}
\frac{f_K}{f_\pi}\frac{m_K^2}{m_\pi^2}\frac{m_u+m_d}{m_u+m_s},\quad
f_\Xi= a_X \frac{c_{u-d}}{c_uc_d}\sqrt{\frac{b_\Xi}{b_X}}\frac{1}{r_\Xi}
\frac{1}{\sqrt{2}}
\frac{f_K}{f_\pi}\frac{m_K^2}{m_\pi^2}\frac{m_u+m_d}{m_u+m_s},\cr
& c_X= a_X \frac{s_u}{c_u}
\frac{f_D}{f_\pi}\frac{m_D^2}{m_\pi^2}\frac{m_u+m_d}{m_c+m_d},\quad
 c_H= a_X (-)\frac{s_d}{c_d}\sqrt{\frac{b_H}{b_X}}\frac{1}{r_H}
\frac{f_D}{f_\pi}\frac{m_D^2}{m_\pi^2}\frac{m_u+m_d}{m_c+m_d},\cr
 & \hskip 1cm
c_\Omega= a_X
\frac{c_{u+d}}{c_uc_d}\sqrt{\frac{b_\Omega}{b_X}}\frac{1}{r_\Omega}
\frac{1}{\sqrt{2}}
\frac{f_D}{f_\pi}\frac{m_D^2}{m_\pi^2}\frac{m_u+m_d}{m_c+m_d},\quad
c_\Xi= a_X (-)\frac{c_{u-d}}{c_uc_d}\sqrt{\frac{b_\Xi}{b_X}}\frac{1}{r_\Xi}
\frac{1}{\sqrt{2}}
\frac{f_D}{f_\pi}\frac{m_D^2}{m_\pi^2}\frac{m_u+m_d}{m_c+m_d},\cr
& d_X= a_X \frac{s_us_d}{c_uc_d}
\frac{f_{D_s}}{f_\pi}\frac{m_{D_s}^2}{m_\pi^2}\frac{m_u+m_d}{m_c+m_s},\quad
 d_H= a_X \sqrt{\frac{b_H}{b_X}}\frac{1}{r_H}
\frac{f_{D_s}}{f_\pi}\frac{m_{D_s}^2}{m_\pi^2}\frac{m_u+m_d}{m_c+m_s},\cr
 & \hskip 1cm
d_\Omega= a_X
\frac{s_{u+d}}{c_uc_d}\sqrt{\frac{b_\Omega}{b_X}}\frac{1}{r_\Omega}
\frac{1}{\sqrt{2}}
\frac{f_{D_s}}{f_\pi}\frac{m_{D_s}^2}{m_\pi^2}\frac{m_u+m_d}{m_c+m_s},\quad
d_\Xi= a_X \frac{s_{u-d}}{c_uc_d}\sqrt{\frac{b_\Xi}{b_X}}\frac{1}{r_\Xi}
\frac{1}{\sqrt{2}}
\frac{f_{D_s}}{f_\pi}\frac{m_{D_s}^2}{m_\pi^2}\frac{m_u+m_d}{m_c+m_s}.
\end{split}
\label{eq:afcd2}
\end{equation}

\subsection{$\boldsymbol{\pi^+ \to \ell^+ \nu_\ell}$}

Combining (\ref{eq:afcd2}) and (\ref{eq:betaeqs}) yields
\begin{equation}
\begin{split}
\beta_X &= \frac{c_u^2 c_d^2}{a_X}=c_uc_d \frac{f_\pi}{v_X},\cr
\beta_H &= \frac{c_u c_d s_u s_d}{a_X}\sqrt{\frac{1-b_H}{1-b_X}}
= s_us_d \frac{f_\pi}{v_X}\sqrt{\frac{1-b_H}{1-b_X}},\cr
\beta_\Omega &=-\frac{1}{\sqrt{2}} \frac{s_{u+d} c_u
c_d}{a_X}\sqrt{\frac{1-b_\Omega}{1-b_X}}
= -\frac{1}{\sqrt{2}}
s_{u+d}\frac{f_\pi}{v_X}\sqrt{\frac{1-b_\Omega}{1-b_X}},\cr
\beta_\Xi &= \frac{1}{\sqrt{2}} \frac{s_{u-d}c_u c_d}{a_X}
\sqrt{\frac{1-b_\Xi}{1-b_X}}
= \frac{1}{\sqrt{2}} s_{u-d}\frac{f_\pi}{v_X} 
\sqrt{\frac{1-b_\Xi}{1-b_X}},
\end{split}
\end{equation}
which replaces (\ref{eq:betasol1}).
According to (\ref{eq:pilept5}) and owing to $b_\Xi=0$,
 this makes the corresponding amplitude controlled by
\begin{equation}
f_\pi\; c_d \left(c_u +s_u\frac{s_d}{c_d} \sqrt{\frac{b_H(1-b_H)}{b_X(1-b_X)}}
-\frac{1}{\sqrt{2}}\frac{s_{u+d}}{c_d}\sqrt{\frac{b_\Omega(1-b_\Omega)}{b_X(1-b_X)}}
\right).
\label{eq:pilept22}
\end{equation}

\subsection{$\boldsymbol{K^+ \to \ell^+ \nu_\ell}$}

Combining (\ref{eq:afcd2}) and (\ref{eq:zetaeqs}) gives
\begin{equation}
\begin{split}
\zeta_X &= \frac{c_u^2 s_dc_d}{a_X F_K}=\frac{c_us_d}{F_K}\frac{f_\pi}{v_X},\cr
\zeta_H &= -\frac{s_uc_u c_d^2}{a_X F_K}\sqrt{\frac{1-b_H}{1-b_X}}
= -\frac{s_uc_d}{F_K}\sqrt{\frac{1-b_H}{1-b_X}}\frac{f_\pi}{v_X},\cr
\zeta_\Omega
&=\frac{1}{\sqrt{2}}\frac{c_{u+d}c_uc_d}{a_XF_K}\sqrt{\frac{1-b_\Omega}{1-b_X}}
=\frac{1}{\sqrt{2}} \frac{c_{u+d}}{F_K}\sqrt{\frac{1-b_\Omega}{1-b_X}}
\frac{f_\pi}{v_X},\cr
\zeta_\Xi &= \frac{1}{\sqrt{2}}\frac
{c_{u-d}c_uc_d}{a_X F_K}\sqrt{\frac{1-b_\Xi}{1-b_X}}
=\frac{1}{\sqrt{2}}\frac{c_{u-d}}{F_K}\sqrt{\frac{1-b_\Xi}{1-b_X}}\frac{f_\pi}{v_X},
\end{split}
\end{equation}
which replaces (\ref{eq:zetasol1}).
According to (\ref{eq:Klept5}) the leptonic decay amplitude of $K^+$ is now
controlled by
\begin{equation}
\frac{f_\pi}{F_K}\; s_d \left(
c_u -s_u\frac{c_d}{s_d} \sqrt{\frac{b_H(1-b_H)}{b_X(1-b_X)}}
+\frac{1}{\sqrt{2}}\frac{c_{u+d}}{s_d}\sqrt{\frac{b_\Omega(1-b_\Omega)}{b_X(1-b_X)}}
\right).
\label{eq:Klept2}
\end{equation}
We recall that $f_\pi/F_K$ is given in (\ref{eq:fpiFK}).

\subsection{Updating $\boldsymbol {b_X, b_H, b_\Omega, \delta}$}
\label{subsec:update}

Since the first term in (\ref{eq:pilept22}) and (\ref{eq:Klept2}) gives a
fair description of leptonic decays of charged pions and kaons, a
cancellation between the other 2 contributions is wished for.
For $\theta_u \ll \theta_d$, this requires, for both pions and kaons
\begin{equation}
\theta_u \approx
\frac{1}{\sqrt{2}}\sqrt{\frac{b_\Omega(1-b_\Omega)}{b_H(1-b_H)}}.
\label{eq:tetaucond}
\end{equation}
The values (see (\ref{eq:bnum}))
 that we obtained at $\theta_u=0$ for the parameters $b_H$ and
$b_\Omega$ appear now grossly erroneous since they yield $\theta_u \approx
4.18$.

Since $\theta_u \ll1$ (see (\ref{eq:tetaunum})),
(\ref{eq:tetaucond}) points out at $b_\Omega \approx0$ or $b_\Omega \approx 1$.
However, the $b$'s being themselves  functions of $\theta_d$ and $\theta_u$ (see
(\ref{eq:bomprec}), (\ref{eq:bxprec}), (\ref{eq:bhprec}) below),
this cannot be settled without, in particular,
 re-investigating the masses of pseudoscalar mesons, for $\theta_u \not=0$.

\begin{figure}
\centering

\includegraphics[width=10 cm, height=6 cm]{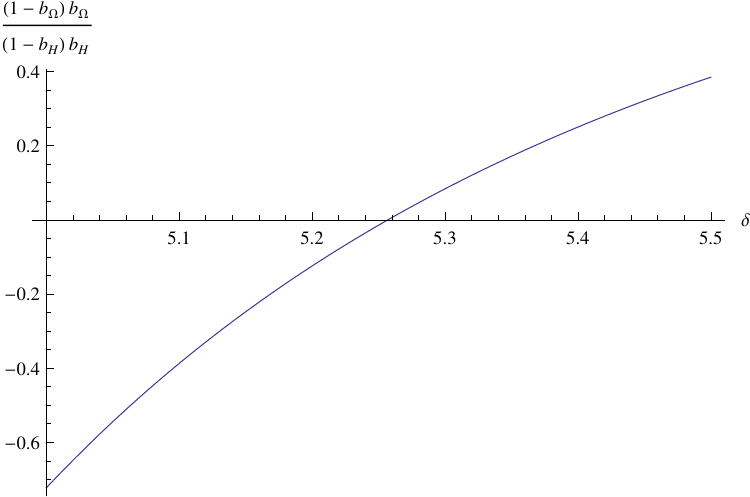}

\caption{$\frac{b_\Omega(1-b_\Omega)}{b_H(1-b_H)}$ as a function of
$\delta$}

\label{figure:deltalept}
\end{figure}

Using $r_1,r_2,r_3,r_4$ defined in (\ref{eq:4rdef}), one gets
\begin{equation}
b_\Omega = 1- \frac{c_{2(u+d)} + c_{2u}c_{2d}}
{-(c_{2(u-d)}+c_{2u}c_{2d}) + \delta\left(
c_{2u}c_{2d}\, r_1 - r_4\right)
},
\label{eq:bomprec}
\end{equation}

\begin{equation}
b_X= 1-\displaystyle\frac{2}{
\displaystyle\frac{\delta r_2}{c_{2d}} +\displaystyle\frac{\delta}{2}\left(
r_1+\displaystyle\frac{r_4}{c_{2u}c_{2d}}\right)
-\displaystyle\frac12\displaystyle\frac{1}{1-b_\Omega}\left(
1-\displaystyle\frac{c_{2(u+d)}}{c_{2u}c_{2d}}\right)
-\displaystyle\frac12\left(1-\displaystyle\frac{c_{2(u-d)}}{c_{2u}c_{2d}}\right)
},
\label{eq:bxprec}
\end{equation}

\begin{equation}
b_H= 1-\displaystyle\frac{2}{
-\displaystyle\frac{\delta r_2}{c_{2d}} +\displaystyle\frac{\delta}{2}\left(
r_1+\displaystyle\frac{r_4}{c_{2u}c_{2d}}\right)
-\displaystyle\frac12\displaystyle\frac{1}{1-b_\Omega}\left(
1-\displaystyle\frac{c_{2(u+d)}}{c_{2u}c_{2d}}\right)
-\displaystyle\frac12\left(1-\displaystyle\frac{c_{2(u-d)}}{c_{2u}c_{2d}}\right)
}.
\label{eq:bhprec}
\end{equation}
We shall take the value  (\ref{eq:tetaunum}) of $\theta_u$ and
the experimental value (\ref{eq:numtetac}) for
 $\theta_d-\theta_u$.
Then, (\ref{eq:tetaucond}) determines the value of $\delta$.
In Fig.~\ref{figure:deltalept} we plot the square of the r.h.s. of
(\ref{eq:tetaucond}) as a function of $\delta$.
Because $2\theta_u^2 \approx .0032 \ll 1$, the solution
is very close to the value at which the
blue curve in Fig.\ref{figure:deltalept} crosses the horizontal axis.
One gets
\begin{equation}
\boxed{
\delta \approx 5.259\,GeV^2
}
\label{eq:deltanum}
\end{equation}

It also corresponds, in agreement with our intuition, to
$b_\Omega \ll1$, more precisely, from (\ref{eq:bomprec}), putting in the
physical values of the charged pseudoscalar mesons masses
\begin{equation}
\boxed{
b_\Omega \approx 7\,10^{-4}
}
\label{eq:bomnum2}
\end{equation}
This limit $b_\Omega \to 0$, that we have only been able to achieve 
at $\theta_u \not= 0$, turns out to also correspond to
suitable leptonic decays of charged pions and kaons when $\theta_u$ was
taken to $0$ (see (\ref{eq:ampilept}) and (\ref{eq:amKlept})).
It however could not be justified, then.

One also gets, respectively from (\ref{eq:bxprec}) and (\ref{eq:bomprec})
\begin{equation}
\boxed{b_X \approx .996564, \quad b_H \approx .27743}
\label{eq:bhbxnum2}
\end{equation}

\subsection{The small value of $\boldsymbol{b_\Omega}$}

Like  $b_H$ at $\theta_u=0$ (see subsection \ref{subsec:smallbh}),
$b_\Omega$ given in (\ref{eq:bomprec})
 does not have either a reliable expansion at the chiral limit $m_\pi
\to 0$; it starts indeed with
\begin{equation}
b_\Omega \stackrel{m_\pi \to 0}{\simeq}
1-\displaystyle\frac{c_{2(d-u)} +
3\,c_{2(d+u)}}{\delta(-2+c_{2(d-u)}+c_{2(d+u)})}\;m_\pi^2 + \ldots
\label{eq:bomexp}
\end{equation}
in which the term between parentheses in the denominator is small.
The best that we can do, like we did for $b_H$,
is to  write $\delta = \delta_{b_\Omega=0} + \zeta m_\pi^2$, in which 
$\delta_{b_\Omega=0}=\displaystyle\frac{4c_{2u}c_{2d}}{c_{2u}c_{2d}r_1-r_4}$
is the value of $\delta$ at which $b_\Omega=0$.
(\ref{eq:bomprec}) then  yields
$b_\Omega \simeq \zeta m_\pi^2\left(\frac{1}{m_K^2}+\frac{1}{m_D^2}\right)$.
Numerically, $\zeta$ is not vanishing because $\theta_u$ is not either and,
numerically, one gets
$\zeta = \displaystyle\frac{\delta -\delta_{b_\Omega=0}}{m_\pi^2}\approx
.093$, such that
\begin{equation}
b_\Omega \approx .093\, m_\pi^2
\left(\frac{1}{m_K^2}+\frac{1}{m_D^2}\right).
\label{eq:bomsmall}
\end{equation}
More comments concerning  $b_\Omega$ will be made in
subsection \ref{subsec:finetune}.

\section{Neutral pseudoscalar mesons}
\label{section:newneutral}

The relations that come out of the orthogonality conditions
among neutral pseudoscalars turn out to be
 the same as for $\theta_u=0$;  like
before, the $\eta$ meson fails to be orthogonal to $K^0+\bar K^0$, and also
now to $D^0 + \bar D^0$.\newline
The  masses of $\pi^0, K^0$ and $D^0$
get now in good agreement with experiment without invoking $\hat b_X >1$.

\subsection{Orthogonality}

We refer to the general eqs.~(\ref{eq:orthog1}). To solve them, we use
the results (\ref{eq:solnu2}), which are valid whatever $\theta_u$ and
$\theta_d$.

$\bullet$\ Use $(a),(c)$
(we recall $\hat b_H=1 \Rightarrow \hat\delta_H=0$).
\begin{equation}
(a)-(c) \Rightarrow
\frac{\delta_\Xi}{\nu_\Xi^4}=\frac{\hat\delta_\Xi}{\hat\nu_\Xi^4},
\label{eq:xihxi}
\end{equation}
\begin{equation}
(a)+(c) \Rightarrow
- c_{2(u-d)} \frac{\delta_\Omega}{\nu_\Omega^4}
+\frac12 s_{2u}s_{2d} \frac{\hat\delta_X}{\hat\nu_X^4}
+ c_{2u}c_{2d} \frac{\hat\delta_\Omega}{\hat\nu_\Omega^4} =0.
\end{equation}

$\bullet$\ Use $(i),(j)$.
\begin{equation}
\begin{split}
& (i) \Rightarrow  \frac12 s_{2u}^2
\frac{\hat\delta_X}{\hat\nu_X^4}
+ c_{2u}^2 \frac{\hat\delta_\Omega}{\hat\nu_\Omega^4}
-\frac{\hat\delta_\Xi}{\hat\nu_\Xi^4}=0,\cr
&(j)\Rightarrow
+\frac12 s_{2d}^2 \frac{\hat\delta_X}{\hat\nu_X^4}
+ c_{2d}^2 \frac{\hat\delta_\Omega}{\hat\nu_\Omega^4}
-\frac{\hat\delta_\Xi}{\hat\nu_\Xi^4}=0.
\end{split}
\end{equation}

$\bullet$ Using $(i),(j)$ and $(a)-(c)$ yields
\begin{equation}
\frac{\hat\delta_\Omega}{\hat\nu_\Omega^4}
=\frac{\delta_\Omega}{\nu_\Omega^4},
\label{eq:omhom1}
\end{equation}
and
\begin{equation}
\frac{\hat\delta_X}{\hat\nu_X^4}=2
\frac{\delta_\Omega}{\nu_\Omega^4}.
\label{eq:omhx1}
\end{equation}
Then, $(a)+(c)$ is satisfied.

$\bullet$\ The results (\ref{eq:xihxi}), (\ref{eq:omhom1}), (\ref{eq:omhx1}),
 together with (\ref{eq:solnu2}) combine accordingly into
\begin{equation}
\boxed{
\frac{b_X(1-b_X)}{\mu_X^6}
=\frac{b_H(1-b_H)}{\mu_H^6}
=\frac{b_\Omega(1-b_\Omega)}{\mu_\Omega^6}
=\frac12\frac{\hat b_X(1-\hat b_X)}{\hat \mu_X^6}
=\frac{\hat b_\Omega(1-\hat b_\Omega)}{\hat \mu_\Omega^6}
}
\label{eq:sumsolnu5}
\end{equation}
like for $\theta_u=0$ (see (\ref{eq:sumsolnu3}) and (\ref{eq:KKperp2})).

$\bullet$\ while
(e) and (g) are verified while (f) and (h) are not: like for $\theta_u=0$,
 $\pi^0$ is
orthogonal to $K^0+\bar K^0$ and $D^0 + \bar D^0$ but  $\eta$ is not.
We check below that, indeed, neither (g) and (h), nor (e) and (f) can be
simultaneously satisfied.

*\  $(g)$ and $(h)$, which correspond respectively to the
orthogonality of $\pi^0$ and $\eta$
\footnote{The interpolating fields of which being defined, as before, as
being proportional respectively to $\bar u\gamma_5 u-\bar d\gamma_5 d$ and
$\bar u\gamma_5 u + \bar d\gamma_5 d$}
 to $K^0+\bar K^0$, cannot be satisfied simultaneously.
\begin{equation}
\begin{split}
& (g)+(h) :
-c_u^2 s_dc_d\frac{\delta_X}{\nu_X^4}
+c_u^2 s_dc_d\frac{\hat\delta_X}{\hat\nu_X^4}
+s_u^2 s_dc_d\frac{\delta_H}{\nu_H^4}
-s_u^2 s_dc_d\frac{\overbrace{\hat\delta_H}^{0}}{\hat\nu_H^4}
+\frac12 s_{2u}c_{2d}\frac{\delta_\Omega}{\nu_\Omega^4}
-\frac12 s_{2u}c_{2d}\frac{\hat\delta_\Omega}{\hat\nu_\Omega^4}=0\cr
& \hskip 3cm\Rightarrow  s_{2(u-d)}\frac{\delta_\Omega}{\nu_\Omega^4}
+c_u^2s_{2d} \frac{\hat\delta_X}{\hat\nu_X^4}
-s_{2u}c_{2d}\frac{\hat\delta_\Omega}{\hat\nu_\Omega^4}=0;\cr
& (g)-(h) :
-c_d^2 s_dc_d\frac{\delta_X}{\nu_X^4}
-c_d^2 s_dc_d\frac{\hat\delta_X}{\hat\nu_X^4}
+s_d^2 s_dc_d\frac{\delta_H}{\nu_H^4}
+s_d^2 s_dc_d\frac{\overbrace{\hat\delta_H}^{0}}{\hat\nu_H^4}
+\frac12 s_{2d}c_{2d}\frac{\delta_\Omega}{\nu_\Omega^4}
+\frac12 s_{2d}c_{2d}\frac{\hat\delta_\Omega}{\hat\nu_\Omega^4}=0\cr
& \hskip 3cm \Rightarrow
c_{2d} \frac{\hat\delta_\Omega}{\hat\nu_\Omega^4}
=c_d^2\frac{\hat\delta_X}{\hat\nu_X^4},
\end{split}
\end{equation}
which entails
\begin{equation}
\frac{\hat\delta_X}{\hat\nu_X^4} =
\frac{1}{c_uc_d}\frac{s_{2(u-d)}}{2s_{u-d}}
\ \frac{\delta_\Omega}{\nu_\Omega^4},
\label{eq:omhx3}
\end{equation}
different from (\ref{eq:omhx1}), and
\begin{equation}
\frac{\hat\delta_\Omega}{\hat\nu_\Omega^4} =
\frac{c_d}{c_uc_{2d}}\frac{s_{2(u-d)}}{2s_{u-d}}
\ \frac{\delta_\Omega}{\nu_\Omega^4},
\label{eq:omhom2}
\end{equation}
different from (\ref{eq:omhom1}).

*\ $(e)$ and $(f)$, which correspond to the
orthogonality of $\pi^0$ and $\eta$ to $D^0 + \bar D^0$, cannot be
satisfied simultaneously either.

\begin{equation}
\begin{split}
& (e)+(f) \Rightarrow
s_{2u}(c_u^2\frac{\hat\delta_X}{\hat\nu_X^4}-c_{2u}\frac{\hat\delta_\Omega}{\hat\nu_\Omega^4})=0,\cr
& (e)-(f) \Rightarrow
\frac{\delta_\Omega}{\nu_\Omega^4}(c_{2d}s_{2u}-s_{2d}c_{2u})
= \frac{\hat\delta_X}{\hat\nu_X^4} s_{2u}c_d^2
-\frac{\hat\delta_\Omega}{\hat\nu_\Omega^4} s_{2d}c_{2u}.
\end{split}
\end{equation}
This yields
\begin{equation}
\frac{\hat\delta_X}{\hat\nu_X^4}=
\frac{c_{2d}s_{2u}-s_{2d}c_{2u}}{s_{2u}c_d^2 -s_{2d}c_u^2}\
\frac{\delta_\Omega}{\nu_\Omega^4},
\label{eq:xhx4}
\end{equation}
different from both (\ref{eq:omhx1}) and (\ref{eq:omhx3})
and
\begin{equation}
\frac{\hat\delta_\Omega}{\hat\nu_\Omega^4}=
\frac{c_u^2}{c_{2u}}\
\frac{c_{2d}s_{2u}-s_{2d}c_{2u}}{s_{2u}c_d^2 -s_{2d}c_u^2}\
\frac{\delta_\Omega}{\nu_\Omega^4},
\label{eq:omhom4}
\end{equation}
different from both (\ref{eq:omhom1}) and (\ref{eq:omhom2})

\subsection{The masses of $\boldsymbol{\pi^0, K^0, D^0}$}
\label{subsec:pikdmass}

We refer to eqs.~(\ref{eq:mpi02}), (\ref{eq:mk02}) and (\ref{eq:md02}).
To get to the following expressions we use the relations (\ref{eq:solnu2})
and the following definitions and tricks.
\begin{equation}
\begin{split}
& \frac{\delta_i}{\nu_i^4} = \delta \hat v_H^2
\frac{b_i(1-b_i)}{2\mu_i^6},\cr
& \hat b_H=1 \Rightarrow \hat\delta_H=0,\cr
& \frac{b_X(1-b_X)}{\mu_X^6}
=\frac{b_H(1-b_H)}{\mu_H^6}
=\frac{b_\Omega(1-b_\Omega)}{\mu_\Omega^6}
=\frac{b_\Xi(1-b_\Xi)}{\mu_\Xi^6}
=\frac12\frac{\hat b_X(1-\hat b_X)}{\hat \mu_X^6}
=\frac{\hat b_\Omega(1-\hat b_\Omega)}{\hat \mu_\Omega^6}
=\frac{\hat b_\Xi(1-\hat b_\Xi)}{\hat \nu_\Xi^6},\cr
& \frac{1}{\nu_i^4}\equiv \hat v_H^2\frac{b_i}{2\mu_i^6}
= \hat v_H^2 \frac{1}{1-b_i}\; \frac{b_i(1-b_i)}{2\mu_i^6},\cr
& \frac{1}{\hat\nu_H^4} \equiv \hat v_H^2 \frac{\hat b_H}{2\hat\mu_H^6}
=\hat v_H^2 \frac{1}{2\mu_X^6}\underbrace{\frac{\mu_X^6}{\hat
\mu_H^6}}_{1/\hat r_H^2}=\hat v_H^2 \frac{1}{b_X(1-b_X)}
\frac{b_X(1-b_X)}{2\mu_X^6}\frac{1}{\hat r_H^2},\cr
& b_\Xi=0=\hat b_\Xi \Rightarrow \frac{1}{1-b_\Xi}=1=\frac{1}{1-\hat
b_\Xi},
\end{split}
\end{equation}
and one cancels $\hat v_H^2$ between numerators and denominators. 
This gives
\begin{equation}
\begin{split}
m_{\pi^0}^2 &= \delta\;\displaystyle\frac
{(c_u^2+c_d^2)^2 + 2(c_u^2-c_d^2)^2 + (s_u^2+s_d^2)^2 +s_{2u}^2 +s_{2d}^2 }
{\displaystyle\frac{(c_u^2+c_d^2)^2}{1-b_X}
+2\displaystyle\frac{(c_u^2-c_d^2)^2}{1-\hat b_X}
+\displaystyle\frac{(s_u^2+s_d^2)^2}{1-b_H}
+\displaystyle\frac{(s_u^2-s_d^2)^2}{b_X(1-b_X)}\displaystyle\frac{1}{\hat
r_H^2}
+\displaystyle\frac12\displaystyle\frac{(s_{2u}+s_{2d})^2}{1-b_\Omega}
+\displaystyle\frac12\displaystyle\frac{(s_{2u}-s_{2d})^2}{1-\hat
b_\Omega}}\cr
& \cr
& \approx
\displaystyle\frac
{4\delta}
{\displaystyle\frac{(c_u^2+c_d^2)^2}{1-b_X}
+2\displaystyle\frac{(c_u^2-c_d^2)^2}{1-\hat b_X}
+\displaystyle\frac{(s_u^2+s_d^2)^2}{1-b_H}
+\displaystyle\frac{(s_u^2-s_d^2)^2}{b_X(1-b_X)}\displaystyle\frac{1}{\hat
r_H^2}
+\displaystyle\frac12\displaystyle\frac{(s_{2u}+s_{2d})^2}{1-b_\Omega}
+\displaystyle\frac12\displaystyle\frac{(s_{2u}-s_{2d})^2}{1-\hat
b_\Omega}},
\end{split}
\label{eq:mpi0prec}
\end{equation}
\begin{equation}
\begin{split}
m_{K^0}^2 &= \delta\;\displaystyle\frac
{\overbrace{2s_d^2 c_d^2 + \displaystyle\frac12 c_{2d}^2 + \frac12}^{1}}
{\displaystyle\frac{s_d^2 c_d^2}{2} \left(
\displaystyle\frac{1}{1-b_X}
+2\displaystyle\frac{1}{1-\hat b_X}
+\displaystyle\frac{1}{1-b_H}
+\displaystyle\frac{1}{b_X(1-b_X)}\displaystyle\frac{1}{\hat r_H^2}
\right) + \displaystyle\frac{c_{2d}^2}{4} \left(
\displaystyle\frac{1}{1-b_\Omega}
+\displaystyle\frac{1}{1-\hat b_\Omega}\right)
+\frac12 }\cr
& \cr
&=\displaystyle\frac
{4\delta}
{\displaystyle\frac{s_{2d}^2}{2} \left(
\displaystyle\frac{1}{1-b_X}
+2\displaystyle\frac{1}{1-\hat b_X}
+\displaystyle\frac{1}{1-b_H}
+\displaystyle\frac{1}{b_X(1-b_X)}\displaystyle\frac{1}{\hat r_H^2}
\right) + c_{2d}^2 \left(
\displaystyle\frac{1}{1-b_\Omega}
+\displaystyle\frac{1}{1-\hat b_\Omega}\right)
+2 },
\end{split}
\label{eq:mK0prec}
\end{equation}
\begin{equation}
\begin{split}
m_{D^0}^2 &= \delta\;\displaystyle\frac
{\overbrace{2s_u^2 c_u^2 + \displaystyle\frac12 c_{2u}^2+\frac12}^{1}}
{\displaystyle\frac{s_u^2 c_u^2}{2} \left(
\displaystyle\frac{1}{1-b_X}
+2\displaystyle\frac{1}{1-\hat b_X}
+\displaystyle\frac{1}{1-b_H}
+\displaystyle\frac{1}{b_X(1-b_X)}\displaystyle\frac{1}{\hat r_H^2}
\right) + \displaystyle\frac{c_{2u}^2}{4} \left(
\displaystyle\frac{1}{1-b_\Omega}
+\displaystyle\frac{1}{1-\hat b_\Omega}\right)
 +\frac12 }\cr
& \cr
&=\displaystyle\frac
{4\delta}
{\displaystyle\frac{s_{2u}^2}{2} \left(
\displaystyle\frac{1}{1-b_X}
+2\displaystyle\frac{1}{1-\hat b_X}
+\displaystyle\frac{1}{1-b_H}
+\displaystyle\frac{1}{b_X(1-b_X)}\displaystyle\frac{1}{\hat r_H^2}
\right) + c_{2u}^2 \left(
\displaystyle\frac{1}{1-b_\Omega}
+\displaystyle\frac{1}{1-\hat b_\Omega}\right)
 +2 }.
\end{split}
\label{eq:mD0prec}
\end{equation}

Inside (\ref{eq:mpi0prec}), (\ref{eq:mK0prec}) and (\ref{eq:mD0prec}),
$b_\Omega, b_X, b_H$ are given by (\ref{eq:bomprec}), (\ref{eq:bxprec}) and
 (\ref{eq:bhprec}) in terms of $\delta, \theta_d, \theta_u$ and of the
masses of the charged pseudoscalar mesons. Since
$\theta_c\equiv\theta_d-\theta_u$ and $\theta_u$
are respectively given by (\ref{eq:numtetac}) and (\ref{eq:tetaunum}), 
$\delta$ by (\ref{eq:deltanum}), $b_H$ and $b_X$ by (\ref{eq:bhprec}) and
(\ref{eq:bxprec}),
the 3 equations (\ref{eq:mpi0prec}), (\ref{eq:mK0prec}) and (\ref{eq:mD0prec})
should be enough to determine $\hat b_X, \hat b_\Omega, \hat r_H$.
Numerical studies show that:\newline
* the masses of $\pi^0$
\footnote{that we may identify with that of $\pi^+$ since we did not introduce
electromagnetism and we know that the $\pi^+-\pi^0$ mass difference is
essentially electromagnetic.}
and $K^0$ can be suitably accounted for at the condition that 
\begin{equation}
\boxed{
\hat r_H
\equiv \frac{<\bar c c-\bar s s>}{<\bar u u+\bar d d>}\geq .945}
\end{equation}
They are practically insensitive to the value of $\hat b_\Omega$ which,
supposedly small, like $b_\Omega$, according to our intuition, can even be
considered to be vanishing;\newline
* the mass of $D^0$ is only off by $\approx 20\,MeV$;\newline
* $\hat b_X$ should be ${\cal O}(1)$ but, unfortunately, one gets too small
a sensitivity to determine this parameter accurately; a more exhaustive
study of the full system of equations is probably necessary for
this.

To give an idea of the precision of the determination,
for $\hat r_H=.96, \hat b_\Omega=0$ and $\hat b_X=.8$, one gets 
\begin{equation}
\begin{split}
& m_{\pi^0} \approx 139.38\,MeV = m_{\pi^+}^{exp} -190\,KeV =
m_{\pi^0}^{exp} + 4.42\,MeV,\cr
& m_{K^0}\approx 496.8\,MeV = m_{K^0}^{exp} + 3.1 MeV, \cr
& m_{D^0}\approx 1.843\,GeV = m_{D^0}^{exp} -22\,MeV.
\end{split}
\label{eq:newneutmass}
\end{equation}

By introducing a non-vanishing $\theta_u$, we have increased $\hat r_H$
from $.57$ (see (\ref{eq:hrhnum})) to $\hat r_H > .95$ and gotten a quite
satisfactory agreement for the masses of neutral pseudoscalar mesons.
So, while the need of a rather large condensate for heavy quarks is
confirmed, staying with $\hat b_X <1$ has become ``much more possible''
than when imposing $\theta_u=0$. 

From (\ref{eq:hvh}), using (\ref{eq:bxprec}),  since $b_\Omega, \hat b_\Omega \ll1$, $\hat b_H=1
\approx b_X$, and taking, like for calculation the masses of $\pi^0, K^0,
D^0$, $\hat b_X \approx .8$, one gets
\begin{equation}
\boxed{
\hat v_H \approx v_X \approx 151\,GeV}
\label{eq:hvhvxnum2}
\end{equation}

\section{The Higgs spectrum}
\label{section:higgsspectrum}

From the values of the $b$ parameters and of $\delta$ that we have determined
in subsection \ref{subsec:update} we get
\begin{equation}
\boxed{
m_{\hat H^3} \approx 3.24\,GeV \approx m_{X^0},\quad
m_{H^0} \approx 1.65\,GeV,\quad m_{\Omega^0}\approx 86\,MeV}
\end{equation}
$\hat\Omega^3$ is, like $\Omega^0$, presumably very light, and so are
$\Xi^0$ and $\hat \Xi^3$, for the same reasons as when $\theta_u=0$ since we
still have $b_\Xi=0=\hat b_\Xi$.
With the value $\hat b_X = .8$ that we used to fit $m_{\pi^0}, m_{K^0},
m_{D^0}$ in subsection \ref{subsec:pikdmass},
one gets $m_{\hat X^3} \approx 2.9\,GeV$. To precisely determine
$\hat b_\Omega$ an extensive study of the fermionic sector is
needed.

\section{Quark condensates} \label{section:qcond}

Let us have finally an estimate of the ratios of some of the quark condensates
like we did in section \ref{section:solution} for the case $\theta_u=0$.
From (\ref{eq:sumsolnu5}) and the values of the $b$'s given in
(\ref{eq:bomnum2}) and (\ref{eq:bhbxnum2}), one gets
\begin{equation}
\boxed{
r_H\equiv \frac{<\bar c c+\bar s s>}{<\bar u u+\bar d d>}
\approx 7.42 ,\quad
r_\Omega\equiv\frac{1}{\sqrt{2}}\frac{<\bar u c+ \bar c u + \bar d s+ \bar
s d>} {<\bar u u+\bar d d>}
\approx .455}
\end{equation}
$<\bar c c>$ and $<\bar s s>$ are given by the l.h.s of (\ref{eq:vevnum}),
which requires the knowledge of $\hat r_H$. Taking the value $\hat r_H=.96$
that we used to fit the masses of neutral pseudoscalars leads to
\begin{equation}
\boxed{
<\bar c c> \approx 5.96 \,\mu_X^3, \quad
<\bar s s> \approx 4.60\,\mu_X^3}
\end{equation}
which are both large negative values (we recall that $\mu_X^3 \equiv
(<\bar u u+\bar d d>)/\sqrt{2}$ is known from the GMOR relation).
 This confirms  that large condensates for heavy quarks are
wished for, which  may me the sign that a 3rd generation is needed.
These large  $<\bar q q>$'s are certainly the sign that our extension of
the GSW model is still incomplete.

\section{Conclusion for the case $\boldsymbol{\theta_d\not=0, \theta_u\not=0}$}
\label{section:conclus2}

\subsection{Generalities}

The values of $\theta_d$ and $\theta_u$ agree with the estimates
$\theta_c \approx \theta_d \simeq
\sqrt{\frac{|m_d|}{m_s}}$ and $\theta_u \simeq \sqrt{\frac{m_u}{m_c}}$ that
were obtained on various other grounds
(see footnote \ref{foot:oldmix}). They are
independent quantities, and the Cabibbo angle $\theta_c=\theta_d-\theta_u$
cannot incorporate all the physics  for 2 generations. This is
in sharp contrast with the genuine GSW
model in which only the difference $\theta_d-\theta_u$ is physically
relevant.\newline
Dealing with a non-vanishing $\theta_u$  provided several improvements to the
fit between this model and experimental data.
Together with charged pseudoscalar mesons, the masses of neutral
pseudoscalars $\pi^0, K^0, D^0$ can now also be described with a good
accuracy.\newline
$b_H$ has increased to $\approx .28$ while
$b_\Omega$ and, presumably $\hat b_\Omega$, too, have become very small. This
agrees with the fact that, at least perturbatively, non-diagonal quark
condensate, which only occur at 2-loops, should be very small.\newline
Leptonic decays of charged pions and kaons can be
correctly accounted for; this goes with  $b_\Omega \to 0$, which was also
wished for when $\theta_u=0$ but could not be, then, argued
for.\newline
$\delta$ has increased from $\delta \approx m_{D_s}^2$ up to $\delta \approx 5.26
GeV^2$; the mass of the ``standard-like'' Higgs boson has gone from
$\sqrt{2}m_{D_s}  =2.78\,GeV$ up to $\sqrt{2\delta}=3.24\,GeV$.
This scaling factor concerns all the Higgs bosons.
The Higgs spectrum has been modified accordingly.
The 2 Higgs bosons $\Omega^0$ and $\hat\Omega^3$,
 which had  intermediate masses for $\theta_u=0$,
are now very light.

\subsection{Rapidly varying functions, slow-converging expansions,
coincidences and fine-tuning} \label{subsec:finetune}

One among the most important issues is certainly the important role of
the very small parameter $\theta_u$, concerning in particular  the
spectrum of the Higgs bosons.
As I will show below, it is the consequence of  the presence of
rapidly varying functions, which often have poles,
 of the extreme care with which the chiral limit
must be implemented \ldots and some ``bad luck'' which positioned the solution,
 in the case  $\theta_u=0$, inside a very special set of values of the
parameters. 

\begin{figure}
\centering

\includegraphics[width=10 cm, height=6 cm]{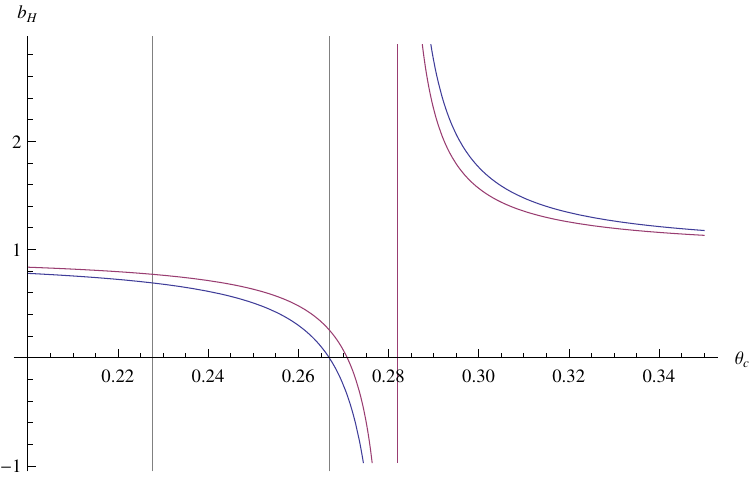}

\caption{$b_H$ at $\theta_u=0$ as a function of $\theta_c$
for $\delta =m_{D_s}^2 \approx
3.87495\,GeV^2$ (blue) and $\delta = 5.26\,GeV^2$ (red) ; the vertical lines
stand at $\theta_c = .227591$ (experimental value) and $\theta_c = .2669$
(value found at $\theta_u=0$)}

\label{figure:bhtetac1}
\end{figure}

The value  $\delta \approx m_{D_s}^2$
at $\theta_u=0$ had been obtained by considering the
$s$ quark mass at the chiral limit, limit at which, unfortunately,
we also mentioned that $b_H$ has no reliable expansion.
$b_H$ depends of $\delta$ and can vary very rapidly with $\theta_c$. The
combination of the two values obtained for $\delta$ and $\theta_c$
determined, then, $b_H$ to be very close to $0$, but also, as shown below,
close to a region where it varies rapidly. It was thus
extreme and unstable fine tuning.
On Fig.\ref{figure:bhtetac1} below we plot $b_H$ at $\theta_u=0$
 as a function of $\theta_c$,
for $\delta \approx m_{D_s}^2$ (blue, corresponding to our result at
$\theta_u=0$) and for $\delta \approx 5.26\,GeV^2$
(red, corresponding to our result (\ref{eq:deltanum}) at $\theta_u \not=0$).
  The 2 vertical lines are drawn -- at the experimental value
(\ref{eq:numtetac}) $\theta_c = .2276$; -- at $\theta_c \approx .2669$
as determined from charged pseudoscalar mesons at $\theta_u=0$ in
(\ref{eq:tetadnum}).
%
%
We see on Fig.\ref{figure:bhtetac1} that the values of $\theta_c=\theta_d$ and $\delta$
that were obtained at $\theta_u=0$ coincide with
$b_H \approx 0$, but that $b_H$ is also in a domain where it varies very
fast. Keeping the same value $\theta_c = .2669$ and still staying at
$\theta_u=0$, we see that varying $\delta$ between  $m_{D_s}^2$ and
$5.26\,GeV^2$ triggers relatively large variations of $b_H$.
One also sees on the same figure that keeping $\delta$ fixed and
varying $\theta_c$ between $.228$ and
$.267$, which is not a big variation, also triggers large
variations of $b_H$.

$b_X$ is found to be very stable, but this is not the case for $b_\Omega$,
that behaves in many respects like $b_H$.

\begin{figure}
\centering

\includegraphics[width=10 cm, height=6 cm]{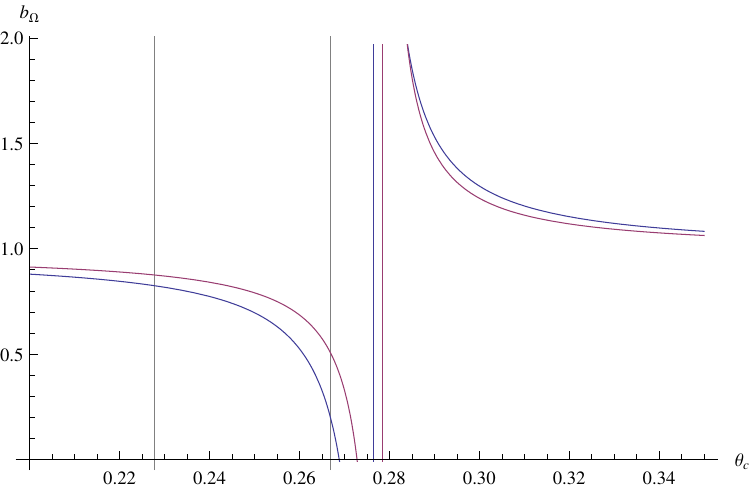}

\caption{$b_\Omega$ at $\theta_u=0$
 as a function of $\theta_c$ for $\delta =m_{D_s}^2\approx
3.87495\,GeV^2$ (blue) and $\delta = 5.26\,GeV^2$ (red); the vertical lines
stand at $\theta_c = .227591$ (experimental value) and $\theta_c = .2669$}

\label{figure:bomtetac1}
\end{figure}

\begin{figure}
\centering

\includegraphics[width=10 cm, height=6 cm]{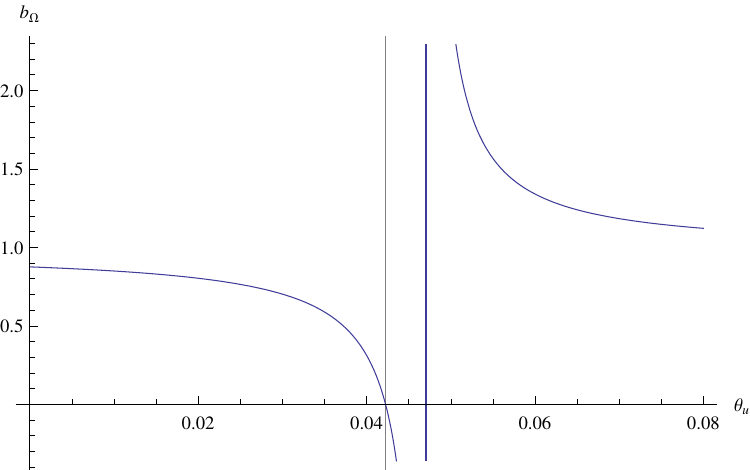}

\caption{$b_\Omega$ at the measured value  $\theta_c=.227591$
and $\delta = 5.529\,GeV^2$  as a function of $\theta_u$; the vertical line
stands at $\theta_u = .04225$}

\label{figure:bomtetau3}
\end{figure}

\begin{figure}
\centering

\includegraphics[width=10 cm, height=6 cm]{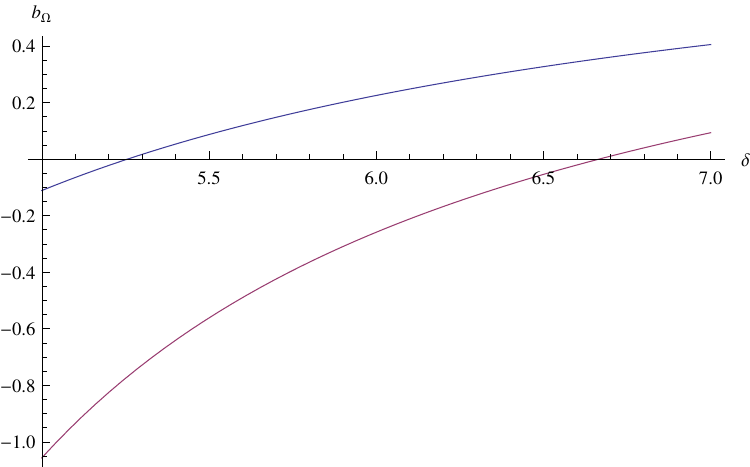}

\caption{$b_\Omega$ at the measured value  $\theta_c=.227591$
  as a function of $\delta$ for $\theta_u=.04225$ (blue) and
$\theta_u=.04436$ (red)}

\label{figure:bomtetau2}
\end{figure}

We plot on Fig.\ref{figure:bomtetac1} $b_\Omega$ at $\theta_u=0$ as a function of $\theta_c$.
The 2 vertical lines are at the same positions as in
Fig.\ref{figure:bhtetac1}.
%
%
For $\theta_c = .2669$ one
recovers $b_\Omega \approx .2$ that we had found at $\theta_u=0$,
 but going to a larger $\delta$ and / or to a
smaller $\theta_c$ seems to increase $b_\Omega$ instead of bringing it
close to $0$ as we found for $\theta_u \not=0$. $b_\Omega$ is therefore very
sensitive to the value of the small parameter $\theta_u$ itself. This is
shown on Fig.\ref{figure:bomtetau3} in which we plot $b_\Omega$ as a function of
$\theta_u$ at the experimental value of $\theta_c$ (\ref{eq:numtetac})
 and at the value (\ref{eq:deltanum}) of $\delta$ that we have determined.
We witness the same phenomenon as the one that occurred for $b_H$ at
$\theta_u=0$ as a function of $\theta_c$:
$b_\Omega$ is very close to $0$ but it is also in a region of
very fast variation with $\theta_u$, close to the pole.
%
%
The value of $\delta$ at which $b_\Omega$ vanishes is also
very sensitive to the value of $\theta_u$. This is seen on
Fig.\ref{figure:bomtetau2}, in which we plot $b_\Omega$ as a function of $\delta$ for
two values of $\theta_u$, $\theta_u=.04225$ as
 given by (\ref{eq:tetaunum}) (blue curve)
and $1.05 \times$ this value, $\theta_u = .04436$ (red curve).
%

Since $b_\Omega$ lies in a domain where it is very
sensitive to the value of other parameters,
the question is of course ``can we trust the value that we have obtained?''
We have no answer except that it corresponds to our intuition
of non-diagonal quark condensates being very small. We recall indeed that
$b_\Omega$ is related to $r_\Omega = \frac{<\bar u c+ \bar c u + \bar d s +
\bar s d>}{\sqrt{2}<\bar u u+\bar d d>}$ by the relation
(\ref{eq:sumsolr1}) $r_\Omega
=\sqrt{b_\Omega(1-b_\Omega)/b_X(1-b_X)}$ and, since $b_X(1-b_X)$ is small,
for $r_\Omega$ to be small one needs $b_\Omega$ to be very small.

We deal with a very tightly entangled series of parameters;
if one changes $\theta_u$ by a very small amount (which is very
conceivable because it has been deduced from the value of $\theta_c$, which
has itself experimental uncertainties, and from
 the masses of charged pseudoscalars,
which have also some small uncertainties), one can change $\delta$ by a
large amount since it roughly corresponds to the value at which
$b_\Omega$ vanishes. This has in turn consequences on $b_X$ (small, because
it is very stable) and on $b_H$. 
Since large variations have been triggered by going from
 $\theta_u=0$ to $\theta_u \not=0$, it is of course necessary to
take our results with care.
It is unfortunate that it is also related with the
existence and properties  of very light scalars, one of the most
interesting but also controversial domain of research of the last decades
\cite{dark} \cite{CLEOBABAR}.\newline
We are undoubtedly dealing with very fine tuned physics,
in which some parameters furthermore stubbornly resist being expanded in
powers of $m_\pi$ (chiral limit) or of other small parameters like
$\theta_d$ or $\theta_u$.\newline

\subsection{A promising way}

This path of investigation was initiated \cite{Machet0}
 because it looked  the most natural and, 
so far, the physics of pseudoscalar
mesons has been remarkably well described.
That there exists one and, seemingly,
only one solution to the sets of equations that match experimental data
gives a fair prejudice that it provides
an optimized set of parameters to fit the physical world, and
good confidence that it is  a good  way to proceed.

We are still clearly far from investigating the whole domain of
 even  the small subset of pseudoscalar mesons.
Later works will consider in particular their semi-leptonic
decays, for example the one of $\pi^+$ into a very light Higgs boson and
leptons, because it could be a  way to detect such an elusive particle.

While, for $\theta_u=0$,
 the determination of the unique mixing angle $\theta_d=\theta_c$
 and of the $b$ parameters
was made {\it ab initio}, this was not the case when
we introduced the two mixing angles $\theta_u$ and $\theta_d$.
We instead relied on the measured value of $\theta_c=\theta_d-\theta_u$ to
get the value of $\theta_u$. The consistency of the method has nevertheless
 proved  to be very satisfying.\newline
When dealing with 2 mixing angles,
we did not go either through the analysis of the fermionic constraints.
The reason for this is that they can no longer
be simplified and easily handled. In particular they now also include the
parameter $\delta_{\Omega\bar\Omega}$, such that a larger set of equations
is needed. It may well be that one is obliged to solve the whole system,
which is not an easy task.\newline
A solution to the puzzling ``maximal'' fermionic mixing that
occurred at $\theta_u=0$ (see section \ref{section:fermix})
looks nevertheless in sight.
We mentioned indeed, that, at $\theta_u=0$,  the value of the Cabibbo angle
that we obtained coincides with a pole of $\frac{\mu_{ds}}{\mu_d-\mu_s}$,
unless $b_\Omega$ and $\hat b_\Omega$ become extremely small.
We have seen that this is precisely what happens
when turning on $\theta_u$, which, despite all the issues concerning fine
tuning, gives now reasonable hope for a matching between
bosonic and fermionic mixing angles.\newline
What is the final sign of the mass of the $d$ quark? Without an exhaustive
study of the whole system of equations we have unfortunately no definitive
answer to give. We have seen that it should be negative for 1 generation
and for 2 generations at $\theta_u=0$. If it turns positive when switching
on $\theta_u$, this may look more ``conventional'', but this it also
 provides one more example of a ``parameter'' which is extremely sensitive
to the small parameter $\theta_u$. Fine tuning is then still more
severe.

We did not seriously analyze  the $\eta$ meson, in particular
concerning its  orthogonality with $K^0+\bar K^0$. Its definition itself
maybe in cause and  the mixing between $\pi^0$ and $\eta$ must probably be
taken into account.

A fairly large condensate of heavy quarks seems consistently
 needed. This goes against common intuition.
However, as we shall study more deeply in the forthcoming work(s),
quark condensation is presumably linked to the existence 
of strong ``interactions'' of quarks. They
 of course interact with gauge bosons (gluons), but, in this extension
of the GSW model, with all scalar (Higgs) and pseudoscalar bosons,
and it is rather easy
 to realize that their couplings are mostly not ``standard'' and that some of
them are strong, which may  provide another origin for quark condensation.

Last, doubts remain concerning the existence and spectrum of very light scalars.
The reason for this is that, as shown in subsection \ref{subsec:finetune},
it is a highly fined tuned sector. Furthermore, 
this issue cannot undoubtedly be settled  without including
the 3rd generations and all the paraphernalia of their  mixing angles (6 + $CP$
violating phases). To add to the complexity,
one knows that it is unfortunately
very dangerous to approximate anyone by $0$. This
results in a very large system of equations to be
solved simultaneously, which goes beyond the limits of this work.


\chapter{Symmetries. Outlook and prospects} \label{chapt:conclusion}

This last chapter is mainly dedicated to the  symmetries that
underlie this model, and to some remarks concerning the normalization of
states that pave the way to forthcoming works. We finally list
the topics to be investigated in the future.

\section{Symmetries}\label{section:symmetries}

\subsection{Entangled symmetries and breakings}\label{subsec:breaking}

Inside the chiral $U(4)_L \times U(4)_R$ group,
we have identified in (\ref{eq:Ts}) the 3 generators of the left-handed
 group $SU(2)_L$ of weak interactions. 
Before being considered as a local group of symmetry, this $SU(2)_L$ and
its mirror group $SU(2)_R$ build up the chiral group $SU(2)_L \times SU(2)_R$.
In the 4-dimensional space of the 4 components
$(\Delta^0, \Delta^3, \Delta^+, \Delta^-)$
of any quadruplet $\Delta$, endowed with the corresponding basis
\begin{equation}
v^0=\left(\begin{array}{c} 1\cr 0\cr 0\cr 0\end{array}\right),\
v^3=\left(\begin{array}{c} 0\cr 1\cr 0\cr 0\end{array}\right),\
v^+ = \left(\begin{array}{c} 0\cr 0\cr 1\cr 0\cr\end{array}\right),\
v^- = \left(\begin{array}{c} 0\cr 0\cr 0\cr
1\cr\end{array}\right),
\label{eq:4basis}
\end{equation}
the three $T_L$ generators write
\begin{equation}
T^3_L = \left(\begin{array}{rrrrr}
0 & -\frac12 & \vline & 0 & 0 \cr
-\frac12 & 0 & \vline  & 0 & 0 \cr
\hline
0 & 0 & \vline  & -\frac12 & 0 \cr
0 & 0 & \vline & 0 & \frac12 \end{array}\right),\quad
T^+_L = \left(\begin{array}{rrrrr}
0 & 0 & \vline & 0 & -1 \cr
0 & 0 & \vline  & 0 & -1 \cr
\hline
-\frac12 & \frac12 & \vline  & 0 & 0 \cr
0 & 0 & \vline & 0 & 0 \end{array}\right),\quad
T^-_L = \left(\begin{array}{rrrrr}
0 & 0 & \vline & -1 & 0 \cr
0 & 0 &\vline  & 1 & 0 \cr
\hline
0 & 0& \vline  & 0 & 0 \cr
-\frac12 & -\frac12 & \vline  & 0 & 0 \end{array}\right);
\label{eq:t3l}
\end{equation}
the three $T_R$ generators write
\begin{equation}
T^3_R = \left(\begin{array}{rrrrr}
0 & \frac12 & \vline  & 0 & 0 \cr
\frac12 & 0 &\vline &  0 & 0 \cr
\hline
0 & 0 & \vline  & -\frac12 & 0 \cr
0 & 0 &\vline  & 0 & \frac12 \end{array}\right),\quad
T^+_R = \left(\begin{array}{rrrrr}
0 & 0& \vline  & 0 & 1 \cr
0 & 0& \vline  & 0 & -1 \cr
\hline
\frac12 & \frac12 & \vline  & 0 & 0 \cr
0 & 0 & \vline  & 0 & 0 \end{array}\right),\quad
T^-_R = \left(\begin{array}{rrrrr}
0 & 0 &\vline  & 1 & 0 \cr
0 & 0 &\vline  & 1 & 0 \cr
\hline
0 & 0 &\vline  & 0 & 0 \cr
\frac12 & -\frac12 &\vline  & 0 & 0 \end{array}\right);
\label{eq:t3r}
\end{equation}
the three $SU(2)$ generators $\vec T= \vec T_L + \vec T_R$ of the diagonal
subgroup are accordingly
\begin{equation}
T^3 = \left(\begin{array}{rrrrr}
0 & 0 & \vline & 0 & 0 \cr
0 & 0 & \vline & 0 & 0 \cr
\hline
0 & 0 & \vline & -1 & 0 \cr
0 & 0 & \vline & 0 & 1 \end{array}\right)= Q,\quad
T^+ = \left(\begin{array}{rrrrr}
0 & 0 & \vline & 0 & 0 \cr
0 & 0 & \vline & 0 & -2 \cr
\hline
0 & 1 & \vline & 0 & 0 \cr
0 & 0 & \vline & 0 & 0 \end{array}\right),\quad
T^- = \left(\begin{array}{rrrrr}
0 & 0 & \vline & 0 & 0 \cr
0 & 0 & \vline & 2 & 0 \cr
\hline
0 & 0 & \vline & 0 & 0 \cr
0 & -1 & \vline & 0 & 0 \end{array}\right),
\label{eq:t3}
\end{equation}
This diagonal $SU(2)$ is the so-called custodial $SU(2)$ group.
Its 3rd generator, $T^3$, can be identified with the electric charge
\cite{Machet0}.

Since by anticommutation the $T$'s close on the identity matrix
$\mathbb I$, it is often natural to consider the larger chiral group $U(2)_L
\times U(2)_R$.

In section \ref{section:quadsym} we also identified an ``generation''
 $U(2)^g_L \times U(2)^g_R$ group of
transformations that moves inside the 8-dimensional
 space of quadruplets. Its generators
commute with the ones of the gauge group, such that the two groups of
transformations are orthogonal.
The equivalent of the custodial $SU(2)$ for horizontal transformations can
be identified with the $SU(2)^g$ ``generation group of symmetry''.
It acts as follows on quarks
\begin{equation}
\begin{split}
& L^+.(c,s)= (u,d),\quad L^-.(u,d)=(c,s),\cr
& L^+.(u,d)=(0,0)= L^-.(c,s),\cr
& L^3.(u,d)=\frac12(u,d),\quad L^3.(c,s)=-\frac12(c,s).
\end{split}
\label{eq:hordiag}
\end{equation}
 and is accordingly seen to be  connected to flipping the
generations.

The vertical and horizontal $U(2)_L \times U(2)_R$ intersect along the
chiral $U(1)_L \times U(1)_R$, which, at the level of generators, are
directly connected with parity (see subsection \ref{subsec:parity}).

Which $SU(2)$ can be identified with the ``flavor'' $SU(2)$? Since it should
survive for 1 generation so as to keep the 3 pions  quasi-degenerate, it
cannot be the generation diagonal $SU(2)$ which, as we have seen,
shrinks to the trivial diagonal $U(1)$ for 1 generation. It can accordingly
only be the custodial $SU(2)$.

We already emphasized in the core of the paper that, at the same time as
the various symmetries are tightly entangled, so are their breaking.
While the breaking of the gauge symmetry is signaled by non-vanishing bosonic
VEV's $v_i, \hat v_i \not= 0$, fermionic condensates $<\bar q_i q_j>\not=0$
are usually considered to trigger chiral symmetry breaking. We have shown,
for example in (\ref{eq:sumsolnu3}), that the  2 sets of VEV's
are indeed not independent.

It is  worth  investigating, inside each quadruplet, which subgroup of the
chiral $U(4)_L\times U(4)_R$ group  is left
unbroken by its vacuum. We have only to focus on the diagonal $U(4)$ since
we know in advance that parity is broken. A $4\times4$ generator of the
diagonal $U(4)$ annihilates the vacuum if and only if it commutes with the
corresponding $\mathbb M$ matrix (see (\ref{eq:SP}) and (\ref{eq:PS})).
Along these steps, one gets the following result: the 4 states of the type $\Delta_i^0$ are
annihilated by generators of the type
$\left(\begin{array}{ccccc}
a &  & \vline & c &  \cr
  & a & \vline & & c \cr
\hline
d & & \vline & b &  \cr
 & d & \vline & & b \end{array}\right)$
while the 4 states of the type $\hat\Delta_i^3$ are annihilated by 
generators of the type
$\left(\begin{array}{ccccc}
a &  & \vline &   &  \cr
  & a & \vline & &   \cr
\hline
  & & \vline & b &  \cr
 &   & \vline & & b \end{array}\right)$.
This shows that the set of 4 $<\Delta_i^0>\not=0$ break the chiral $U(4)_L \times
U(4)_R$ down to the diagonal $U(2)$ (which contains in particular
$U(1)_{em}$) while the set of 4
$<\hat\Delta_i^3>\not=0$ break it down to $U(1)\times U(1)_{em}$, in which
the first $U(1)$ is the trivial one. In the presence of all types of VEV's,
we thus conclude that the only group that is left unbroken is $U(1)\times
U(1)_{em}$. Of course this paves a straightforward way for the introduction of the
photon as a gauge particle. 

The generators that we displayed above annihilate sets of vacuum states. If
one considers a single state, for example $X^0$ or $H^0$,
 all generators of the
type $\left(\begin{array}{ccccc}
a &   & \vline & e &   \cr
  & b & \vline &   & f\cr
\hline
g &   & \vline & c &   \cr
  & h & \vline &   & d 
\end{array}\right)$,
with 8 independent real entries annihilate them.
Likewise, $\hat X^3$ gets annihilated by
$\left(\begin{array}{ccccc}
a &   & \vline &   &   \cr
  & b & \vline &   & f\cr
\hline
  &   & \vline & c &   \cr
  & h & \vline &   & d 
\end{array}\right)$ 
while 
$\hat H^3$ is annihilated by
$\left(\begin{array}{ccccc}
a &   & \vline & e &   \cr
  & b & \vline &   &  \cr
\hline
 g &   & \vline & c &   \cr
  &   & \vline &   & d 
\end{array}\right)$,
both having only 6 independent entries.
The last four are annihilated by generators with 8 independent entries
(they include generators of the diagonal ``generation'' group $SU(2)^g$ with
generators $\vec L$):
$\Omega^0$ is annihilated by
$\left(\begin{array}{ccccc}
a & b & \vline & c & d \cr
b & a & \vline & d & c\cr
\hline
 e & f & \vline & g & h \cr
f & e & \vline & h & g 
\end{array}\right)$,
$\Xi^0$ by
$\left(\begin{array}{ccccc}
a & b & \vline & c & d \cr
-b & a & \vline & -d & c\cr
\hline
 e & f & \vline & g & h \cr
-f & e & \vline & -h & g 
\end{array}\right)$,
$\hat\Omega^3$ by
$\left(\begin{array}{ccccc}
a & b & \vline & c & d \cr
b & a & \vline & -d & -c\cr
\hline
 e & f & \vline & g & h \cr
-f & -e & \vline & h & g 
\end{array}\right)$,
and $\hat\Xi^3$ by
$\left(\begin{array}{ccccc}
a & b & \vline & c & d \cr
-b & a & \vline & d & -c\cr
\hline
 e & f & \vline & g & h \cr
f & -e & \vline & -h & g 
\end{array}\right)$.
We see that, while for individual quadruplets the little groups can be
quite large (8 dimensional), their intersection is only $U(1)\times
U(1)_{em}$. In particular the chiral ``generation'' group $U(2)^g_L \times
U(2)^g_R$ gets totally broken.

Since the chiral/gauge $SU(2)_L\times SU(2)_R$ groups are of special
relevance to us, let us determine their action on the vacuum inside each
quadruplet. From the general results above, we see that all $\Delta_i^0$
states are annihilated by the diagonal $U(2)$ while all $\hat\Delta^3$
states are only annihilated by $U(1)\times U(1)_{em}$.
Leaving aside the trivial $U(1)$, the situation concerning the custodial
group can be summarized by
\begin{equation}
\vec T^\pm.(X^0,H^0,\Omega^0,\Xi^0)=0,\quad T^3.(X^0,H^0,\Omega^0,\Xi^0,
\hat X^3, \hat H^3, \hat\Omega^3,\hat\Xi^3)=0.
\label{eq:Taction}
\end{equation}
$SU(2)_L$, which is to become local, is totally broken, which
yields 3 Goldstone bosons inside each quadruplet. There cannot be more,
since any Higgs cannot be a Goldstone, such that, even when the breaking is
stronger, like for example in $\hat X^3$ or $\hat H^3$ where the chiral
$U(4)_L \times U(4)_R$ with 32 generators gets broken down to a set of 6
independent generators, only 3 Goldstones are generated.

Last, for the diagonal  $SU(2)^g$ group with generators $\vec L$:
\begin{equation}
L^3.(X^0,\hat X^3, \hat H^3, H^0)=0,\quad
L^1.(\Omega^0,\hat\Omega^3)=0,\quad L^2.(\Xi^0,\hat\Xi^3)=0.
\label{eq:Laction}
\end{equation}

The invariance of the scalar potential and Yukawa Lagrangian are worth some
remarks. Would all normalizing factors $v_i/\sqrt{2}\mu_i^3, \hat
v_i/\sqrt{2}\hat\mu_i^3$ be identical, $V$ in (\ref{eq:Vgen}) would be invariant
by the whole chiral group $U(4)_L \times U(4))_R$ acting according to
(\ref{eq:group}) on quarks. This invariance is broken because the
normalizations are different. The only invariance left is the chiral $SU(2)_L
\times SU(2)_R$ because the quadruplets are complex doublets of both left
and right $SU(2)$. This is why this chiral/gauge group and its breaking
is the most  important in this kind of physics. When one evokes ``chiral
symmetry breaking'' one generally thinks of this chiral group; its breaking
by quark condensates down to the custodial $SU(2)$ generates 3 Goldstones
inside each quadruplet which, for 1 generation, are the pions.

The Yukawa Lagrangian (\ref{eq:Lyuk0}) is by construction invariant by $SU(2)_L$ 
but, because $\delta_{i\hat\imath}\not=\kappa_{\hat\imath i}$, it is not
invariant by the chiral $SU(2)_L \times SU(2)_R$. This is why it is fitted
to give soft masses to the Goldstones of the broken chiral symmetry which
are not the 3 Goldstones of the broken gauge symmetry.

\subsection{Left or right?}\label{subsec:lr}

Another important point is that such an extension of the GSW model can be a
right-handed $SU(2)_R$ as well as left-handed $SU(2)_L$ gauge theory. This
because all quadruplets that we have built are stable by both groups and
are isomorphic to complex $SU(2)_L$ by the laws of transformations
 (\ref{eq:ruleL}) and/or $SU(2)_R$ doublets by the laws of transformation
 (\ref{eq:ruleR}).

Then, why not a $SU(2)_L \times SU(2)_R$ spontaneously broken
 gauge theory? For 1 generation,
the answer is clear: with only 4 pseudoscalar and 4 scalars
 one has not enough degrees of freedom to provide at
the same time 6 Goldstones and 3 physical pions: the latter should vanish
from the spectrum, together with the $\eta$ and 2 charged scalars.
However this argument is no longer valid for more generations. We have
already mentioned the fact that $\eta$ being a Goldstone is very unphysical
in the case of 1 generation, but that for 2 generations, the role is played
by another diagonal pseudoscalar meson etc \ldots
 The only
necessity is to provide 6 true Goldstones that vanish from the spectrum.
For 2 generations is seems hard to achieve, but this may be kept in mind
for 3 generations, where in particular the physics of $\bar q(\gamma_5) t$
bound states, with expected strong coupling, could be very rich (and at the
same time difficult to handle).

At this stage one can only state that, by its scalar structure,
 the extension that we propose has the potential
to also be extended to a left-right gauge theory.

\subsection{Parity and its breaking}\label{subsec:parity}

The action of the $U(1)_L$ and $U(1)_R$ {\em generators} ${\mathbb I}_L$
and ${\mathbb I}_R$ on the quadruplets
is interesting since, like for 1 generation,
it is directly connected with parity. Indeed, for any pair $\Delta_i,
\hat\Delta_i, i\in [X,H,\Omega,\Xi]$ of quadruplets, one has, like for 1
generation (see (\ref{eq:par1gen})),
\begin{equation}
{\mathbb I}_L. \frac{\sqrt{2}\mu_i^3}{v_i}\Delta_i =
-\frac{\sqrt{2}\hat\mu_i^3}{\hat v_i}\hat\Delta_i ,\quad
{\mathbb I}_R. \frac{\sqrt{2}\mu_i^3}{v_i}\Delta_i =
+\frac{\sqrt{2}\hat\mu_i^3}{\hat v_i}\hat\Delta_i .
\label{eq:parity}
\end{equation}

The group $U(1)_L \times U(1)_R$ is a subgroup of both the vertical and
horizontal  $U(2)_L \times U(2)_R$ chiral groups.
Its diagonal subgroup is the trivial
 $U(1)$ with generator the
$4\times 4$ identity matrix $\mathbb I$, which is just multiplying $\psi$
by a phase.

One of the most conspicuous aspect of parity violation is that the 2
charged longitudinal $W^\pm_\parallel$ are scalars while the neutral
$W^3_\parallel$ is pseudoscalar.

Would parity be an unbroken symmetry, the VEV's of parity transformed
doublets $<{\mathfrak s}^0 + {\mathfrak p}^3>$ and $<{\mathfrak p}^0 +
{\mathfrak s}^3>$ would be identical, which is not the case.

\subsection{The generation group of transformations}\label{subsec:gentrans}

We are going here to give more information about the chiral
 generation group $SU(2)^g_L \times SU(2)^g_R$ with generators $\vec L$ given
in (\ref{eq:Ls}).

The action of the generators of its diagonal subgroup on quarks has been
given in (\ref{eq:hordiag}).

At the level of bilinear quark operators, using the laws of transformations
(\ref{eq:trans2}), one gets the following results (we forget below about
the normalizations of the quadruplets).

By the action of  $SU(2)^g_L$, the 8 quadruplets split
into the 4 following doublets
\begin{equation}
\left(\begin{array}{c}
X-\hat X   \cr \Omega-\hat\Omega +(\Xi-\hat\Xi)
\end{array}\right),\quad
\left(\begin{array}{c}
\Omega -\hat\Omega-(\Xi-\hat\Xi)\cr H-\hat H
\end{array}\right),\quad
\left(\begin{array}{c}
H+\hat H   \cr \Omega+\hat\Omega +(\Xi+\hat\Xi)
\end{array}\right),\quad
\left(\begin{array}{c}
\Omega+\hat\Omega-(\Xi+\hat\Xi)   \cr X+\hat X
\end{array}\right),
\label{eq:gentransL}
\end{equation}
while by  $SU(2)^g_R$ they split into the  4 following ones
\begin{equation}
\left(\begin{array}{c}
X+\hat X   \cr \Omega+\hat\Omega +(\Xi+\hat\Xi)
\end{array}\right),\quad
\left(\begin{array}{c}
\Omega +\hat\Omega-(\Xi+\hat\Xi)\cr H+\hat H
\end{array}\right),\quad
\left(\begin{array}{c}
H-\hat H   \cr \Omega-\hat\Omega +(\Xi-\hat\Xi)
\end{array}\right),\quad
\left(\begin{array}{c}
\Omega-\hat\Omega-(\Xi-\hat\Xi)   \cr X-\hat X
\end{array}\right),
\label{eq:gentransR}
\end{equation}
which contain states of mixed $1\pm\gamma_5$ parity.
In both (\ref{eq:gentransL}) and (\ref{eq:gentransR}) the upper and lower
 components have respectively $+\frac12$ and $-\frac12$ quantum numbers
with respect to $L^3_L$ and $L^3_R$. One moves inside each doublet by the
action of the raising/lowering operators $L^+, L^-$. Moving inside the
whole 8-dimensional space of quadruplets requires sequences of alternate
left and right transformations.

With respect to the diagonal $SU(2)^g$ the 8 multiplets split into
2 singlets and 2 triplets which contain states of given parity
\begin{equation}
(X+H),\quad (\hat X+\hat H),\quad
\left(\begin{array}{c}
\Omega+\Xi \cr
X-H\cr
\Omega-\Xi
\end{array}\right),\quad
\left(\begin{array}{c}
\hat\Omega+\hat\Xi\cr
\hat X-\hat H\cr
\hat\Omega-\hat\Xi\end{array}\right).
\label{eq:gentransdiag}
\end{equation}
It should be clear that, in (\ref{eq:gentransL}), (\ref{eq:gentransR}) and
(\ref{eq:gentransdiag}) above, we are dealing with multiplets of
quadruplets.

The generation group reproduces therefore, in the space of quadruplets,
the symmetry pattern that the chiral/gauge group created inside each
quadruplet: doublets for left and right groups, singlet + triplet for the
diagonal group.
 
In (\ref{eq:gentransdiag}), let us remark that the neutral pseudoscalar
entries of $\Omega\pm\Xi$, $\hat\Omega\pm\hat\Xi$ entries match $K^0 \pm
\bar K^0$ and $D^0 \pm \bar D^0$, but that their charged pseudoscalar
entries are $K^\pm \pm D^\pm$. This is to be contrasted with $\Omega, \Xi,
\hat\Omega, \hat\Xi$ which separate charged $D$ and $K$ mesons but mix the
neutral ones.

\section{The mass pattern of neutral scalars}
\label{section:scalpattern}

Unlike pseudoscalar mesons which are built from quark mass eigenstates,
 these states are by construction
flavor $\bar q_i q_j$ eigenstates.

Despite our lack of information concerning the $\Xi^0$ and $\hat \Xi^3$
Higgs bosons, they are presumably light and, at least from perturbative
arguments, quasi-degenerate. So,  the mass pattern
of neutral scalars found in section \ref{section:higgsspectrum}
 clearly exhibits a splitting into
1 heavy triplet $(\hat H^3, X^0, \hat X^3)$, 2 light doublets  $(\Xi^0,
\hat \Xi^3), (\Omega^0, \hat\Omega^3)$ and 1 singlet $H^0$
with intermediate mass. We  also notice that, going from $\theta_u=0$ to
$\theta_u\not=0$, 2 mass scales have been swapped, but that the structure
into 1 triplet, 2 doublets and 1 singlet has been seemingly preserved.
Some $SU(2)$ symmetry can therefore be suspected to be at work.

One one side, this comes as  a surprise because of the apparent anarchy
in the spectrum of scalar mesons that appears in experimental data
\cite{PDG}.  On the
other side, there should a priori be no reason why $SU(2)$ or $SU(3)$
{\em ``rotated''} flavor symmetry operates on pseudoscalar
 $\bar q^i_m \gamma_5 q^j_m$ bound states of quark mass eigenstates
while their  unrotated scalar
partners keep insensitive to a similar symmetry.
A noticeable difference is however that this result came out only after
rather long calculations and could hardly have been guessed from the start.

The question that arises is the nature of this symmetry. As far as
pseudoscalar mesons are concerned it is usual to consider that it is the
diagonal-custodial $SU(2)$. Indeed, for 1 generation, the breaking of
$SU(2)_L \times SU(2)_R$ down to the diagonal $SU(2)$ is invoked to yield
the 3 pions as Goldstone bosons. It is comforted by the fact that
the 4 components of any quadruplet split into a neutral singlet plus a
triplet of the custodial $SU(2)$, which perfectly fits the pion case; the 3
states correspond to the quantum numbers $(-1,0,+1)$ of $T^3$. $T^\pm$ are
the raising-lowering operators that move inside the pion triplet. For 1
generation, the diagonal flavor $SU(2)$ coincides with the strong $SU(2)$
of isospin which operates, at the quark level, in the $(u,d)$ space.

As far as scalar mesons are concerned, the situation looks more
intricate. Indeed, the components of the doublets and the triplet
into which they split belong to different quadruplets:
$(X, \hat X, \hat H)$ for the triplet, $(\Omega, \hat\Omega)$ and $(\Xi,
\hat\Xi)$ for the doublets. So, obviously, pairs of parity-transformed
quadruplets are involved, and, accordingly, the chiral $U(1)_L \times
U(1)_R$ symmetry which we have shown to be  tightly related to parity plays
a role.
Would the quasi-standard Higgs $\hat H^3$ be in reality a singlet and its
degeneracy with $X^0$ and $\hat X^3$ purely accidental, one could think
that the 3 doublets include pairs of bosons belonging to parity transformed
quadruplets, which one could interpret as a left over of parity symmetry. But
this does not fit $H^0$ and $\hat H^3$  which seem largely split, at least
for 2 generations.

The solution may lie in (\ref{eq:Laction}). While {\em the} vacuum of the
theory is defined by the 8-set of VEV's $\{<X^0>, <\hat X^3>, <H^0>, <\hat
H^3>, <\Omega^0>, <\hat\Omega^3>, <\Xi^0>, <\hat\Xi^3>\}$, the effective
scalar potential $V_{eff}$ that we minimize to get the Higgs masses is
split into 4 terms involving only pairs of parity-transformed
quadruplets.  We expect accordingly that the symmetry group in relation
with
any such pair of Higgs bosons  is the one leaving invariant the corresponding
part of $V_{eff}$. The  unbroken subgroup corresponds therefore
to the generators that annihilate the relevant pair of vacuum states:
$L^1$ for $(\Omega, \hat\Omega)$, $L^2$ for $(\Xi, \hat\Xi)$,
$L^3$ for both $(X,\hat X)$ and $(H, \hat H)$.
The Higgs bosons are then expected to split into
$SU(2)^g$ multiplets labeled by the  quantum numbers of $L^1$, $L^2$ and
$L^3$. For the first two pairs, they can be only doublets (or singlets), for the
last two pairs, the corresponding
 4 Higgs bosons could split into singlets, doublets of triplet.
Our calculations show that they seemingly fall into 1 singlet + 1 triplet.
Therefore, the mass pattern of Higgs boson appears as  the left imprint
of the broken generation symmetry.

Now, why such a mass pattern has not been detected?\newline
 For the 3 heaviest
states, because they correspond to 3 quasi-degenerate Higgs bosons at the
same mass as the ``quasi-standard'' Higgs. First, for 2 generations, it is
not yet heavy enough and our results are non-physical, yet. Secondly, as we
shall see in the next work, the 2 partners of $\hat H^3$ are much more
weakly coupled and therefore hard to detect.\newline
As far as the lightest scalars are concerned, as we shall also show in a
subsequent work, they are also certainly very difficult to detect because
they are lighter than the pions, extremely weakly coupled to leptons.
Furthermore, they are plagued with non-perturbative couplings
to hadronic matter, which makes theoretical predictions hazardous.\newline
One is left with 1 neutral scalar at an intermediate mass, around $1.6\,GeV$.
Though the spectrum will certainly be modified when a 3rd generation is
added, one can already notice that some states like $f^0$ of $K^{*0}$
fall in this mass range. 

One must also be careful not to draw too fast conclusions because the
 mass pattern that we have obtained may also be
 the consequence of our choice for the scalar potential
and for the Yukawa Lagrangian. We recall that we did not consider the most
general,  but, for the latter, we wanted to cancel  from the start FCNC's,
and for the former we wished to satisfy general requirements concerning, in
particular chiral symmetry. The most general expressions would include a
dramatically large number of parameters, trigger many unwanted processes, and their
systematic treatment would certainly be, in practice, unfeasible.
So, maybe  we have only shown  that such a construction is feasible, and that it
leads to interesting consequences that have not been obtained in other frameworks.
After all, it nature is not simple, what can we do ?

\section{Outlook}
\label{section:conclusion}

\subsection{Miscellaneous remarks} \label{subsec:comments}

$\bullet$\ This work is nothing more than a fit of (a small part of)
 the physical world at
low energy (pseudoscalar mesons, gauge bosons) by an extension of
the GSW model which is the smallest one that can naturally
incorporate all known mesons. It was not evident from the
start that a solution existed and could be found rather simply, but it does
exist,
and we have reasonable hope that the very few problems which are left
(orthogonality between some neutral mesons, large quark condensates)
will find a solution when one more generation is added.
Arguments that led, in the past, to the introduction of heavy fermions and
technicolor-like theories \cite{Susskind} become void;
heavy fermions are not needed and enough energy
scales and VEV's are available to describe low energy physics up to the
electroweak scale.

$\bullet$\ The cornerstone of the whole construction is 
 the one-to-one correspondence demonstrated in section
\ref{section:bijection}
between the complex Higgs doublet of the GSW model and some
very specific quadruplets of bilinear quark operators that mix states of
different parities (scalars and pseudoscalars). This set of quadruplets
split into two parity transformed subsets of the type $({\mathfrak s},
\vec{\mathfrak p})$ and $({\mathfrak p}, \vec{\mathfrak s})$.
This type of configuration was never investigated before but comes
out naturally as soon as the bijection is uncovered. In
particular, even for 1 generation, 2 Higgs doublets are necessary, which
are parity-transformed of each other. For $N$
generations, this number grows to $2N^2$.

$\bullet$\ The underlying operating symmetries are all subgroups of the
chiral $U(2N)_L \times U(2N)_R$. 
Parity has been tightly linked to the chiral $U(1)_L \times U(1)_R$ group,
which is broken to the trivial diagonal $U(1)$. A ``generation''
$SU(2)^g_R\times SU(2)^g_L$  group has also been identified, which operates in
the space of quadruplets; its diagonal part flips generations at
the quark level.

$\bullet$\ The  Higgs bosons are composite states that fit into the family
of $J=0$ mesons built with 2 quarks. They exhibit quasi-degeneracies, which
appear as the imprint of the broken symmetry among generations.
The mass of the ``quasi-standard'' Higgs boson increases like that of the
heaviest pseudoscalar meson. This puts it at the order of the $D_s$ scale for 2
generations.

$\bullet$\ Naturalness.\newline
The point of view that we adopt is that a theory is natural if and only if it
describes nature as we observe it.  Our extension of the GSW model has been
tailored to this purpose. The only assumptions that have been made
concern the scalar potential and Yukawa Lagrangian.
Since they do not spoil the description of
physical reality, we consider them as natural. More general potentials of
Yukawa Lagrangian are conceivable, but at the price of loosing
much elegance and simplicity.

$\bullet$\ Light Higgses and dark matter.\newline
An important  phenomenon is the appearance of  light scalars. This shows
that there are other motivations as the ones exposed for example exposed
in \cite{Barbieri} to look for such particles. It is also
a highly fine tuned sector. 
Only a careful study of their couplings can tell whether they can have
escaped detection up to now. If yes, they role as one of the possible
components of dark matter \cite{dark} should of course be
carefully scrutinized.

$\bullet$\ Quark condensation and strongly coupled light
scalars.\newline
Light scalars happen to be extremely weakly coupled to
everything except to hadronic matter, to which they are strongly coupled
\cite{Machet5}. This is likely (see \ref{subsec:normalization})
to be the signal of the existence of  massive
hadronic bound state(s). Then, their normalization plays a crucial role
in their physical coupling to the rest of the world.
The mass of hadronic bound states, if we think for example of pions, goes
 along with
chiral symmetry breaking and quark condensation,  phenomena  commonly
attributed to strong interactions, that is, in the absence of no other
candidate, to gluons. Now, in the case where quarks get also strongly
coupled to light scalars, it is  worth investigating whether this
could  be another source of  vacuum
instability and quark condensation. We have of course specially in mind the
condensation of heavy quarks, which seem to be needed.

\subsection{A prelude to forthcoming works : nature and normalization of
 asymptotic states}
\label{subsec:normalization}

After the whole set of equations is solved, we can in principle calculate
all couplings that occur  in the Lagrangian as functions of physical
masses and measured parameters.
However, the {\em physical} couplings, in particular between mesons (which
includes Higgs bosons)  and quarks,
crucially depend on the normalization of the former.

To make this easily understandable,
let us take once again (see section \ref{section:intro}) the example of the
charged pion, which occurs inside the $X$ quadruplet (\ref{eq:Xq2gen}).
The kinetic terms (\ref{eq:kin2gen})
 for $X^\pm$ are normalized to $1$ and the corresponding 
Yukawa couplings, given by (\ref{eq:Lyuk0}), include
 $\delta_X\; X^- \left( \frac{v_X}{\sqrt{2}\mu_X^3}
\sqrt{2} \bar u\gamma_5 d\right) + \ldots$ with $\sqrt{2}\mu_X^3 = <\bar u
u+\bar d d>$.  
 $\delta_X=\delta(1-b_X)$ by (\ref{eq:delrel1}),
$v_X \approx 151\,GeV$ and $\mu_X^3$, given by the GMOR relation,
close to $\sqrt{2}f_\pi^2 m_{\pi^+}^2/(m_u+m_d)$. This yields, using
(\ref{eq:bxprec}) a  coupling 
$\delta \frac{b_X(1-b_X) \hat v_H (m_u+m_d)}{\sqrt{2}f_\pi^2 m_\pi^2} X^-
\bar u\gamma_5d \approx -29\; X^- \bar u\gamma_5 d$
which we do not know a priori how to handle
\footnote{Of course, it is also fine-tuned and depends on how close $b_X$
is to $1$.}.
More information is however available from Current Algebra
 because  $X^-$ is connected to the
 charged pion by (\ref{eq:example2}) and that,
accordingly, the Yukawa coupling rewrites
$\delta_X \frac{v_X}{f_\pi}(\cos\theta_c \pi^-
+\ldots)\frac{v_X}{\sqrt{2}\mu_X^3}(\cos\theta_c \sqrt{2} \bar u_m \gamma_5 d_m
+\ldots)$ inducing a coupling of $\pi^-$ to quarks equal to
$\delta(1-b_X)\frac{v_X^2}{f_\pi \mu_X^3} \cos^2\theta_c \pi^- \bar u_m\gamma_5 d_m$.
It is also $\gg1$, but 
we did not yet take into account  the normalization of the charged
pions. The kinetic terms (\ref{eq:kin2gen}) for the $X$ Higgs multiplet
 induce indeed  pion kinetic terms\break
$\partial_\mu(\frac{v_X}{f_\pi}\cos\theta_c\pi^-) \partial ^\mu(
\frac{v_X}{f_\pi}\cos\theta_c\pi^+)$
and normalizing the pions back to  $1$ amounts
to dividing the kinetic + Yukawa Lagrangian by $v_X^2\cos^2\theta_c/f_\pi^2$.
This rescales the coupling between  charged pions and quarks to\break
 $\delta(1-b_X) \frac{f_\pi}{\mu_X^3}  \pi^- \bar u_m \gamma_5
d_m \approx \frac{\delta(1-b_X)(m_u+m_d)}{\sqrt{2}f_\pi m_\pi^2}\pi^-\,\bar
u\gamma_5 d \approx -.018\, \pi^=\,\bar u\gamma_5 d$
 which is much smaller (in modulus) that $1$
\footnote{In the simple case
of 1 generation (see chapter \ref{chapt:1gen})
 $\theta_c=0$,  $\delta_X=m_\pi^2$ and $v_X=f_\pi$: the
normalization factor $v_X/f_\pi$ is $1$ and the Yukawa couplings
exactly match  the mass term $m_\pi^2 \pi^+ \pi^-$ for  charged pions
(fulfilling the PCAC and GMOR low energy relations), to which only the $X$
quadruplet contributes. For 2 generations,  Yukawa
couplings rewrite in their bosonised form
$(\frac{v_X\cos\theta_c}{f_\pi})^2 \delta(1-b_X)\pi^+\pi^-$ while kinetic
terms rewrite $(\frac{v_X\cos\theta_c}{f_\pi})^2\partial_\mu\pi^+
\partial^\mu \pi^-$; $\sqrt{\delta(1-b_X)}\approx.134\,GeV$ 
is now only close to the pion mass since the pion receives contributions
to its mass from several quadruplets.}. The {\em physical} coupling of pion
to quarks is smaller than $1$, but we could only calculate it with the help
of low energy relations.\newline
This simple example shows that, if the asymptotic
 bosonic states are not correctly identified
(like we did first, considering that $X^\pm$ are asymptotic states), or if
they are not suitably normalized (like we did in the second case when we
ignored the normalization of the pions), their couplings to quarks that
occur in the genuine Yukawa Lagrangian are not the physical ones.
The normalization plays  a specially important role  because the
kinetic terms (\ref{eq:kin2gen}) are quadratic in the mesonic fields
while the genuine Yukawa couplings (\ref{eq:Lyuk0}) are linear. \newline
It can therefore occur that the physical coupling of a meson 
to quarks can be much smaller than the one naively read on the 
Yukawa Lagrangian.
For pions, luckily enough we  know the normalization factor from
elementary Current Algebra considerations, but, for scalar mesons, the situation
is more intricate
\footnote{The estimates of the decays
$V\to \gamma + light\  scalar$ in \cite{VHgamma} have to be re-investigated since
``standard'' Yukawa couplings between scalars and quarks have been used.
At the end of its
paper, Wilczek warns  that it is only valid in the strict framework of the
GSW model with a light Higgs boson.}.

\section{Prospects}\label{section:prospects}

Very natural composite extensions of the
GSW model exist, with no extra fermion,
 which are very likely to suitably describe known
physics up to the electroweak scale. At least they provide an optimal set
of parameters to fit basic physics from the chiral scale
up to the electroweak scale. The Higgs spectrum is much
richer than in the GSW model and  exhibits  quasi-degeneracies.
It the trend observed for 1 and 2 generations persists,
the mass of the quasi-standard Higgs boson will continue to grow like that of the
heaviest pseudoscalars.
A general phenomenon seems also to be the occurrence of light neutral
scalars. All this makes  the Higgs sector  a privileged place to look
for new physics \cite{Machet5}. Since, for example, 
$W$'s and quarks get their masses from several Higgs bosons,  their couplings 
cannot be standard any more.
It is in particular important to  re-investigate in this framework
the so-called ``decoupling theorem'' \cite{decoupling}.
It is indeed disquieting that, in the GSW model, very heavy fermions, which
should behave like quasi-classical objects, do no decouple and even
have large quantum effects (loops), which manifest themselves, for example,
in the decay $Higgs \to \gamma\gamma$.
At the same time, the GSW model should be recovered in a suitable limit (probably,
according to what we learned up to now, at the price of a lot of fine tuning).

For many reasons, it is  highly desirable to introduce
the 3rd generation and
to look for  a solution of the system of equations that would determine
the $18$ bosonic VEV's, $18$ fermionic VEV's, $3\times 18$ Yukawa
couplings, more than $6$ mixing angles, the masses of the 18 Higgs bosons {\em etc}.
This is a very hard task, all the more as none of the 6 mixing angles
(forgetting $CP$ violating phases) can safely be turned to $0$.
The example of 2 generations has moreover shown, by successive trials and errors,
 that these equations can only be solved in a carefully devised order,
and that no global / numerical solution can be trusted. Reasons for this
is that the equations involve fast varying functions of potentially
complex variables, with poles here and there  which make
calculations highly unstable, and, on a general ground, that
solving consistently a system of more  than 30 equations including trigonometric
functions is highly non-trivial. It looks therefore hopeless to solve them
like we did for  2 generations.

We end this incomplete work by an (also incomplete) list of  goals to achieve:\newline
* find the spectrum of Higgs bosons for 3 generations and all their couplings;
 confirm in particular that one of them weights around 125 GeV;\newline
* confirm (or infirm) the presence of light scalars;\newline
* find reasonable arguments for  a large $<\bar t t>$
condensate, presumably in relation with the strong coupling of light Higgses to
quarks;\newline
* have a closer look at the $\eta$ (and so \ldots)
 mesons and their orthogonality to other neutral pseudoscalars;\newline
* calculate all mixing angles in terms  of bosonic and fermionic
quantities; check the matching (or not) between the two sets of relations;\newline
* include $CP$ violating effects;\newline
* investigate whether and how light scalars could be detected,
actualize the corresponding
 calculation of a vector particle $V$ decaying into $\gamma$
+ a light scalar \cite{VHgamma};\newline
* re-analyze the data of radiative $J/\Psi$ and $\Upsilon$ decays
\footnote{Analysis have been mostly done for a $CP$-odd light scalar in the
final state in connexion with supersymmetric extensions of the GSW model.}
\cite{CLEOBABAR};\newline
* study these light scalars as possible components of dark matter
\cite{dark};\newline
* study their roles in other phenomena, like the 1/2 leptonic decays of
$\pi^+$, the  $g-2$ of the muon
\cite{g-2} or the proton radius \cite{atomic};\newline
* after getting the physical couplings of quarks to gauge and
to Higgs bosons,
revisit the limit of very heavy fermions \cite{decoupling}
  and compare with the GSW model;\newline
* recover the GSW model at a certain limit.

\vskip .5cm
{\em \underline{Acknowledgments}: It is a pleasure to thank V.A. Novikov and M.I.
Vysotsky, who always ask the right questions at the right time. I also
thank O.~Babelon, S.~Davidson,  G.~Moultaka and P.~Slavich for their constructive remarks.}


\begin{em}

\end{em}


\newpage

\appendix


\section{Appendix : Flavor quark bilinears expressed in terms of  quark
mass eigenstates}\label{section:flavmass}

We write below the formul{\ae} expressing the flavor bilinear quark
operators in terms of bilinears of quark mass eigenstates and of the 2
mixing angles $\theta_u$ and $\theta_d$. They are largely used throughout
the paper. The formul{\ae} for scalar and pseudoscalar bilinears are of
course similar.

\begin{equation}
\begin{split}
\ast\ &\bar u\gamma_5 d = c_uc_d\, \bar u_m\gamma_5 d_m +
c_us_d\, \bar u_m\gamma_5 s_m +
s_uc_d\, \bar c_m\gamma_5 d_m + s_us_d\, \bar c_m\gamma_5 s_m,\cr
\ast\ &\bar u\gamma_5 u -\bar d\gamma_5 d
= \frac{c_u^2+c_d^2}{2}(\bar u_m \gamma_5 u_m-\bar d_m \gamma_5 d_m)
 +\frac{c_u^2-c_d^2}{2}(\bar u_m \gamma_5 u_m+\bar d_m \gamma_5 d_m)\cr
&\hskip 1cm +\frac{s_u^2+s_d^2}{2}(\bar c_m \gamma_5 c_m-\bar s_m \gamma_5
s_m)
+\frac{s_u^2-s_d^2}{2}(\bar c_m \gamma_5 c_m+ \bar s_m \gamma_5 s_m)\cr
&\hskip 1cm + s_uc_u(\bar u_m \gamma_5 c_m + \bar c_m \gamma_5 u_m)
- s_dc_d(\bar d_m \gamma_5 s_m + \bar s_m \gamma_5 d_m)\cr
\ast\ &\bar u\gamma_5 u +\bar d\gamma_5 d
= \frac{c_u^2+c_d^2}{2}(\bar u_m \gamma_5 u_m+\bar d_m \gamma_5 d_m)
 +\frac{c_u^2-c_d^2}{2}(\bar u_m \gamma_5 u_m-\bar d_m \gamma_5 d_m)\cr
&\hskip 1cm +\frac{s_u^2+s_d^2}{2}(\bar c_m \gamma_5 c_m+\bar s_m \gamma_5
s_m)
+\frac{s_u^2-s_d^2}{2}(\bar c_m \gamma_5 c_m- \bar s_m \gamma_5 s_m)\cr
&\hskip 1cm + s_uc_u(\bar u_m \gamma_5 c_m + \bar c_m \gamma_5 u_m)
+ s_dc_d(\bar d_m \gamma_5 s_m + \bar s_m \gamma_5 d_m).
\end{split}
\end{equation}

\begin{equation}
\begin{split}
\ast\ &\bar c\gamma_5 s =s_us_d\, \bar u_m\gamma_5 d_m - s_uc_d\,
\bar u_m\gamma_5 s_m -
c_us_d\, \bar c_m\gamma_5 d_m + c_uc_d\, \bar c_m\gamma_5 s_m,\cr
\ast\ &\bar c\gamma_5 c -\bar s\gamma_5 s
= \frac{c_u^2+c_d^2}{2}(\bar c_m \gamma_5 c_m-\bar s_m \gamma_5 s_m)
 +\frac{c_u^2-c_d^2}{2}(\bar c_m \gamma_5 c_m+\bar s_m \gamma_5 s_m)\cr
&\hskip 1cm +\frac{s_u^2+s_d^2}{2}(\bar u_m \gamma_5 u_m-\bar d_m \gamma_5
d_m)
+\frac{s_u^2-s_d^2}{2}(\bar u_m \gamma_5 u_m+ \bar d_m \gamma_5 d_m)\cr
&\hskip 1cm - s_uc_u(\bar u_m \gamma_5 c_m + \bar c_m \gamma_5 u_m)
+ s_dc_d(\bar d_m \gamma_5 s_m + \bar s_m \gamma_5 d_m),\cr
\ast\ &\bar c\gamma_5 c +\bar s\gamma_5 s
= \frac{c_u^2+c_d^2}{2}(\bar c_m \gamma_5 c_m+\bar s_m \gamma_5 s_m)
 +\frac{c_u^2-c_d^2}{2}(\bar c_m \gamma_5 c_m-\bar s_m \gamma_5 s_m)\cr
&\hskip 1cm +\frac{s_u^2+s_d^2}{2}(\bar u_m \gamma_5 u_m+\bar d_m \gamma_5
d_m)
+\frac{s_u^2-s_d^2}{2}(\bar u_m \gamma_5 u_m- \bar d_m \gamma_5 d_m)\cr
&\hskip 1cm - s_uc_u(\bar u_m \gamma_5 c_m + \bar c_m \gamma_5 u_m)
- s_dc_d(\bar d_m \gamma_5 s_m + \bar s_m \gamma_5 d_m).
\end{split}
\end{equation}

\begin{equation}
\begin{split}
\ast\ &\bar u \gamma_5 s + \bar c \gamma_5 d
= -s_{u+d}(\bar u_m \gamma_5 d_m - \bar c_m \gamma_5 s_m)
+ c_{u+d}(\bar u_m \gamma_5 s_m + \bar c_m \gamma_5 d_m),\cr
\ast\ &( \bar u\gamma_5 c + \bar c\gamma_5 u)-(\bar d\gamma_5 s
+\bar s\gamma_5 d)
= c_{2u} (\bar u_m \gamma_5 c_m + \bar c_m\gamma_5 u_m)
- c_{2d}(\bar d_m\gamma_5 s_m + \bar s_m\gamma_5 d_m)\cr
& -\frac{s_{2u}+s_{2d}}{2}(\bar u_m\gamma_5 u_m-\bar
d_m\gamma_5 d_m)
 -\frac{s_{2u}-s_{2d}}{2}(\bar u_m\gamma_5 u_m+\bar d_m\gamma_5 d_m)
+s_{2u}\bar c_m\gamma_5 c_m -s_{2d}\bar s_m\gamma_5 s_m,\cr
\ast\ &( \bar u\gamma_5 c + \bar c\gamma_5 u)+(\bar d\gamma_5 s +\bar s\gamma_5 d)
=  c_{2u} (\bar u_m \gamma_5 c_m + \bar c_m\gamma_5 u_m)
+ c_{2d}(\bar d_m\gamma_5 s_m + \bar s_m\gamma_5 d_m)\cr
& -\frac{s_{2u}-s_{2d}}{2}(\bar u_m\gamma_5 u_m-\bar
d_m\gamma_5 d_m)
 -\frac{s_{2u}+s_{2d}}{2}(\bar u_m\gamma_5 u_m+\bar d_m\gamma_5 d_m)
+s_{2u}\bar c_m\gamma_5 c_m +s_{2d}\bar s_m\gamma_5 s_m.
\end{split}
\end{equation}

\begin{equation}
\begin{split}
\ast\ &\bar u \gamma_5 s - \bar c \gamma_5 d
= s_{u-d}(\bar u_m \gamma_5 d_m +\bar c_m \gamma_5 s_m)
+ c_{u-d}(\bar u_m \gamma_5 s_m-\bar c_m \gamma_5 d_m),\cr
\ast\ &(\bar u\gamma_5 c -\bar c\gamma_5 u)- (\bar d\gamma_5 s -\bar s\gamma_5 d)
= (\bar u_m\gamma_5 c_m -\bar c_m\gamma_5 u_m)- (\bar d_m\gamma_5 s_m
-\bar s_m\gamma_5 d_m),\cr
\ast\ &(\bar u\gamma_5 c -\bar c\gamma_5 u)+ (\bar d\gamma_5 s -\bar s\gamma_5 d)
= (\bar u_m\gamma_5 c_m -\bar c_m\gamma_5
u_m)+ (\bar d_m\gamma_5 s_m -\bar s_m\gamma_5 d_m).
\end{split}
\end{equation}

\vskip 1cm

{\baselineskip 5pt
\centerline{\rule{7cm}{1pt}}
}

\end{document}